\documentclass[aps, prd, reprint, notitlepage, a4paper, floats, amsmath, amssymb, amsfonts,superscriptaddress,showpacs, showkeys,nofootinbib,longbibliography]{revtex4-2}

\pdfoutput=1

\usepackage{graphicx}
\usepackage[usenames,dvipsnames]{xcolor}
\usepackage[normalem]{ulem}

\definecolor{dodgerblue}{HTML}{1E90FF}
\definecolor{viennared}{HTML}{DA0A14}
\usepackage[
	bookmarksnumbered, bookmarksopen, bookmarksopenlevel=2,
	breaklinks=true,
	colorlinks=true, filecolor=viennared, citecolor=dodgerblue,
	linkcolor=viennared, urlcolor=dodgerblue, menucolor=viennared,
]{hyperref}
\usepackage[english]{babel}
\allowdisplaybreaks[1]
\usepackage[utf8]{inputenc}
\usepackage{lmodern}
\usepackage{booktabs}
\usepackage{diagbox}
\usepackage{mathrsfs}
\usepackage{mathtools}
\usepackage{multirow}

\allowdisplaybreaks

\graphicspath{{fig/}}

\def\mF{\mathcal{F}}

\def\pphi{p_\varphi}
\def\pphidot{\dot{p}_\varphi}

\newcommand{\AEI}{\affiliation{Max Planck Institute for Gravitational Physics (Albert Einstein Institute), Am M\"uhlenberg 1, Potsdam 14476, Germany}}
\newcommand{\Maryland}{\affiliation{Department of Physics, University of Maryland, College Park, MD 20742, USA}}
\newcommand{\COG}{\affiliation{Center of Gravity, Niels Bohr Institute, Blegdamsvej 17, 2100 Copenhagen, Denmark}}
\newcommand{\URI}{\affiliation{Department of Physics and Center for Computational Research, University of Rhode Island, Kingston, RI 02881, USA}}  
\newcommand{\UMassDPhy}{\affiliation{Department of Physics and Center for Scientific Computing and Data Science Research, University of Massachusetts, Dartmouth, MA 02747, USA}}

\def\mF{\mathcal{F}}

\newcommand{\hamp}[1]{ \mathsf{h}_{#1}}
\newcommand{\dhamp}[1]{ \dot{\mathsf{h}}_{#1}}
\newcommand{\ddhamp}[1]{ \ddot{\mathsf{h}}_{#1}}

\begin{document}

\title{Modeling the merger-ringdown of an eccentric test-mass inspiral into a Kerr black hole using the effective-one-body framework}

\author{Guglielmo Faggioli}\email{guglielmo.faggioli@aei.mpg.de}
\AEI
\author{Alessandra Buonanno}
\AEI
\Maryland
\author{Maarten van de Meent}
\COG
\AEI
\author{Gaurav Khanna}
\URI
\UMassDPhy

\date{\today}

\begin{abstract}
We characterize and phenomenologically model the merger-ringdown of gravitational waves emitted by a small compact object that plunges and merges into a Kerr black hole from equatorial-eccentric inspirals. The waveforms are generated employing a time-domain Teukolsky code sourced with trajectories computed using the effective-one-body framework. 
We span values of the Kerr spin $a\in[-0.9, 0.9] $, eccentricity at the last stable orbit (LSO) $ e_{\rm LSO} \in [0,0.9] $, and relativistic anomaly $ \xi_{\rm LSO} \in [0 , 2 \pi]$.
We characterize the last peak of the waveform and ringdown features across the parameter space, finding that the eccentricity mainly affects the last peak features, while it has a smaller impact on the ringdown signal. In contrast, the relativistic anomaly measured at the LSO influences the morphology of the last peak in a restricted portion of the parameter space and has no impact on the ringdown part. 
We perform the analysis for all the spin-weighted spherical harmonic modes normally included in the \texttt{SEOBNR} family of models,  $(\ell,m)\in\{ (2,2), (3,3), (4,4), (5,5), (2,1), (3,2),  (4,3)\}$. 
Finally, we introduce a merger--ringdown model for \texttt{SEOBE-TML}, a forthcoming inspiral--merger--ringdown waveform model for eccentric spin-aligned binary black holes in the test-mass limit, whose features can be extended to comparable-mass regimes.
The model also accounts for quasinormal mode mixing during the ringdown. It provides a first step toward incorporating the impact of residual eccentricity close to merger into spin-aligned effective-one-body merger--ringdown models for binary black holes.
\end{abstract}

\maketitle

\section{Introduction}
The first observation of gravitational waves (GWs)~\cite{LIGOScientific:2016aoc} made by the LIGO-Virgo collaboration~\cite{LIGOScientific:2018mvr, LIGOScientific:2019lzm, LIGOScientific:2020ibl, LIGOScientific:2021usb} in 2015 marked a major milestone in the field of GW astronomy. Since then, the LIGO-Virgo-KAGRA collaboration~\cite{KAGRA:2013rdx, LIGOScientific:2021djp} has reported more than two hundred detections~\cite{LIGOScientific:2025slb} originating from the mergers of compact binary systems, such as stellar-mass black holes (BHs) and neutron stars~\cite{KAGRA:2021vkt}.
While the rate of detections grows steadily, GW astronomy is entering an era defined not merely by increased number of detections, but also by precision, enabling tests of general relativity~\cite{LIGOScientific:2019fpa, LIGOScientific:2020tif, LIGOScientific:2021sio} and studies on astrophysical formation channels~\cite{Mandel:2009nx, Stevenson:2017tfq, Rodriguez:2018rmd, Fragione:2018vty, Zevin:2021rtf}.

Moreover, forthcoming experiments such as the Einstein Telescope~\cite{Punturo_2010,Abac:2025saz}, Cosmic Explorer~\cite{Evans:2021gyd}, and LISA~\cite{LISA,LISA:2024hlh} are expected to deliver not only a larger number of observations, but also qualitatively new ones, probing lower frequencies, high masses and higher mass ratios. As detector sensitivity improves, the observational reach will extend to rarer and dynamically richer sources that fall outside the assumptions underlying many current waveform models. One clear signature of this increased complexity is orbital eccentricity. While binary BHs (BBHs) formed through isolated binary evolution~\cite{Stevenson:2017tfq} are expected to circularize efficiently before entering the observational band~\cite{Peters:1963ux, Peters:1964zz},
systems that form via dynamical interactions in dense stellar environments like globular clusters
or galactic nuclei~\cite{PortegiesZwart:1999nm, Miller:2001ez}, or through the Kozai-Lidov mechanism in triple systems~\cite{Kozai:1962zz, LIDOV1962719} can retain measurable eccentricity at detection~\cite{Wen:2002km, Samsing:2013kua, VanLandingham:2016ccd, Zevin:2018kzq, Zevin:2021rtf}. Hence, the identification of such systems would provide direct insight into the astrophysical environments in which compact binaries form. Beyond its astrophysical relevance, eccentricity also plays an important role in waveform modeling: even moderate departures from circularity introduce additional structure in the signal, increase the dimensionality of the parameter space, and generate new characteristic frequencies. If not properly accounted for, these effects can bias parameter estimation~\cite{Gupte:2024jfe, Divyajyoti:2023rht, Favata:2013rwa, Ramos-Buades:2019uvh, Cho:2022cdy, Guo:2022ehk, GilChoi:2022waq, Das:2024zib} and limit the accuracy of tests of general relativity~\cite{Saini:2022igm, Saini:2023rto, Narayan:2023vhm, Gupta:2024gun, Shaikh:2024wyn, Bhat:2022amc, Bhat:2024hyb}.
These considerations place increasing demands on waveform modeling.
To fully exploit the scientific potential of forthcoming detections, waveform models must be both computationally efficient and accurate in a high-dimensional parameter space that includes eccentricity, mass-ratio asymmetry, and spin-induced precession.

In recent years, the GWs modeling community has already devoted effort to extending the quasi-circular (QC) state-of-the-art inspiral--merger--ringdown (IMR) waveform models to accurately describe eccentric BBHs, including the effects of BH spins, in different directions.
The semi-analytical waveform models based on the effective-one-body (EOB) formalism~\cite{Buonanno:1998gg, Buonanno:2000ef} have been extended to incorporate eccentricity effects. This class of models stand out for their ability to combine analytical approximation methods with NR results, achieving high accuracy and computational efficiency for the QC systems~\cite{Damour:2008gu, Pan:2010hz, Pan:2013rra, Taracchini:2013rva, Bohe:2016gbl, Nagar:2018zoe, Cotesta:2018fcv, Babak:2016tgq, Ossokine:2020kjp, Nagar:2018gnk, Nagar:2020pcj, Riemenschneider:2021ppj, Pompili:2023tna}, and for eccentric systems~\cite{Khalil:2021txt, Ramos-Buades:2021adz, Gamboa:2024hli, Gamboa:2024imd, Bini:2012ji, Chiaramello:2020ehz, Nagar:2021gss, Albanesi:2021rby, Placidi:2021rkh, Nagar:2021xnh, Albanesi:2022ywx, Albanesi:2022xge, Nagar:2022fep, Albanesi:2023bgi, Placidi:2023ofj, Nagar:2024dzj, Nagar:2024oyk, Hinderer:2017jcs, Cao:2017ndf, Liu:2019jpg, Liu:2021pkr, Liu:2023dgl}. These models currently incorporate eccentricity only in the inspiral segment of the waveforms and employ QC merger-ringdown (MR) models.
At the same time, the \texttt{IMRPhenom} models~\cite{Husa:2015iqa, Khan:2015jqa, London:2017bcn, Pratten:2020fqn, Garcia-Quiros:2020qpx, Estelles:2021gvs, Estelles:2020twz, Estelles:2020osj, Planas:2025feq, Ramos-Buades:2026kbq} have recently incorporated eccentricity effects in the inspiral regime, both in the time-domain (TD) for the dominant $(2,2)$ and higher-order modes~\cite{Planas:2025feq} and in the frequency-domain for the $(2,\pm2)$ modes~\cite{Ramos-Buades:2026kbq}.
Similarly, the \texttt{NRSurrogate} models~\cite{Blackman:2015pia, Varma:2018mmi, Varma:2019csw, Yoo:2023spi, Rifat:2019ltp, Islam:2022laz, Rink:2024swg, Nee:2025nmh}, which interpolate between numerical relativity (NR) datasets, have recently been extended to include eccentricity effects for the dominant $(2,2)$ mode of non-spinning BBHs over the full IMR waveform~\cite{Nee:2025nmh}.

Regarding the EOB-based waveform models, in the QC case, the merger-ringdown portion of the waveform is phenomenologically modeled by fitting an ansatz~\cite{Damour:2014yha}, or variation of it, to NR merger-ringdown data and hierarchically fitting the resulting coefficients as functions of the system’s intrinsic parameters, such as the symmetric mass ratio and the BBH spins. The merger-ringdown is then attached to the inspiral-plunge segment, computed using resummed post-Newtonian (PN) expressions for the waveform modes~\cite{Damour:2007xr, Damour:2008gu, Pan:2010hz}. 
The phenomenological parameters are fixed by imposing continuity conditions around the merger to ensure a smooth connection between the inspiral, plunge and merger-ringdown descriptions. To extend the validity of these models to a wide range of mass ratios, EOB-based models commonly combine NR information with results from BH perturbation theory (BHPT), exploiting the test-mass (TM) limit to bridge regions of parameter space that are currently inaccessible to NR simulations alone.

For spinning BBHs in the small-mass-ratio regime, merger and post-merger properties are thus studied using BHPT. In this framework, GWs are computed by solving the Teukolsky equation~\cite{Teukolsky:1973ha} for a TM orbiting a Kerr BH. Results in the TM limit (TML) have been widely used to inform waveform models across the full mass-ratio spectrum, playing a central role in the development of EOB-based models~\cite{Nagar:2006xv, Damour:2007xr, Damour:2008gu, Barausse:2011kb, Taracchini:2013wfa, Taracchini:2014zpa, Albanesi:2021rby, Albanesi:2022ywx, Albertini:2022rfe, Albertini:2022dmc, vandeMeent:2023ols, Albertini:2023aol, Albanesi:2023bgi, Albertini:2024rrs, Faggioli:2024ugn, Albanesi:2024fts, Leather:2025nhu, Faggioli:2025hff, Nagni:2025cdw, Nishimura:2026nse}.  
Among these works, Refs.~\cite{Barausse:2011kb, Taracchini:2013wfa, Taracchini:2014zpa, Albanesi:2021rby, Albanesi:2022ywx, Albanesi:2023bgi, Faggioli:2025hff} highlighted the importance of accurately modeling the merger and post-merger stages in the TML. In particular, Refs.~\cite{Barausse:2011kb, Taracchini:2014zpa} analysed merger and post-merger properties from equatorial QC trajectories, providing key input for improving the MR sector of EOB models in the TML and connecting it to the comparable-mass regime. Recently, Ref.~\cite{Nishimura:2026nse} proposed \texttt{SEOB-TML}, a new model in the TML within the \texttt{SEOB} framework that improves the accuracy of the waveforms across the entire IMR by introducing a resummed description of the fluxes at infinity and at the horizon, a refined treatment of the late inspiral--plunge part of the modes, and an improved MR model in the equatorial quasi-circular trajectories in Kerr spacetime.
A similar effort is now required for eccentric systems.

Over the past few years, several studies have investigated the MR properties of eccentric BBHs in both the comparable- and small-mass-ratio regimes, highlighting the importance of extending current QC phenomenological MR models to account for residual eccentricity close to merger. In the comparable-mass case, Refs.~\cite{Carullo:2023kvj, Carullo:2024smg} analysed MR features of eccentric waveforms using eccentric NR simulations from the RIT catalog~\cite{Healy:2022wdn} and the SXS collaboration~\cite{Chu:2009md, Lovelace:2010ne, Lovelace:2011nu, Buchman:2012dw, Hemberger:2013hsa, Scheel:2014ina, Blackman:2015pia, Lovelace:2014twa, Mroue:2013xna, Kumar:2015tha, Chu:2015kft, Boyle:2019kee}, complemented by non-spinning numerical results from BHPT~\cite{Albanesi:2023bgi}, while Ref.~\cite{Nee:2025zdy} explored the impact of the relativistic anomaly measured in the early inspiral on merger properties for comparable-mass eccentric BBHs with non-spinning BHs.

Parallel efforts have also been pursued in the small-mass-ratio regime. In particular, Ref.~\cite{Albanesi:2023bgi} developed a non-spinning EOB ringdown model in the TML that incorporates residual eccentricity at merger, using EOB trajectories to source the Regge-Wheeler-Zerilli equation~\cite{Regge:1957td, Zerilli:1970se} to generate the waveforms. In addition, Ref.~\cite{Becker:2025zzw} investigated merger properties and quasi-normal-modes (QNMs) excitation of waveforms generated sourcing the Teukolsky equation with equatorial-eccentric trajectories whose inspiral is evolved using numerical fluxes computed through a frequency-domain Teukolsky code~\cite{Hughes:2021exa}, and which transition to plunge via an extended Ori-Thorne procedure~\cite{Becker:2024xdi}. Similarly, in Ref.~\cite{Faggioli:2025hff}, some of the authors of this article, investigated the MR features of different GWs modes by sourcing the Teukolsky equation with equatorial critical plunge geodesics of the Kerr metric that starts from the unstable-circular-orbit (UCO)~\cite{Mummery:2023hlo, Dyson:2023fws}, while other works~\cite{Albanesi:2023bgi, DeAmicis:2024eoy, Islam:2024vro, Islam:2025wci} investigated the effects of eccentricity on the late-time tails~\cite{Price:1971fb, Price:1972pw}. Finally, Ref.~\cite{DeAmicis:2025xuh} proposed a new analytically driven approach to extract the QNM amplitudes excited by a TM plunging into a non-spinning central BH, using EOB eccentric inspirals from Ref.~\cite{Albanesi:2023bgi} and exploiting a Green's function framework~\cite{Leaver:1985ax}. Also the gravitational-self-force community started to extend its current spinning QC framework to construct merger and post-merger parts of waveforms~\cite{Compere:2021zfj, Compere:2021iwh, Kuchler:2024esj, Kuchler:2025hwx, Honet:2025dho, Piovano:2026wpz} by analytically investigating the nature of the eccentric inspiral-plunge transition~\cite{Lhost:2024jmw}. We also mention a recent work~\cite{DellaRocca:2025zbe}, which developed a framework to compute QNM excitation amplitudes from particle plunges into Kerr BH, showing enhanced excitation of overtones and subdominant modes at high spins; while they focused on QC equatorial plunges, the formalism is general and extendable to generic Kerr orbits.

In this article we characterize and phenomenologically model the MR of GWs emitted by a TM that inspirals, plunges and merges in the equatorial plane of a Kerr BH. We work in the context of the EOB paradigm, which intrinsically provides the inspiral-plunge transition for QC and eccentric-aligned systems. Specifically, we generate trajectories of the TM by evolving its Hamilton equations of motion equipped with an EOB radiation-reaction (RR) force~\cite{Khalil:2021txt, Faggioli:2024ugn}, which takes into account the dissipative effects due to the emission of gravitational radiation. This RR force extends the resummed QC force of the state-of-the-art \texttt{SEOBNRv5HM} model~\cite{Pompili:2023tna} including eccentric corrections up to 3PN order in its non-spinning sector and up to 2PN in the spinning sector and its leading-order in symmetric mass-ratio part has been assessed with Teukolsky fluxes in Ref.~\cite{Faggioli:2024ugn}. We use these trajectories to source a TD Teukolsky code~\cite{Sundararajan:2007jg,Sundararajan:2008zm,Sundararajan:2010sr,Zenginoglu:2011zz,Field:2020rjr} and generate the GW modes emitted by the TM. We characterize the MR features and fundamental QNM excitation amplitudes as functions of the spin of the central BH, of the eccentricity and relativistic anomaly measured at the LSO. Through this characterization we are able to provide a phenomenological MR model for the most dominant GW modes which extends, in the small-mass-ratio regime, the current QC MR model employed in the \texttt{SEOBNR} waveform model family~\cite{Pompili:2023tna, Gamboa:2024imd} for eccentric-aligned BBHs. While the characterization is performed varying the eccentricity and relativistic anomaly at LSO, this extended MR model employs the effective impact parameter as a fitting variable as proposed in Ref.~\cite{Albanesi:2023bgi, Carullo:2023kvj, Carullo:2024smg}.

This article is structured as follows. In Sec.~\ref{Sec.: methodology} we describe the methodology of our work. Section~\ref{sec.:trajectories} details how we compute equatorial eccentric trajectories of a TM that inspirals, plunges and merges into a Kerr BH within the EOB paradigm, while Sec.~\ref{sec.:waveforms computation} explains how we compute the GWs emitted by the TM using a TD Teukolsky code. In Sec.~\ref{sec.:merger-ringdown model}, we introduce our spin-eccentric MR waveform model. The main results are presented in Sec.~\ref{Sec.: results}. In Sec.~\ref{sec.:the importance of an eccentric merger-ringdown model}, we quantify the impact of including eccentricity effects in MR models, while Sec.~\ref{sec.:waveforms characterization} provides a full characterization of MR features as functions of the spin, eccentricity, and relativistic anomaly of the system. Section~\ref{sec.:the spin-eccentric merger-ringdown model} is dedicated to the spin-eccentric MR model developed in this work. Finally, in Sec.~\ref{Sec: conclusions} we conclude and summarize the main results, and outline possible directions for future work. In the Appendices we provide further details and supplementary discussions that complement the analysis presented in the main text. Finally, we provide the results of the fitted coefficients of our MR model in the Supplemental Material.

\subsection*{Notations}
In this work we adopt geometric units $G = c = 1$ and consider a non-spinning small mass $\mu = \nu M$ orbiting a Kerr BH of mass $M$ with dimensionless spin $a = J/M^2$. The Kerr metric is expressed in Boyer-Lindquist (BL) coordinates $\{\tilde{t}, \tilde{r}, \theta, \varphi \}$ and we restrict our analysis to the equatorial plane, $\theta = \pi/2$.
The dynamics of the small mass is described by canonical coordinates $\{ \tilde{r}, \varphi, P_{\tilde{r}}, P_{\varphi} \}$.
Throughout this article, we consider rescaled dimensionless variables
\begin{gather}
\label{Mscaling}
  t = \frac{\tilde{t}}{M}, \quad r = \frac{\tilde{r}}{M}, \quad p_r = \frac{P_{\tilde{r}}}{\mu}, \quad 
  p_{\varphi}=\frac{P_\varphi}{M\mu}.
\end{gather}
The Hamiltonian $H$, and RR force $\mF = (\mF_r, \mF_{\varphi})$
are rescaled by the small mass $\mu$.
For the ease of notation we set the mass of the Kerr BH to $M=1$.

\section{Methodology} \label{Sec.: methodology}
We now describe the trajectories that
we consider to generate the waveforms and
how these waveforms are computed at future-null infinity
by sourcing the time-domain Teukolsky code. Finally we also describe how we model the MR part of the waveforms.

\subsection{Trajectories} \label{sec.:trajectories}
We consider bound equatorial eccentric EOB trajectories of a small mass $\mu$ inspiraling, plunging and merging into a Kerr BH. Before going into the description of the EOB computation of these trajectories, we provide a summary of the bound equatorial geodesics of the Kerr metric and the Keplerian parametrization, which we employ to label our trajectories' dataset.
\subsubsection{The Kerr metric and its bound geodesics}
\label{sec.: Kerr metric and trajectories' parametrization}
In dimensionless BL coordinates, the Kerr metric~\cite{Kerr:1963ud} in the equatorial plane takes the form
\begin{equation} \label{eq: Kerr metric}
    ds^2 = g_{tt} dt^2 + g_{rr} dr^2 + g_{\varphi \varphi} d\varphi^2 + 2g_{t \varphi} dt d\varphi \ ,
\end{equation} 
where, the non-vanishing components of the metric $g_{\mu \nu}$ have the form
	\begin{subequations}
    \begin{align}
    g_{tt} &= -1 + \frac{2}{r} \ , \\
    g_{rr} &= \frac{r^2}{\Delta} \ , \\
    g_{\varphi \varphi} &= r^2 + a^2 + \frac{2 a^2}{r} \ , \\
    g_{t \varphi} &= - \frac{4 a}{r} \ ,
    \end{align}
    \end{subequations}
with 
	\begin{equation} \label{eq:delta_def}
    \Delta = r^2 - 2r + a^2 \ .
    \end{equation} 

Notably, the inspiral motion is governed by the emission of gravitational radiation which occurs on a time scale that is of order $ (1/ \nu)$. As a consequence, the evolution of the inspiral can be represented as a sequence of stable bound geodesics of the Kerr metric, with decreasing energy $\mathcal{E}$ and angular momentum $\mathcal{L}$. In the equatorial case, the dynamics of these geodesics is described by the standard Kerr geodesic equations in (dimensionless) BL coordinates~\cite{Carter:1968rr}
\begin{subequations}\label{eq:geoeom}
\begin{align}
	\label{eq:radialeom}
		\left(\frac{dr}{d\lambda}\right)^2 & = (\mathcal{E}(r^2+a^2)-a\mathcal{L})^2-\Delta(r^2+(a\mathcal{E}-\mathcal{L})^2), \\
\label{eq:timeeom}
		\frac{dt}{d\lambda} & = \frac{(r^2+a^2)}{\Delta}(\mathcal{E}(r^2+a^2) a\mathcal{L})+a(\mathcal{L}-a\mathcal{E}),
\intertext{and}
\label{eq:azimuthaleom}
		\frac{d\varphi}{d\lambda} & = \frac{a}{\Delta}(\mathcal{E}(r^2+a^2)-a\mathcal{L})+\mathcal{L}-a\mathcal{E},
\end{align}
\end{subequations}
where $\lambda$ is the Mino time variable, which is related to proper time $\tau$ through 
\begin{equation}
	\frac{d\tau}{d\lambda} = r^2.
\end{equation}
As a matter of fact, the right hand side of Eq.~\eqref{eq:radialeom} is a 4th order polynomial in the variable $r$, and can always be expressed in the form~\cite{Fujita:2009bp}
\begin{equation}
	\label{eq:radialeom with roots}
\left(\frac{dr}{d\lambda}\right)^2=(1-\mathcal{E}^2)(r_1-r)(r_2-r)(r_3-r)r = R(r),
\end{equation}
with $r_1$, $r_2$ and $r_3$ being roots of $R(r)$.
As mentioned above, the evolution of the inspiral can be represented as a sequence of stable bound geodesics of Kerr with decreasing energy $\mathcal{E}$ and angular momentum $\mathcal{L}$. For these geodesics the polynomial $R(r)$ in Eq.~\eqref{eq:radialeom with roots} admits three non-vanishing real roots which satisfy
\begin{equation}
\label{eq.:roots hierarchy bound geo}
	\frac{1}{r_1} < \frac{1}{r_2} < \frac{1}{r_3} < \frac{1}{r_{+}} \ ,
\end{equation}
where $r_{+} = 1+\sqrt{1-a^2}$ is the radius of the outer horizon of the Kerr metric.
In this situation, the roots $r_1$ and $r_2$ are respectively named \textit{apocenter} and \textit{pericenter} and the geodesic solution can be described with the Keplerian parametrization in terms of the eccentricity~$e$, the semilatus rectum~$p$, and the relativistic anomaly~$\xi$~\cite{Darwin:1959,Darwin:1961}
\begin{align} \label{eq.: e, p, xi def}
	e &= \frac{r_1 - r_2}{r_1 + r_2}, \quad  
	p = \frac{2r_1 r_2}{r_1 + r_2}, \quad  
	\cos \xi = \frac{p - r}{e r}. 
\end{align}
During the inspiral, the TM evolves adiabatically through a sequence of these geodesics, and the Keplerian quantities in Eq.~\eqref{eq.: e, p, xi def} also evolve with time $t$. This holds until the system reaches the last-stable-orbit (LSO) configuration, which happens when the real roots $r_2$ and $r_3$ in Eq.~\eqref{eq.:roots hierarchy bound geo} coincide, i.e. $r_2 = r_3 = r_{\rm UCO}$, where $r_{\rm UCO}$ is the radius of the unstable-circular-orbit (UCO). 
At the LSO, the energy $\mathcal{E}_{\rm LSO}$ and angular momentum $\mathcal{L}_{\rm LSO}$ of the system equal the energy $\mathcal{E}_{\rm UCO}$ and angular momentum $\mathcal{L}_{\rm UCO}$ of the UCO~\cite{Levin:2008yp,Faggioli:2025hff}
\begin{subequations} \label{eq:LSO E and L}
\begin{align}
\label{eq:energy critical}
	\mathcal{E}_{\rm LSO} &= \mathcal{E}_{\rm UCO} = \frac{(r_{\rm UCO}-2)\sqrt{r_{\rm UCO}}+a}{\sqrt{(r_{\rm UCO}-3)r_{\rm UCO}^2+2a r_{\rm UCO}^{3/2}}}, \\
\label{eq:angular momentum critical}
	\mathcal{L}_{\rm LSO} &= \mathcal{L}_{\rm UCO} = \frac{r_{\rm UCO}^2-2 a \sqrt{r_{\rm UCO}}+a^2}{\sqrt{(r_{\rm UCO}-3)r_{\rm UCO}^2+2a r_{\rm UCO}^{3/2}}}.
\end{align}
\end{subequations}
From Eqs.~\eqref{eq:LSO E and L} it follows that, for a fixed value of the spin $a$, the energy of the UCO is implicitly related to its angular momentum through the radius $r_{\rm UCO}$. In other words, $\mathcal{E}_{\rm UCO}$ can be regarded as a function of $\mathcal{L}_{\rm UCO}$, i.e. $\mathcal{E}_{\rm UCO} = \mathcal{E}_{\rm UCO}(\mathcal{L}_{\rm UCO})$, once the spin of the Kerr BH is specified. After the LSO crossing, the TM continues its transition to plunge and then it performs a geodesic plunge to finally merge into the central Kerr BH.
\subsubsection{Effective-one-body trajectories}
\label{sec.:EOB trajectories}
To fully describe the dynamics of the small mass we work within the EOB framework~\cite{Buonanno:1998gg,Buonanno:2000ef} and consider the Kerr Hamiltonian restricted to equatorial orbits ($\theta = \pi/2$, $p_{\theta} = 0$)~\cite{Barausse:2011kb, Taracchini:2014zpa}: 
  \begin{equation} \label{Eq: Kerr_Hamiltonian}
    H = \Lambda^{-1} \left(2 a p_{\varphi}+\sqrt{\Delta  p_{\varphi}^2 r^2 + \Delta^2 \Lambda  \frac{p_r^2}{r}+\Delta  \Lambda  r} \right) \ , 
  \end{equation}
  where $\Lambda$ is 
  \begin{equation}
    \Lambda = r^3 + 2 a^2 + a^2 r \ ,
  \end{equation}
  and $\Delta$ is defined in Eq.~\eqref{eq:delta_def}.
Instead of the radial momentum $p_r$ we consider $p_{r_{*}}$, which is the momentum conjugate to the tortoise radial coordinate $r_{*}$. 
  The tortoise coordinate is related to the Boyer-Lindquist coordinate $r$ by:
  \begin{subequations}
    \begin{align}
    & dr_{*} = \frac{r^2 + a^2}{\Delta} dr \ , \\
    & p_{r_{*}} = \frac{\Delta}{r^2 + a^2} \, p_r \ .
    \end{align}
  \end{subequations}
This is a general practice~\cite{Damour:2007xr,Pan:2009wj} that is done to improve the numerical stability of the dynamical evolution, since $p_r$ diverges at the horizon while $p_{r_{*}}$ does not.
The evolution of the dynamics is computed by numerically evolving the Hamilton equations
  \begin{subequations} \label{Ham_EOM}
    \begin{align}
      & \dot{r} = \frac{\Delta}{r^2 + a^2} \frac{\partial H}{\partial p_{r_{*}}}(r, p_{r_{*}}, p_{\varphi}) \ , \label{Ham_EOM_1} \\
      & \dot{\varphi} =  \Omega = \frac{\partial H}{\partial \pphi} (r, p_{r_{*}}, p_{\varphi}) \ , \label{Ham_EOM_2} \\ 
      & \dot{p}_{r_*} = -  \frac{\Delta}{r^2 + a^2} \frac{\partial H}{\partial r}(r, p_{r_{*}}, p_{\varphi}) + \mF_r \ , \label{Ham_EOM_3} \\
      & \pphidot = \mF_{\varphi} \ , \label{Ham_EOM_4}
    \end{align}
  \end{subequations} 
where the dot symbol represents a total derivative with respect to the dimensionless BL time coordinate $t$, $\Omega$ is the orbital frequency, scaled by the total mass, and $\mF=(\mF_r, \mF_{\varphi})$ corresponds to the RR force connected to the emission of GWs for generic equatorial orbits.

The RR force components, $\mF_r$ and $\mF_{\varphi}$, are resummed versions of the RR force originally computed in Ref.~\cite{Khalil:2021txt} and extended to 3PN order in the non-spinning eccentric sector~\cite{Gamboa:2024imd, Faggioli:2024ugn}. In Ref.~\cite{Faggioli:2024ugn} the leading order in $\nu$ part of this resummed RR force has been compared with Teukolsky fluxes~\cite{vandeMeent:2015lxa} and it showed a discrepancy $< 5\% $ with the numerical fluxes for values of the spin $a = [-0.99, 0.99 ] $ and eccentricity values $e = [ 0, 0.7 ]$ in the mild-field regime (defined by $\langle \Omega \rangle \le 0.014$)~\cite{Faggioli:2024ugn}.

The expressions for $\mF_r$ and $\mF_{\varphi}$ are given by
  \begin{subequations} \label{eq:full RR force}
  \begin{align}
  & \mF_r = \mF_r^{\text{QC}}\mF_r^{\text{ecc}} \ , \\
  & \mF_{\varphi}=\mF_{\varphi}^{\text{QC}}\mF_{\varphi}^{\text{ecc}} \ ,
  \end{align}
  \end{subequations}
where $\mF_r^{\text{QC}}$ and $\mF_{\varphi}^{\text{QC}}$ are the radial and azimuthal components of the QC prescription of the RR force of \texttt{\texttt{\texttt{SEOBNRv5HM}}} waveform model~\cite{Pompili:2023tna}, while $\mF_{r}^{\text{ecc}}$ and $\mF_{\varphi}^{\text{ecc}}$ are two multiplicative corrections that contain the eccentric PN corrections. The complete expressions in the TML of these eccentric corrections can be found in Appendix~A of Ref.~\cite{Faggioli:2024ugn}.

The QC prescription of the RR force $\mF^{\text{QC}}=(\mF_r^{\text{QC}}, \mF_{\varphi}^{\text{QC}})$ is provided by the expressions
  \begin{subequations} \label{QC force}
  \begin{align} 
    & \mF_{\varphi}^{\text{QC}} = - \frac{\Omega}{8 \pi} \sum_{\ell = 2}^{8} \sum_{m = 1}^{\ell} m^2 |d_{\rm L} h_{\ell m}^{\rm F}|^2  \ , \label{azimuthal QC force} \\
    & \mF_{r}^{\text{QC}} = \frac{p_{r_{*}}}{p_{\varphi}} \mF_{\varphi}^{\text{QC}} \ \label{radial QC force},
  \end{align}
  \end{subequations}
where $d_{\rm L}$ is the luminosity distance of the binary to the observer and $h_{\ell m}^{\rm F}$ are the PN GW spin-weighted spherical harmonic modes resummed in a factorized form~\cite{Damour:2007yf, Damour:2008gu, Damour:2007xr, Pan:2010hz}, given by
  \begin{equation}
    h_{\ell m}^{\rm F} = h_{\ell m}^{\rm (N, \epsilon)} \hat{S}_{\rm eff}^{(\epsilon)} T_{\ell m} f_{\ell m} e^{i \delta_{\ell m}} \label{fact_modes}\ .
  \end{equation}
  Here, $\epsilon$ is the parity of the multipolar waveform mode, such that $\epsilon = 0$ for even $\ell+m$, and $\epsilon = 1$ for odd $\ell+m$. The leading term in Eq.~\eqref{fact_modes}, $h_{\ell m}^{\rm (N, \epsilon)}$ is the Newtonian contribution
  \begin{equation} \label{Newtprefactor}
    h_{\ell m}^{\rm (N, \epsilon)} = \frac{\nu}{d_L}n_{\ell m}^{(\epsilon)}c_{\ell+\epsilon}(\nu) v_\Omega^{\ell} Y^{\ell - \epsilon, -m}\left(\frac{\pi}{2}, \phi \right) \ ,
  \end{equation}
  where $Y^{\ell - \epsilon, -m}(\theta, \phi)$ are the scalar spherical harmonics, $n_{\ell m}^{(\epsilon)}$ and $c_{\ell+\epsilon}(\nu)$ are functions given in Eqs. (28) and (29) of Ref.~\cite{Pompili:2023tna}, and $v_\Omega$ is given by
  \begin{equation}
    v_{\Omega}=\Omega^{1/3}.
  \end{equation} 
We remark that in \texttt{\texttt{\texttt{SEOBNRv5HM}}} a different quantity is employed instead of $v_{\Omega}$~\cite{Pompili:2023tna}, however when applying eccentric corrections $\mathcal{F}_{r, \varphi}^{\rm ecc}$ to the QC RR force~\eqref{QC force}, one has to consider $v_{\Omega}$, as explained in detail in Refs.~\cite{Gamboa:2024hli, Gamboa:2024imd, Faggioli:2024ugn}.
  
The function $\hat{S}_{\rm eff}^{(\epsilon)}$ in Eq.~\eqref{fact_modes} is the effective source term, which is given by 
  \begin{equation}
    \hat{S}_{\rm eff}^{(\epsilon)} = \begin{cases}
      H(r, p_{r_{*}}, p_{\varphi}), \ \ \epsilon = 0 \\
      p_{\varphi} v_\Omega, \ \ \ \ \epsilon = 1 \ .
  \end{cases}
  \end{equation}
  The factor $T_{\ell m}$ resums the leading order logarithms of tail effects and corresponds to
  \begin{equation}
    T_{\ell m} = \frac{\Gamma(\ell + 1 -2i\hat{k})}{\Gamma(\ell + 1)}e^{\pi \hat{k}} e^{2 i \hat{k} \ln{2 m \Omega r_0}} \ ,
  \end{equation}
  where $\Gamma$ is the Euler gamma function, $\hat{k} = m \Omega$ in the TML and $r_0 = 2/\sqrt{e}$.
  The remaining part of the factorized modes \eqref{fact_modes} is expressed as an amplitude $f_{\ell m}$ and a phase $\delta_{\ell m}$, 
  which are computed such that the expansion of $h_{\ell m}$ agrees with the PN expanded modes.
  We point the reader to Appendix B of Ref.~\cite{Pompili:2023tna} for the explicit expressions of the different $f_{\ell m}$ and $\delta_{\ell m}$ terms. 
  In this work we consider the TML of Eqs.~\eqref{eq:full RR force}, \eqref{QC force} and~\eqref{fact_modes}, which is obtained by setting the symmetric mass ratio $\nu$ to zero in the expressions, except for the leading $\nu$ term in the Newtonian prefactor~\eqref{Newtprefactor}.

To set the initial conditions for the dynamics evolution, we proceed differently from the standard prescriptions employed in Refs.~\cite{Ramos-Buades:2021adz, Gamboa:2024hli, Gamboa:2024imd}, which initialize the evolution before the LSO and at large separations. In our work, we instead require full control of the trajectory configuration at the LSO. For a given BH with spin $a$, fixing the eccentricity $e_{\rm LSO}$ at the LSO, uniquely determines the semilatus-rectum $p_{\rm LSO}$~\cite{Stein:2019buj}, energy $\mathcal{E}_{\rm LSO}$ and angular momentum  $\mathcal{L}_{\rm LSO}$ at LSO (see Eqs.~\eqref{eq:LSO E and L}). Then, the only free parameter to be fixed at the LSO is the relativistic anomaly $\xi_{\rm LSO}$. Specifying $(a, e_{\rm LSO}, \xi_{\rm LSO})$ allows us to invert the relations $\mathcal{L}_{\rm LSO} = p_{\varphi, \rm LSO}$ and $\mathcal{E}_{\rm LSO} = H(r_{\rm LSO}, p_{r_* , \rm LSO}, p_{\varphi, \rm LSO})$ and thus obtain the corresponding EOB phase-space coordinates at the LSO. Starting from this point, we integrate backward in time the EOB equations of motion~\eqref{Ham_EOM} to generate a prescribed number of radial cycles before the LSO (we choose typically $4$–$5$ cycles), and use the earliest phase-space point of this backward evolution as the initial condition for the forward integration through the transition to plunge and merger.

As a final remark we mention that during the evolution of the dynamics, we do not employ the RR force until merger, but we follow a similar procedure as in Ref.~\cite{Taracchini:2014zpa}. Specifically, after the LSO crossing and when the radial coordinate approaches the radius of the UCO, i.e. when $\mathcal{E} \ge \mathcal{E}_{\rm UCO}$ and $r \approx r_{\rm UCO}$, we smoothly switch off the effects of the multiplicative eccentric corrections $\mathcal{F}_{r, \varphi}^{\rm ecc}$ in Eqs.~\eqref{eq:full RR force} by imposing $\mathcal{F}_{r, \varphi}^{\rm ecc} = 1$. Practically, we implement this condition by replacing the factors $\mathcal{F}_{r, \varphi}^{\rm ecc}$ in Eqs.~\eqref{eq:full RR force} with
\begin{equation}
\tilde{\mathcal{F}}_{r, \varphi}^{\rm ecc} = 1 + S(r)\,[\mathcal{F}_{r, \varphi}^{\rm ecc}-1] \, ,
\end{equation}
where $S(r) = 1/ [ 1 + \exp{ [ -(r-r_{\rm UCO})/ \sigma ] } ]$, with $\sigma = 0.05$.  We adopt this procedure because, during our analysis, we found that these eccentric corrections behave unphysically during the plunge phase (already after the transition to plunge), in a way similar to what was reported in Ref.~\cite{Nagar:2021gss}. 
This choice is also physically consistent with the fact that, in the TML, the LSO transition is characterized by a transient QC orbital evolution of the small mass on the UCO before plunging: this means that we suppress the effects of the eccentric corrections $\mathcal{F}_{r, \varphi}^{\rm ecc}$ in a regime where they are already negligible, since by construction $\mathcal{F}_{r, \varphi}^{\rm ecc} = 1$ on circular orbits~\cite{Khalil:2021txt, Faggioli:2024ugn}.
After the transition to plunge, as the small mass approaches the inner light-ring, we smoothly suppress the entire RR force $\mathcal{F}_{r, \varphi}$ in Eqs.~\eqref{eq:full RR force} by a factor $1/ [ 1 + \exp{ [ -(r-r_{\rm LR})/ \sigma ] } ]$, with $\sigma = 0.05$ and $r_{\rm LR} = 2+2 \cos{[\frac{2}{3} \arccos{(-a)}]}$ the radius of the inner light-ring. This is the same procedure employed in Ref.~\cite{Taracchini:2014zpa} and is motivated by the fact that in this part of the plunge the motion is geodetic to a good approximation, and is not affected by the details of the GW fluxes.

In this work, we characterize the planar orbits through the spin $a$ of the central Kerr BH, the eccentricity $e_{\rm LSO}$ and relativistic anomaly $\xi_{\rm LSO}$ measured at the LSO, i.e. the trajectories are labelled with the set $\{a, e_{\rm LSO}, \xi_{\rm LSO} \}$. As mentioned above, fixing $e_{\rm LSO}$ automatically fixes $p_{\rm LSO}$, which becomes a redundant parameter.
\begin{figure}[tp!]
\includegraphics[width=1.\linewidth]{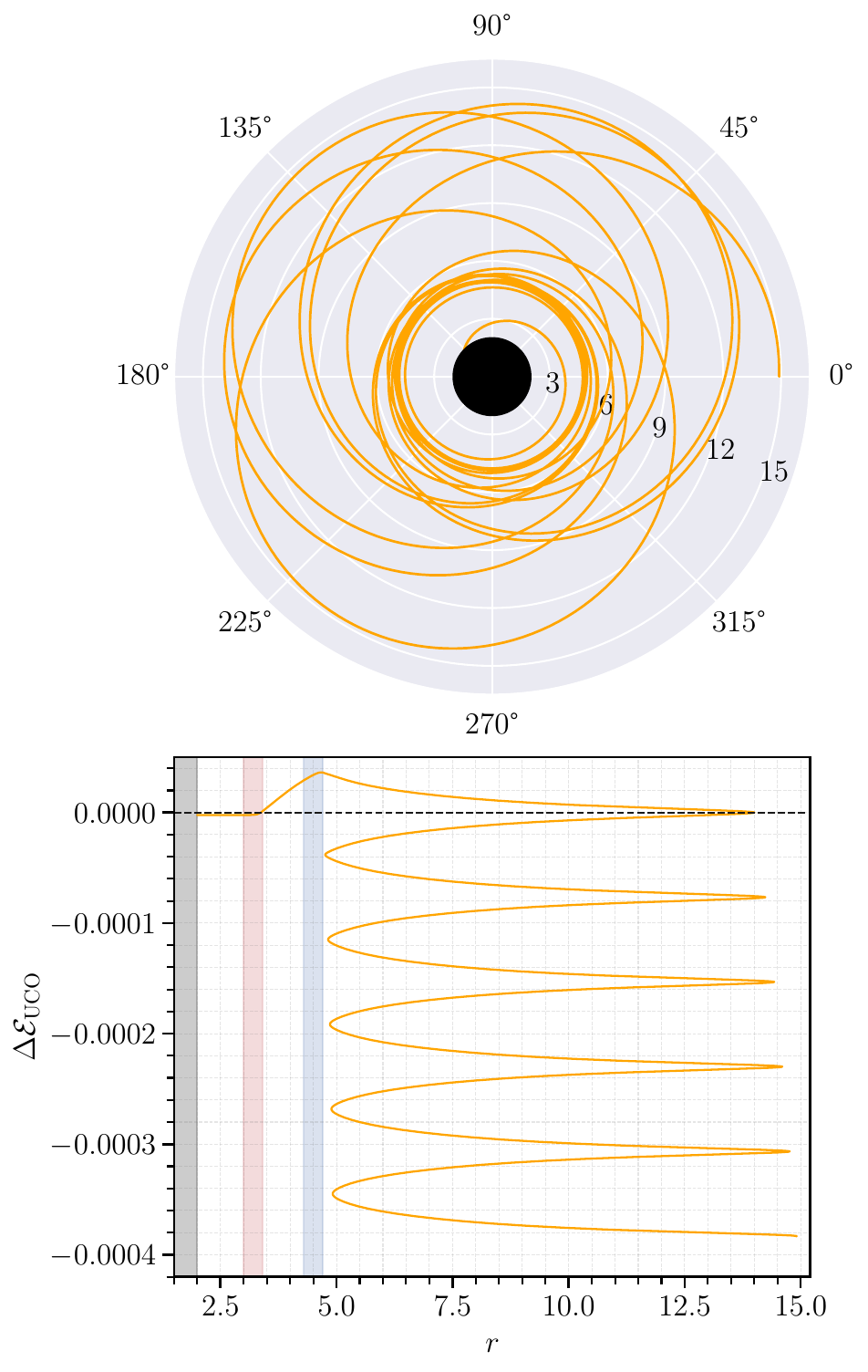}
\caption{Example of an eccentric trajectory of a small mass $\mu = 10^{-3}$. 
In the top panel we show the trajectory in the equatorial plane of the Kerr BH (black disk), evolved for five radial cycles before the LSO and then plunging and merging in the BH.  The trajectory is characterized by a spin \(a = 0\) of the central BH, an eccentricity \(e_{\rm LSO} = 0.5\) and a relativistic anomaly \(\xi_{\rm LSO} = \pi\) at the LSO. In the bottom panel we show the evolution along the trajectory of \(\Delta \mathcal{E}_{\rm UCO}\), defined in Eq.~\eqref{eq.: DeltaEUCO def}, with respect to the radial coordinate $r$. During the inspiral phase \(\Delta \mathcal{E}_{\rm UCO}<0\), consistent with the bound motion; the LSO crossing occurs when \(\Delta \mathcal{E}_{\rm UCO}=0\) (dashed horizontal line), as expressed in Eqs.~\eqref{eq:LSO E and L}. The vertical blue band (around $r\approx 4.5$) marks the interval where the effects of the eccentric corrections to the RR force in Eqs.~\eqref{eq:full RR force} are smoothly switched off, while the vertical red band (around $r\approx 3.2$) indicates the region where the full RR force is smoothly suppressed.}
\label{fig:traj_plot}
\end{figure}

In Fig.~\ref{fig:traj_plot}, we show an example of an eccentric inspiral-plunge-merger trajectory of a small mass $\mu = 10^{-3}$ computed as explained above. 
The top panel shows the trajectory in the equatorial plane of the Kerr BH, evolved for five radial cycles prior to the transition to plunge and subsequent merger with the central BH, indicated by the black disk. The trajectory is characterized by a spin value \(a = 0\) of the central BH, \(e_{\rm LSO} = 0.5\) and \(\xi_{\rm LSO} = \pi\). 
The bottom panel shows the evolution of the quantity \begin{equation} \label{eq.: DeltaEUCO def}
\Delta \mathcal{E}_{\rm UCO} = \mathcal{E} - \mathcal{E}_{\rm UCO}(\mathcal{L}_{\rm UCO} = p_{\varphi}) \, ,
\end{equation}
with respect to the radial coordinate $r$. During the inspiral phase, \(\Delta \mathcal{E}_{\rm UCO} < 0\), consistently indicating bound motion. 
The crossing of the LSO is identified, by definition of LSO (see Eq.~\eqref{eq:energy critical}), by the vanishing of \(\Delta \mathcal{E}_{\rm UCO}\), highlighted by a black dashed horizontal line, after which the system undergoes plunge and merger (we refer to Appendix A of Ref.~\cite{Faggioli:2025hff} for more details on the evolution of $\Delta \mathcal{E}_{\rm UCO}$). 
The vertical blue band (around $r \approx  4.5$) denotes the interval over which the eccentric corrections $\mathcal{F}_{r, \varphi}^{\rm ecc}$ to the RR force in Eqs.~\eqref{eq:full RR force} are smoothly switched off as explained above, while the vertical red band (around $r \approx 3.2$) indicates the subsequent region where the full RR force is smoothly suppressed. In this work we consider a small mass $\mu = 10^{-3}$, values of the spin of the central Kerr BH $-0.9 \le a < 0.9$, eccentricity at the LSO $0.0 \le e_{\rm LSO} \le 0.90$ and relativistic anomaly at the LSO $0 \le \xi_{\rm LSO} < 2 \pi$. For the orbits with $a = 0.9$ we consider a small mass $\mu = 10^{-4}$ in order to have at least $4$ radial cycles in the inspiral before the LSO crossing for all the considered values of $e_{\rm LSO}$.

\subsection{Waveform computation} \label{sec.:waveforms computation}
The trajectories we consider in this article correspond to the world lines of a small mass $\mu$ on a Kerr BH of mass $M$ and can be interpreted as perturbations of the Kerr metric. Hence, we can use BH perturbation theory as the framework to produce the gravitational waveforms sourced by these trajectories, numerically.
We compute the waveforms by solving the Teukolsky master equation~\cite{Teukolsky:1973ha}, which in Boyer-Lindquist coordinates reads
  \begin{equation} \label{Teukolsky equation}
    \begin{aligned}
      & - \left[ \frac{(r^2 + a^2)^2}{\Delta} -a^2\sin{\theta}^2 \right]\partial_{tt}\Psi - \frac{4ar}{\Delta} \partial_{t \varphi} \Psi \\
      & - 2s \left[ r - \frac{(r^2 - a^2)}{\Delta} + ia\cos{\theta} \right] \partial_{t} \Psi + \Delta^{-s} \partial_{r}(\Delta^{s+1} \partial_r \Psi) \\
      & + \frac{1}{\sin{\theta}} \partial_{\theta} (\sin{\theta} \partial_{\theta} \Psi) + \left[ \frac{1}{\sin{\theta}^2} - \frac{a^2}{\Delta} \right]\partial_{\varphi \varphi} \Psi \\
      & + 2s \left[ \frac{a(r - 1)}{\Delta} + \frac{i\cos{\theta}}{\sin{\theta}^2} \right] \partial_{\varphi} \Psi - (s^2 \cot{\theta}^2 - s) \Psi \\
      & = -4 \pi (r^2 + a^2 \cos{\theta}^2) T \ .
    \end{aligned}
  \end{equation}
  This equation describes the evolution of scalar, vector, and tensor perturbations of a Kerr BH.
The parameter $s$ is the \textit{spin weight} of the field. In particular, when $s = \pm 2$ the equation describes radiative degrees of freedom, and for $s = -2$ it is $\Psi = (r - ia\cos{\theta})^4\psi_4$, where 
  $\psi_4$ is the Weyl curvature scalar that describes outgoing GWs.
  
  A system composed of a TM orbiting a Kerr BH is interpreted as a perturbed Kerr metric, and within this interpretation the source term $T$ in the right-hand side of Eq.~\eqref{Teukolsky equation}
  describes a TM moving in the Kerr spacetime. 
  The source term $T$ of a TM orbiting a Kerr BH is constructed from Dirac-delta functions of the variables $r$ and $\theta$, as well as first and second derivatives of the delta functions in these variables. 
  These terms are sourced at the location of the TM, hence the source $T$ depends on the trajectory that the TM follows in the Kerr spacetime. Full details on construction of the TD source term can be found in Refs.~\cite{Sundararajan:2008zm, Sundararajan:2010sr}.
In this analysis, the trajectories used to source the term $T$ are the EOB trajectories introduced in Sec.~\ref{sec.:EOB trajectories}.
  
To solve Eq.~\eqref{Teukolsky equation}, we make use of the finite difference TD code developed in Refs.~\cite{Sundararajan:2007jg, Sundararajan:2008zm,Sundararajan:2010sr,Zenginoglu:2011zz,Field:2020rjr}, which employs hyperboloidal time slicing to extract the GW modes at future-null infinity.  
  At future null infinity, the Weyl scalar $\psi_4$ and the waveform strain $h = h_{+} - i h_{\times}$ are related by the expression
  \begin{equation} \label{Eq.: psi_4 definition}
    \psi_4 = \frac{1}{2}\frac{\partial^2 h}{\partial t^2} = \frac{1}{2} \left( \frac{\partial^2 h_{+}}{\partial t^2} - i\frac{\partial^2 h_{\times}}{\partial t^2} \right) \ ,
  \end{equation}
  and following standard practices, the waveform strain is decomposed in $-2$-spin-weighted spherical harmonics
  \begin{equation} \label{Eq.: strain spherical harmonic decomposition}
    h(t, \iota, \phi) = \sum_{\ell m} {}_{-2}Y_{\ell m}(\iota, \phi)\,h_{\ell m}(t)  \ ,
  \end{equation} 
  where $\iota$ denotes the binary’s inclination angle with respect to the orbital plane, $\phi$ is the azimuthal direction to the observer and $h_{\ell m}$ are the ($-2$ spin-weighted spherical harmonic) gravitational-waveform modes\footnote{Here, $t$ is the retarded time and $h$ is normalized by a factor $\nu$.}.
More details on the numerical accuracy of the TD Teukolsky code can be found in Refs.~\cite{Sundararajan:2007jg, Sundararajan:2008zm, Sundararajan:2010sr,Zenginoglu:2011zz, Barausse:2011kb, Field:2020rjr}.
\subsection{Merger-ringdown model} \label{sec.:merger-ringdown model}
\subsubsection{Anatomy of the ringdown} \label{sec.:Anatomy of the ringdown}
In general relativity, the final remnant of a BBH merger can be described as a perturbed Kerr BH that relaxes to equilibrium by emitting GWs~\cite{Vishveshwara:1970cc, Press:1971wr, Teukolsky:1973ha, Chandrasekhar:1975zza}. This final stage of the waveforms is referred to as the ringdown part and is characterized by three main contributions: (i) a prompt response, which dominates at very early times (ii) the QNMs contribution, which govern the intermediate ringdown and (iii) the power-law tails, which characterize the late-time decay~\cite{Price:1971fb, Price:1972pw, Teukolsky:1973ha, Leaver:1986gd, Berti:2025hly}.

After the prompt response part of the ringdown, when the QNMs' excitation has finished~\cite{Andersson:1995zk, Andersson:1996cm, Chavda:2024awq}, the total GW strain $h(t)$ can be expressed as a sum of QNMs and tail contributions
\begin{equation}
h(t, \iota, \phi) = h^{\rm QNM}(t, \iota, \phi)+h^{\rm tail}(t, \iota, \phi) \ ,
\end{equation}
and can be decomposed as a superposition of spin-weighted spherical harmonics as in Eq.~\eqref{Eq.: strain spherical harmonic decomposition}
\begin{equation} \label{eq: QNM and tail strain sum in spherical harmonic decomp}
h(t, \iota, \phi) = \sum_{\ell m} \left[ h^{\rm QNM}_{\ell m}(t) + h^{\rm tail}_{\ell m}(t) \right] \, {}_{-2}Y_{\ell m}(\iota, \phi) \ ,
\end{equation}
where the single waveform modes can be expressed as 
\begin{equation} \label{eq: QNM and tail spherical modes}
h_{\ell m}(t) = h^{\rm QNM}_{\ell m}(t)+h^{\rm tail}_{\ell m}(t) \ .
\end{equation}

The total strain QNM contribution $h^{\rm QNM}(t, \iota, \phi)$ can be expressed in terms of spin-weighted \textit{spheroidal} harmonic functions as~\cite{Teukolsky:1973ha, Berti:2005ys, Lim:2019xrb, Hughes:2019zmt}
\begin{align} \label{eq: QNM spheroidal decomp}
h^{\mathrm{QNM}}(t,\iota, \phi)
& = \sum_{\mathfrak{l}mnp} 
 A_{\mathfrak{l}mnp} e^{-i \sigma_{\mathfrak{l}mnp}(t - t_0) }\, {}_{-2}S^{a\sigma_{\mathfrak{l}mnp}}_{\mathfrak{l}mn}(\iota, \phi) 
\end{align}
where $(\mathfrak{l}, m)$ are angular harmonic numbers (with $\mathfrak{l} \ge 2$ and $-\mathfrak{l} \le m \le \mathfrak{l}$), $n \ge 0 $ is the overtone number (with $n = 0$ being the longest-lived, fundamental mode) and $p = \{+1, -1 \}$ is the prograde/retrograde index, labeling, respectively, the ordinary mode and its mirror mode.
The quantities $A_{\mathfrak{l}mnp}$ are time independent complex amplitudes, $\sigma_{\mathfrak{l}mnp}$ are the QNMs complex frequencies, $t_0$ is a reference time during the ringdown and ${}_{-2}S^{a\sigma_{\mathfrak{l}mnp}}_{\mathfrak{l}mn}(\iota, \phi)$ are $-2$ spin-weighted spheroidal harmonic functions~\cite{Teukolsky:1973ha, Press:1973zz}. The careful reader might have noticed that, in Eq.~\eqref{eq: QNM spheroidal decomp}, the QNM contribution to the total strain $h^{\mathrm{QNM}}(t, \iota, \phi)$ is expressed in terms of a different functional basis with respect to the functional basis employed in the decomposition of the total strain in Eq.~\eqref{Eq.: strain spherical harmonic decomposition}. To express the spin-weighted spherical harmonic modes of the QNM contribution $h^{\rm QNM}_{\ell m}(t)$ in Eq.~\eqref{eq: QNM and tail spherical modes} in terms of the spin-weighted spheroidal harmonics functional basis, one has first to express the latter functional basis in terms of the former one
\begin{equation} \label{eq: spheroidal in terms of spherical}
{}_{-2}S^{a\sigma_{\mathfrak{l} mn p }}_{\mathfrak{l}mn}(\iota, \phi)
= \sum_{\ell} \mu_{m \ell \mathfrak{l} n}(a\sigma_{\ell m n p})\,
{}_{-2}Y_{\ell m}(\iota, \phi) \ .
\end{equation}
In Eq.~\eqref{eq: spheroidal in terms of spherical} the mixing coefficients $\mu_{m \ell \mathfrak{l} n}$ are the components of the basis decomposition and depend on the quantity $a\sigma_{\ell m n p}$. We also remark the fact that the summation is performed only on the (spherical) angular harmonic number $\ell$~\cite{Buonanno:2006ui, Kelly:2012nd, Berti:2014fga, Cook:2014cta}.

By plugging the decomposition of ${}_{-2}S^{a\sigma_{\mathfrak{l} mn p }}_{\mathfrak{l}mn}(\iota, \phi)$ in Eq.~\eqref{eq: spheroidal in terms of spherical} into Eq.~\eqref{eq: QNM spheroidal decomp} and comparing with the single QNM contribution in Eq.~\eqref{eq: QNM and tail strain sum in spherical harmonic decomp}, one obtains~\cite{Lim:2019xrb}
\begin{equation}\label{eq.:QNMs spherical in QNMs spheroidal}
h^{\rm QNM}_{\ell m}(t) =
\sum_{\mathclap{\substack{\mathfrak{l} \ge \max(2,|m|) \\ n,p}}}
A_{\mathfrak{l}mnp}\,
\mu_{m \ell \mathfrak{l} n}(a\sigma_{\mathfrak{l} m n p})\,
e^{-i \sigma_{\mathfrak{l}mnp}(t - t_0)} \, .
\end{equation}
which provides the expression for the spherical harmonic modes of the QNM contribution in terms of a sum over the different QNMs spheroidal harmonic modes. 

For the sake of completeness, we also mention that at late-times, the spin-weighted spherical harmonic modes of the tail contribution in Eq.~\eqref{eq: QNM and tail strain sum in spherical harmonic decomp} can be expressed as
\begin{equation} \label{eq:tail MR contribution}
 h^{\rm tail}_{\ell m}(t) = A^{\rm tail}_{\ell m}(t + c^{\rm tail}_{\ell m})^{p^{\rm tail}_{\ell m}} e^{i \phi^{\rm tail}_{\ell m}} \ ,
\end{equation}
where $A^{\rm tail}_{\ell m}$ is the time-independent amplitude, $p^{\rm tail}_{\ell m}$ is the decay exponent governing the power-law falloff, $c^{\rm tail}_{\ell m}$ is an effective time offset corresponding to the onset of the tail-dominated regime, and $\phi^{\rm tail}_{\ell m}$ provides the complex phase. 
However, since in this work we only focus on the early and intermediate time of the ringdown signal, we do not model the tail contributions, which contribution is negligible in this stage of the ringdown. For more details on the tail phenomenology and modeling we refer the reader to Refs~\cite{DeAmicis:2024not, Islam:2024vro, DeAmicis:2024eoy, Islam:2025wci}.

As we will show in more detail in the following section, the current families of waveform models based on the EOB paradigm~\cite{Pompili:2023tna, Gamboa:2024hli, Gamboa:2024imd, Riemenschneider:2021ppj, Nagar:2024dzj, Albanesi:2025txj}, employ a phenomenological approach to model the MR part of the waveforms. This approach was originally proposed in Ref.~\cite{Damour:2014yha}, which introduced a multiplicative decomposition of the MR strain $h^{\rm MR}(t)$ as the product of the fundamental QNM with a remaining time-dependent complex factor whose amplitude and phase are separately fitted on numerical waveforms.

However, current waveform models, in particular the models of the \texttt{TEOBResum}, \texttt{IMRPhenom} and \texttt{SEOBNR} families, still model the ringdown with a QC ansatz, which means that the ringdown model features are still modeled with information coming from spinning QC numerical simulations. In this work we aim to extend these phenomenological MR models using spinning-eccentric numerical simulations computed within BHPT.

\subsubsection{Spin-eccentric merger-ringdown model} \label{Subsec: Spin-eccentric merger-ringdown model}
In order to consider the effects of orbital eccentricity in the ringdown, in this work we consider a modified version of the MR ansatz of the \texttt{\texttt{\texttt{SEOBNRv5HM}}}~\cite{Pompili:2023tna} and \texttt{SEOBNRv5EHM}~\cite{Gamboa:2024hli, Gamboa:2024imd} models, which employ a MR model fitted on QC waveforms. As in \texttt{\texttt{\texttt{SEOBNRv5HM}}} and \texttt{SEOBNRv5EHM}, we consider the ansatz for the MR segment of the spin-weighted spherical harmonic modes defined as
\begin{equation} \label{eq.:MR ansatz}
h^{\mathrm{MR}}_{\ell m}(t_{\ell m}) = \tilde{A}_{\ell m}(t_{\ell m}) 
e^{-i\tilde{\phi}_{\ell m} (t_{\ell m}) }
e^{-i \sigma_{\ell m 0 1} t_{\ell m}},
\end{equation}
where $t_{\ell m}$ is the time coordinate at future-null infinity with respect to the matching time $t_{\rm match}^{\ell m}$
\begin{equation}
t_{\ell m} = t - t_{\rm match}^{\ell m} \ ,
\end{equation}
which corresponds to the time instant where the MR model is attached to the inspiral segment of the waveform. Differently from \texttt{SEOBNRv5HM} and \texttt{SEOBNRv5EHM}, where the matching time is defined as the peak of the $h_{22}$ mode for all the considered modes \footnote{With the only exception of the $h_{55}$ mode, for which it is $t_{\text {match }}^{55} = t_{\text {peak }}^{22}-10$ }, in our new MR model, we define the different matching times $t_{\text {match }}^{\ell m}$ as
\begin{equation}
	\label{eq:t_match}
	t_{\text {match }}^{\ell m}=
	\begin{cases}
	t_{\text {peak }}^{\ell m}, &(\ell, m) = (2,2),(3,3), \\
	& \qquad\quad\,\ \; (4,4), (5,5) \\
	t_{\text {peak }}^{\ell m} - \Delta t_{\rm QC}^{\ell \ell},  &(\ell, m) =(2,1),  \\
	& \qquad\quad\,\ \; (3,2), (4,3),\end{cases}
\end{equation}
where $\Delta t_{\rm QC}^{\ell \ell}$ corresponds to the time distance between the peak of the non-diagonal $(\ell, m)$ mode and the peak of the corresponding diagonal $(\ell, \ell)$ mode directly measured from the Teukolsky modes in the QC case, i.e. for $e_{\rm LSO}=0$. By definition, this quantity depends solely on the spin $a$, and we show the result of the fits of this quantity as a function of $a$ in Appendix~\ref{sec: hierarchical fit results}.
The choice of the attachment time for the non-diagonal modes requires particular care in the presence of eccentricity. A natural option would be to attach the MR model at the peak of the $(2,2)$ mode, as done in the current QC prescriptions of \texttt{SEOBNRv5HM} model. However, in the TML this choice becomes problematic as the eccentricity increases, since it would lead to fitting the MR segment too early with respect to the peak of the non-diagonal modes, preventing an accurate description of the waveform phenomenology around the peak.
Another possibility would be to follow the prescription adopted in Ref.~\cite{Nishimura:2026nse}, where the attachment time is defined using orbital information, namely the time instant where the orbital frequency $\Omega$ becomes negative, for negative spins, and the peak of the non-diagonal $(\ell, m)$ modes for positive spins. In the present work we deliberately avoid this option, as our goal is to construct a procedure that can be extended to the comparable-mass regime. In current \texttt{SEOBNR} models, such as \texttt{SEOBNRv5HM} and \texttt{SEOBNRv5EHM}, the orbital frequency does not always exhibit a well-defined peak in the plunge phase across the parameter space, especially for comparable-mass configurations, making prescriptions based on orbital features less robust.
A further possibility would be to use the peak of the respective $(\ell,\ell)$ modes. However, as the eccentricity increases the separation between the peak of the non-diagonal $(\ell,m)$ modes and the corresponding $(\ell,\ell)$ modes grows. This would again lead to attaching the MR model too early with respect to the natural peak of the non-diagonal modes, preventing the model from accurately capturing the waveform features close to merger across the explored parameter space.
For these reasons, we instead define the attachment time through the offset $\Delta t_{\ell\ell}^{\rm QC}$ measured in the QC case. By construction, this offset remains constant as the eccentricity varies, providing a stable reference for defining the MR attachment time of these modes.

The quantities $\tilde{A}_{\ell m}(t)$ and $\tilde{\phi}_{\ell m}(t)$ introduced in Eq.~\eqref{eq.:MR ansatz} are, respectively, a time-dependent real amplitude and phase, while $\sigma_{\ell m 0 1}$ is the complex frequency of the prograde least-damped QNM of the remnant BH, introduced in Eq.~\eqref{eq: QNM spheroidal decomp}. For the ease of notation from now on we do not write the overtone and prograde index, i.e. we use the notation
\begin{equation} \label{eq.:ease of notation QNMs}
\sigma_{\ell m} = \sigma_{\ell m 0 1} = \sigma_{\ell m}^{\rm R} - i \sigma_{\ell m}^{\rm I}.
\end{equation}
Within this notation, by exploiting a well known symmetry of the Teukolsky equation, the generic mirror mode frequency $\sigma_{\ell m 0 -1}$ can always be expressed as~\cite{Teukolsky:1973ha, Press:1973zz, Leaver:1985ax, Berti:2003jh}
\begin{equation} \label{eq.:ease of notation mirror QNMs}
\sigma_{\ell m 0 -1} = -\sigma^{*}_{\ell -m \, 0 1} = -\sigma^{*}_{\ell -m} \, .
\end{equation}
The QNM frequencies are obtained for each $(\ell, m)$ mode using the \texttt{qnm} Python package~\cite{Stein:2019mop} by setting the BH’s final mass $M=1$.

The expressions for the time-dependent real amplitude $\tilde{A}_{\ell m}(t)$ and phase $\tilde{\phi}_{\ell m}(t)$ are given by
\begin{subequations} \label{eq.:amplitude and phase ansatz}
\begin{align}
\tilde{A}_{\ell m}(t_{\ell m}) & = \Biggl[ c^{\ell m}_{1,c} 
\tanh{\Bigl( c^{\ell m}_{1,f} \ t_{\ell m} 
+ c^{\ell m}_{2,f} \Bigr)} 
+ c^{\ell m}_{2,c} \Biggr]^{1/c^{\ell m}_{3,c}}, \label{eq.:amplitude ansatz} \\
\tilde{\phi}_{\ell m}(t_{\ell m}) & = \phi_{\mathrm{match}}^{\ell m} 
- d^{\ell m}_{1,c}\log{\left[
\frac{1 + d^{\ell m}_{2,f} 
 e^{-d^{\ell m}_{1,f} t_{\ell m}}}
{1 + d^{\ell m}_{2,f}}
\right]}. \label{eq.:phase ansatz}
\end{align}
\end{subequations}
While the expression for the phase $\tilde{\phi}_{\ell m}(t)$ is the same as in \texttt{\texttt{SEOBNRv5HM}} and \texttt{SEOBNRv5EHM}, the amplitude ansatz in Eq.~\eqref{eq.:amplitude ansatz} contains an additional constrained exponent $1/c^{\ell m}_{3,c}$, originally introduced in Ref.~\cite{Albanesi:2021rby} and subsequently incorporated into the \texttt{SEOB-TML} waveform family in Ref.~\cite{Nishimura:2026nse}. This modification was motivated by the observation that, in the TP limit and for large spin values ($a \gtrsim 0.8$), the waveform amplitude tends to flatten in the vicinity of its peak, leading to inaccuracies when modeling the MR amplitude using only the amplitude and its first time derivative at the matching time. The inclusion of the constrained exponent $1/c^{\ell m}_{3,c}$ allows the amplitude ansatz to capture the local curvature of the waveform near the peak, thereby improving the robustness of the fit in the large-spin regime.

The coefficients $c^{\ell m}_{1,c}$, $c^{\ell m}_{2,c}$, $c^{\ell m}_{3,c}$ and $d^{\ell m}_{1,c}$ are obtained by imposing continuous differentiability conditions between the Teukolsky waveform and the amplitude and phase ansatzes~\eqref{eq.:amplitude and phase ansatz} at the matching time $t^{\ell m}_{\text{match}}$, and their expressions are
\begin{subequations}
\begin{align}
c^{\ell m}_{1,c} & =
\frac{\hamp{\ell m} ^{-1 + c^{\ell m}_{3,c}}}{c^{\ell m}_{1,f}\, c^{\ell m}_{3,c}}
\left( \dhamp{\ell m} 
+ \sigma^{\rm I}_{\ell m} \hamp{\ell m} \right)
\cosh^{2}\!\bigl(c^{\ell m}_{2,f}\bigr)  \label{eq.: c1c}\, , 
\\
c^{\ell m}_{2,c} &= \hamp{\ell m}^{c^{\ell m}_{3,c}}
- c^{\ell m}_{1,c} \, \tanh\!\bigl(c^{\ell m}_{2,f}\bigr) \label{eq.: c2c} \, ,
\\
c^{\ell m}_{3,c} &=  
\frac{
	\dhamp{\ell m}^2
-
\hamp{\ell m} \, \ddhamp{\ell m}
}{
\left( \dhamp{\ell m}+ \sigma^{\rm I}_{\ell m} \hamp{\ell m} \right)^{2}
} 
-
\frac{
	2 \, c^{\ell m}_{1,f}
	\hamp{\ell m}
	\tanh( c^{\ell m}_{2,f} ) 
}{
	 \dhamp{\ell m}+ \sigma^{\rm I}_{\ell m} \hamp{\ell m}
}  
 , \label{eq.: c3c}
\end{align}
\end{subequations}
for the constrained coefficients of the amplitude ansatz in Eq.~\eqref{eq.:amplitude ansatz}, and
\begin{align} 
d^{\ell m}_{1,c} = \left[ \omega^{\text{match}}_{\ell m} - \sigma^{\rm I}_{\ell m} \right] 
\frac{1 + d^{\ell m}_{2,f}}{d^{\ell m}_{1,f}\, d^{\ell m}_{2,f}} , \label{eq.:d1c}
\end{align}
for the constrained coefficients of the phase ansatz in Eq.~\eqref{eq.:phase ansatz}.
The quantities $\hamp{\ell m}$, $\dhamp{\ell m}$, and $\ddhamp{\ell m}$  are, respectively, the values of the amplitude of the $h_{\ell m}$ mode, its first and second time derivative at the matching time $t^{\ell m}_{\text{match}}$,
while $\omega^{\text{match}}_{\ell m}$ is the frequency of the $h_{\ell m}$ mode at the matching time. 
In EOB models these values are commonly referred as the input-values (IVs) and are extracted from the inspiral-plunge waveforms modes at the matching time with the MR model, after applying the non-quasi-circular corrections to the inspiral waveforms~\cite{Pompili:2023tna, Gamboa:2024hli, Gamboa:2024imd, Riemenschneider:2021ppj, Nagar:2024dzj, Albanesi:2025txj}. 
In our work we directly extract the IVs from the Teukolsky waveforms.

\subsubsection{Merger-ringdown model for the $(3,2)$ and $(4,3)$ modes} \label{sec.:merger-ringdown model for the (3,2) and (4,3) modes}
The MR of the $(3, 2)$ and $(4, 3)$ modes shows postmerger oscillations~\cite{Buonanno:2006ui, Kelly:2012nd}, mostly related to the mismatch between the spherical harmonic basis used for extraction in NR simulations, and the spheroidal harmonics adapted to the perturbation theory of Kerr BHs. Because of this, it is not possible to use the same ansatz of Eqs.~\eqref{eq.:MR ansatz}, \eqref{eq.:amplitude ansatz} and \eqref{eq.:phase ansatz} straightforwardly for the positive spin cases, where these oscillations are particularly relevant. Hence, in this work we thus adopt the same strategy employed in Ref.~\cite{Pompili:2023tna, Nishimura:2026nse} for the $a>0$ scenarios. Equation~\eqref{Eq.: strain spherical harmonic decomposition} can be formulated in terms of -2 spin-weighted spheroidal harmonics as:
\begin{equation}
h^{\mathrm{MR}}(t,\iota, \phi) = \sum_{\mathfrak{l}mnp} 
 {}^{S}h_{\mathfrak{l}mnp}(t) \, {}_{-2}S^{a\sigma_{\mathfrak{l}mnp}}_{\mathfrak{l}mn}(\iota, \phi) \, ,
\end{equation}
and after expanding the spheroidal functions in terms of the spherical functions as in Eq.~\eqref{eq: spheroidal in terms of spherical}, one obtains
\begin{equation} \label{eq.:strain modes in spheroidal decomp}
h^{\rm MR}_{\ell m}(t) = \sum_{\mathclap{\substack{\mathfrak{l} \ge \max(2,|m|) \\ n,p}}} {}^{S}h_{\mathfrak{l}mnp}(t) \, \mu_{m \ell \mathfrak{l} n}(a\sigma_{\mathfrak{l} m n p}) \ ,
\end{equation}
Starting from Eq.~\eqref{eq.:strain modes in spheroidal decomp}, we can directly model the mode-mixing behaviour to obtain monotonic functions that can be fitted by the ansatz already used for the other modes. Since it is not feasible to sum over all the spheroidal modes to get each spherical mode, we make a few approximations valid for the $0<a \le 0.9$ cases. First, we neglect the overtone $(n > 0)$ contributions in the right-hand side of Eq.~\eqref{eq.:strain modes in spheroidal decomp}. Second, for a given $(\ell, m)$ mode, we neglect the contributions from the spheroidal modes with $\mathfrak{l}>\ell$ since their amplitudes are subdominant compared to the $(\mathfrak{l}, m, 0)$ mode. With these approximations, we can rewrite Eq.~\eqref{eq.:strain modes in spheroidal decomp} as
\begin{equation}
h^{\rm MR}_{\ell m}(t) \simeq \sum_{\mathfrak{l} \le \ell} {}^{S}h_{\mathfrak{l}m01}(t) \, \mu_{m \ell \mathfrak{l} 0}(a\sigma_{\mathfrak{l} m 0 1}) \, .
\end{equation}
Writing it explicitly for the modes of interest,
\begin{align}
h_{22}(t) &\simeq \mu_{2220}\, {}^{S}h_{220}(t), \\
h_{33}(t) &\simeq \mu_{3330}\, {}^{S}h_{330}(t), \\
h_{32}(t) &\simeq \mu_{2320}\, {}^{S}h_{220}(t) + \mu_{2330}\, {}^{S}h_{320}(t), \\
h_{43}(t) &\simeq \mu_{3430}\, {}^{S}h_{330}(t) + \mu_{3440}\, {}^{S}h_{430}(t).
\end{align}
From these equations, we can solve for the ${}^{S}h_{\ell m 0}$ modes to obtain
\begin{subequations} \label{eq.: h320 and h430 spheroidal}
\begin{align}
{}^{S}h_{320}(t) &\simeq 
\frac{
h_{32}(t)\mu_{2220} - h_{22}(t)\mu_{2320}
}{
\mu_{2330}\mu_{2220}
}, \label{eq.: h320 spheroidal}\\
{}^{S}h_{430}(t) &\simeq 
\frac{
h_{43}(t)\mu_{3440} - h_{33}(t)\mu_{3430}
}{
\mu_{3330}\mu_{3440}
}. \label{eq.: h430 spheroidal}
\end{align}
\end{subequations}
Thus, we model the spheroidal ${}^{S}h_{\ell m 0}$ modes using the ansatz of Eq.~\eqref{eq.:MR ansatz}, where in Eq.~\eqref{eq.:phase ansatz} $\phi^{\rm match}_{\ell m}$ is replaced by ${}^{S}\phi^{\rm match}_{\ell m 0}$, which is the phase of ${}^{S}h_{\ell m 0}$ at $t=t^{\rm match}_{\ell m}$. In Eqs.~\eqref{eq.: c1c}, \eqref{eq.: c2c} and \eqref{eq.: c3c} we replace $\hamp{\ell m}$ by ${}^{S}\hamp{\ell m 0}$, and in Eq.~\eqref{eq.:d1c} we replace $\omega^{\rm match}_{\ell m}$ by ${}^{S}\omega^{\rm match}_{\ell m 0}$. Once we have a model for ${}^{S}h_{320}$ and ${}^{S}h_{430}$, it is straightforward to obtain the $(3,2)$ and $(4,3)$ modes by combining them with the $(2,2)$ and $(3,3)$ ones previously obtained by inverting Eqs.~\eqref{eq.: h320 spheroidal} and \eqref{eq.: h430 spheroidal}. In our work we directly extract the IVs of the spheroidal modes ${}^{S}h_{320}(t)$ and ${}^{S}h_{430}(t)$ from the Teukolsky waveforms.

\subsubsection{Fit of the merger-ringdown coefficients} \label{sec.:hierarchical fit}

In order to extend the QC fitting strategy adopted in current \texttt{SEOB} models~\cite{Pompili:2023tna, Nishimura:2026nse}, to model the MR of equatorial spin-eccentric waveforms over the parameter space we employ a procedure which is similar to that of Refs.~\cite{Albanesi:2023bgi, Carullo:2023kvj}. In general, equatorial trajectories of a small mass orbiting a Kerr BH are described by a three-dimensional parameter space. In the Keplerian parametrization~\eqref{eq.: e, p, xi def}, the parameter space can be described by the set $\{a, e_{\rm ref}, \xi_{\rm ref}\}$, where $a$ is the spin of the central BH and $e_{\rm ref}$ and $\xi_{\rm ref}$ denote, respectively, the eccentricity and relativistic anomaly measured at a chosen reference point in the inspiral. As we will show in Sec.~\ref{sec.:impact of relativistic anomaly on the merger-ringdown}, when the reference point is the LSO, i.e. $e_{\rm ref} = e_{\rm LSO}$ and $\xi_{\rm ref} = \xi_{\rm LSO}$, and when $\nu \le 10^{-3}$, we find that the dependence of the waveform features on the relativistic anomaly is negligible almost everywhere in the parameter space $\{a, e_{\rm ref}, \xi_{\rm ref}\}$. Note that, this result is not in contradiction with previous findings~\cite{Nee:2025zdy, Becker:2024xdi}, since in those works the relativistic anomaly is defined with respect to a different reference point in the early inspiral and not at the LSO. In our work, where the LSO is taken as the reference point, we find that $\xi_{\rm LSO}$ does not play a relevant role in shaping the MR features, and we therefore do not include it as an independent fitting variable in the hierarchical fits.
In practice, we perform a hierarchical fit of the free coefficients $c^{\ell m}_{1,f}$, $c^{\ell m}_{2,f}$, $d^{\ell m}_{1,f}$, and $d^{\ell m}_{2,f}$ in the amplitude and phase ansatzes in Eq.~\eqref{eq.:amplitude and phase ansatz} as functions of the spin $a$ and of a parameter encoding the eccentricity, measured at a specific reference point. Following Ref.~\cite{Albanesi:2023bgi}, instead of using the eccentricity $e_{\rm LSO}$, which is gauge dependent, we parametrize eccentric effects in the MR model through the gauge-invariant quantity $\tilde{b} = \mathcal{L}/\mathcal{E}$, corresponding to the impact parameter of the trajectory. To improve the performance of the fitting procedure, we work with its offset from the quasi-circular value,
\begin{equation}
b = \tilde{b} - \tilde{b}_{\rm QC}.
\end{equation}
Instead of measuring $b$ at the LSO, we measure it at the peak of the $h_{22}$ mode, i.e. at $t = t^{22}_{\rm peak}$. This choice is motivated by the fact that this point represents a physical reference time that can be inferred directly from the waveform, in view of an extension of this methodology using NR waveforms characterized by higher values of $\nu$ with respect to the waveforms employed in this work. We pick the reference time $t^{22}_{\rm peak}$ from the waveform, but, for simplicity, we still measure the value of $b$ from the trajectories evolution at that time instant. We remark that, as shown in Ref.~\cite{Carullo:2023kvj}, the quantity $b$ can be extracted directly from the numerical waveform, thus through a procedure that does not rely on measurement done on the trajectory. In the future, the availability of such a procedure will be particularly important, since extending the fits to include the symmetric mass ratio $\nu$ as an additional dimension of the parameter space, by combining waveforms generated with BHPT and NR simulations into a single dataset, will require a unique and well-defined methodology to extract the quantities labeling the parameter space dimensions.

The hierarchical fit is carried out in two steps. First, the MR ansatz in Eq.~\eqref{eq.:MR ansatz} is fitted to each individual Teukolsky waveform represented by a point in the parameter space $(b, a)$. In the second step, the coefficients $c^{\ell m}_{1,f}$, $c^{\ell m}_{2,f}$, $d^{\ell m}_{1,f}$, and $d^{\ell m}_{2,f}$, obtained during the first step, are fitted as functions of $b$ and $a$. 
The ansatz we use for the hierarchical fit of the free parameters $\theta^{\ell m}_{i,f}(b,a) = \{ c^{\ell m}_{i,f}, d^{\ell m}_{i,f} \}$ is provided by the rational function
\begin{equation} \label{eq.:hierarchical fit ansatz}
\theta^{\ell m}_{i,f}(b,a) = \frac{C^{\ell m}_{0}(a) + C^{\ell m}_{1}(a) b + C^{\ell m}_{2}(a) b^{2}}{C^{\ell m}_{3}(a) + C^{\ell m}_{4}(a) b} \, ,
\end{equation}
where the coefficients $C^{\ell m}_{i}(a)$ are second-order polynomials in the spin $a$
\begin{align}
C^{\ell m}_{0}(a) &= C^{\ell m}_{00} + C^{\ell m}_{01}a + C^{\ell m}_{02}a^2 + C^{\ell m}_{03}a^3 \, , \\
C^{\ell m}_{i}(a) &= C^{\ell m}_{i0} + C^{\ell m}_{i1}a + C^{\ell m}_{i2}a^2 \, \, , \, \,  \text{for } i>0 \, .
\end{align}
We fit the ansatz~\eqref{eq.:hierarchical fit ansatz} using a least-squares method over the $(b, a)$ parameter space. Appendix~\ref{sec: hierarchical fit results} provides further technical details on the hierarchical fitting procedure, and we provide the explicit expressions of the obtained fits in the Supplemental Material.

\subsubsection{QNMs mixing modeling}
\label{sec.:QNMs mixing modeling}
As we mentioned in Sec.~\ref{sec.:Anatomy of the ringdown}, and explicitly shown in Eq.~\eqref{eq.:QNMs spherical in QNMs spheroidal}, the main contribution to the ringdown after the prompt response comes from a superposition of damped oscillating functions, the QNMs. If we consider only the fundamental modes (i.e. neglecting all the terms with $n>0$) and expand over the prograde/retrograde index $p$ in Eq.~\eqref{eq.:QNMs spherical in QNMs spheroidal} we get
\begin{align} \label{eq.:QNM spherical wo overtones}
h^{\rm QNM}_{\ell m}(t) & = \sum_{\mathfrak{l} \ge \max(2,|m|)} \Biggl[ A_{\mathfrak{l}m01} \ \mu_{m \ell \mathfrak{l} 0}(a\sigma_{\mathfrak{l} m 0 1}) \ e^{-i \sigma_{\mathfrak{l}m01}(t - t_0)  } 
\nonumber 
\\ 
& 
+ A_{\mathfrak{l}m0 \, -1} \ \mu_{m \ell \mathfrak{l} 0}(a\sigma_{\mathfrak{l} m 0 -1}) \ e^{-i \sigma_{\mathfrak{l}m0-1}(t - t_0)} \Biggr]
\nonumber 
\\ 
& 
= \sum_{\mathfrak{l} \ge \max(2,|m|)} \Biggl[ A_{\mathfrak{l}m01} \ \mu_{m \ell \mathfrak{l} 0}(a\sigma_{\mathfrak{l} m}) \ e^{-i \sigma_{\mathfrak{l}m}(t - t_0)} 
\nonumber 
\\ 
& 
+ A_{\mathfrak{l}m0 \, -1} \ \mu_{m \ell \mathfrak{l} 0}(-a\sigma^{*}_{\mathfrak{l} -m}) \ e^{i \sigma^{*}_{\mathfrak{l}-m}(t - t_0)} \Biggr]
\, ,
\end{align}
where in the last equality we adopted the ease of notation introduced in Eqs.~\eqref{eq.:ease of notation QNMs} and~\eqref{eq.:ease of notation mirror QNMs} for the QNM frequencies.

In this work, we phenomenologically model the ringdown as in Eq.~\eqref{eq.:MR ansatz}. By construction, this ansatz has time dependent amplitude $\tilde{A}_{\ell m}(t_{\ell m})$ and phase $\tilde{\phi}_{\ell m}(t_{\ell m})$ that behave as activation functions such that at intermediate times in the ringdown they settle, respectively, to the values of the amplitude $A_{\ell m01} \, \mu_{m \ell \ell 0}(a \sigma_{\ell m})$ and phase $\sigma^{\rm R}_{\ell m}$ of the $(\ell, m, 0, 1)$ QNM. This means that at intermediate times after the peak, when the QNMs contribution is dominating, we are considering only the prograde $(\ell, m, 0, 1)$ least-damped QNM contribution to the ringdown and neglecting all the other QNMs contributions. As a matter of fact some of these additional QNM contributions are particularly evident and relevant when considering systems with a small mass ratio, like the ones considered in this work. In order to consider further QNM contributions we extend the same strategy developed in previous works~\cite{Taracchini:2014zpa, Albanesi:2023bgi, Nishimura:2026nse}. However, due to the different impact the mixing has, depending on the modes, we split the QNM mixing modeling by distinguishing between the $\ell =m$  and the $\ell \ne m$ spherical modes. 

For the $\ell = m$ modes and the $(2,1)$ mode we account for the mixing by adding to the ansatz in Eq.~\eqref{eq.:MR ansatz} the contributions of the retrograde $(\ell, m, 0, -1)$ least-damped QNM and of the prograde $(\ell+1, m, 0, 1)$ least-damped QNM as
\begin{align}
& h^{\mathrm{MR}}_{\ell m} (t_{\ell m}) = \tilde{A}_{\ell m}(t_{\ell m}) 
e^{-i\tilde{\phi}_{\ell m} (t_{\ell m}) } \Biggl[
e^{-i \sigma_{\ell m} t_{\ell m}}
\nonumber \\
& + \mathscr{S}(t_{\ell m}) \frac{A_{\ell m0 \, -1} \, \mu_{m \ell \ell 0}(-a\sigma^*_{\ell -m})}{A_{\ell m 0 1}\, \mu_{m \ell \ell 0}(a\sigma_{\ell m})}e^{i \sigma^{*}_{\ell -m} t_{\ell m}} +
\nonumber \\ 
& + \mathscr{S}(t_{\ell m}) \frac{A_{\ell+1 m 0 1} \, \mu_{m \ell \ell+1 \, 0}(a\sigma_{\ell+1 \, m})}{A_{\ell m 0 1}\, \mu_{m \ell \ell 0}(a\sigma_{\ell m})}e^{-i \sigma_{\ell +1 m} t_{\ell m}} \Biggr],\label{eq.:QNM mixing ansatz}
\end{align}
where $\mathscr{S}(t)$ is an activation function defined as
\begin{align} \label{eq.: def activation function}
& \mathscr{S}(t) = e^{i \gamma \, \mathrm{sech} \left(\frac{t}{\tau_p}\right)} \nonumber \\ 
& \times
\left[
\frac{
\tau_s \tanh \left(\frac{t_s}{\tau_s}\right)
- \mathrm{sech}^2 \left(\frac{t_s}{\tau_s}\right)\tanh(t)
+ \tau_s \tanh \left(\frac{t - t_s}{\tau_s}\right)
}{
\tau_s
- \mathrm{sech}^2 \left(\frac{t_s}{\tau_s}\right)
+ \tau_s \tanh \left(\frac{t_s}{\tau_s}\right)
}
\right] \, ,
\end{align}
which has been introduced in Ref.~\cite{Nishimura:2026nse} and that activates the QNM mixing contributions in the QNM dominated regime. This refined form of $\mathscr{S}(t)$ differs from the one introduced in Ref.~\cite{Taracchini:2014zpa} in two important aspects. First, it satisfies $\mathscr{S}(0) = 0$ and $\mathscr{\dot{S}}(0) = 0$, so that at the attachment point the waveform reduces smoothly to the pure $(\ell, m, 0, 1)$, thus ensuring the differentiability of the MR ansatz. Second, the prefactor $e^{i \gamma \, \mathrm{sech} \left(\frac{t}{\tau_p}\right)}$ provides additional phase flexibility, which becomes especially important for large negative spins. In this regime ($a \le -0.8$), the early ringdown contains also other QNMs that are absent from our explicit model in Eq.~\eqref{eq.:QNM mixing ansatz}. The extra phase term therefore plays a phenomenological role, compensating for this missing physics. The parameters $(t_s, \tau_s, \tau_p, \gamma)$ are tuned for each mode to control the onset and steepness of the activation and they are tuned to be $t_s=20$, $\tau_s = 7.5$ and $\tau_p = 7.5$ for all the considered $\ell = m$ modes, and $t_s=13.5$, $\tau_s = 7.5$ and $\tau_p = 7.5$ for $(2,1)$ mode for all the considered spins $a \ge -0.9$. The parameter $\gamma$ is obtained by optimizing the mismatch 
\begin{equation} \label{eq.: def mismatch}
\mathcal{M}_{\ell m} = 1 - \frac{(h^{\rm MR}_{\ell m} \, | \, h^{\rm Teuk}_{\ell m})}{\sqrt{(h^{\rm MR}_{\ell m} \, | \, h^{\rm MR}_{\ell m})(h^{\rm Teuk}_{\ell m} \, | \, h^{\rm Teuk}_{\ell m})}} \, ,
\end{equation}
of the full MR model of the $h^{\rm MR}_{\ell m}$ mode defined in Eq.~\eqref{eq.:QNM mixing ansatz} with respect to the Teukolsky waveforms $h^{\rm Teuk}_{\ell m}$. In Eq.~\eqref{eq.: def mismatch}, the inner product $(\cdot \mid \cdot)$ is defined as
\begin{equation} \label{eq.:def inner prod}
(h_1 \mid h_2) = \frac{1}{T}
\left|
\int_{t_{0}}^{t_{0}+T} h_1(t)\, h_2^{*}(t)\, dt
\right| \, .
\end{equation}
In Eq.~\eqref{eq.:def inner prod}, $t_0$ denotes the matching time of each $h_{\ell m}$ mode, $t_0 = t^{\ell m}_{\rm match}$. We choose $T=100$ for waveforms with central black-hole spin $a \le 0.7$, and $T=150$ for $a>0.7$, in order to capture the longer post-merger evolution observed in highly prograde configurations~\cite{Taracchini:2014zpa}.

For the $(3,2)$ and $(4,3)$ modes, we adopt a different strategy, and adapt our model to what is proposed in Ref.~\cite{Nishimura:2026nse}. For negative spins $a \le 0.0$ we proceed exactly as done for the $\ell = m$ modes, i.e. we model the spherical MR modes considering Eq.~\eqref{eq.:QNM mixing ansatz}. However, instead of adding the prograde $(\ell + 1, m, 0, 1)$ QNM, we add the $(\ell - 1, m, 0, 1)$ mode. Hence for the spherical $(3,2)$ mode, we consider the contribution of the $(2,2,0,1)$ QNM, while for the $(4,3)$ mode we consider the $(3,3,0,1)$ QNM. On the other hand, for strictly positive spins $a > 0.0$ we proceed as described in Sec.~\ref{sec.:merger-ringdown model for the (3,2) and (4,3) modes}. In these cases, since the $(\ell - 1, m, 0, 1)$ contributions are already modeled as described in Sec.~\ref{sec.:merger-ringdown model for the (3,2) and (4,3) modes}, by explicitly incorporating the mixing with spheroidal harmonics (see Eqs.~\eqref{eq.: h320 and h430 spheroidal}), our phenomenological MR fit only requires the additional contribution from the least-damped mirror $(\ell,m,0,-1)$ QNM, specifically we model them as
\begin{align}
& h^{\mathrm{MR}}_{\ell m} (t_{\ell m}) = \tilde{A}_{\ell m}(t_{\ell m}) 
e^{-i\tilde{\phi}_{\ell m} (t_{\ell m}) } \Biggl[
e^{-i \sigma_{\ell m} t_{\ell m}}
\nonumber \\
& + \mathscr{S}(t_{\ell m}) \frac{A_{\ell m 0 \, -1} \, \mu_{m \ell \ell 0}(-a\sigma^*_{\ell -m})}{A_{\ell m 0 1}\, \mu_{m \ell \ell 0}(a\sigma_{\ell m})}e^{i \sigma^{*}_{\ell -m} t_{\ell m}} \Biggr], \label{eq.:QNM mixing ansatz h32 h43}
\end{align}
where $\mathscr{S}(t)$ is the activation function defined in Eq.~\eqref{eq.: def activation function}. For these modes we tune the parameters $(t_s, \tau_s, \tau_p)$ to be $(t_s=13.5, \tau_s=7.5, \tau_p=7.5)$ for the $(2,1)$ mode, $(t_s=16, \tau_s=7.5, \tau_p=7.5)$ for the $(3,2)$ mode and $(t_s=19, \tau_s=7.5, \tau_p=7.5)$ for the $(4,3)$ mode \footnote{Note that these values are different from the values in Ref.~\cite{Nishimura:2026nse} because we employ different attaching times for these modes.}. The parameter $\gamma$ is optimized as described for the $\ell = m$ case, i.e. by minimizing the mismatch in Eq.~\eqref{eq.: def mismatch}. We find that $\gamma$ does not significantly depend on the eccentricity at the LSO, hence we fit it, for the different modes, as a function of the spin $a$. We provide more details in Appendix~\ref{sec: hierarchical fit results} and the results of the fit in the Supplemental Material.

In order to extract the QNM amplitudes $A_{\mathfrak{l}mnp}$ from the Teukolsky waveforms we employ the \texttt{qnmfinder} code~\cite{Mitman:2025hgy}, which implements a systematic TD extraction of QNM by fitting the waveform over the full two-sphere using a nonlinear least-squares procedure. The algorithm performs a reverse-in-time search for QNM dominance and assesses the robustness of the extracted amplitudes through stability criteria evaluated over timescales comparable to each mode’s damping time, enabling a consistent identification of the excited QNMs and their overtones. We employ the code \texttt{qnmfinder} on all the Teukolsky waveforms and extract the corresponding amplitudes $A_{\mathfrak{l}mnp}$. By default, the code refers to the values of the different $A_{\mathfrak{l}mnp}$ at the peak of the total news $\dot{h}(t)$, but we rescale and refer to the extracted amplitudes at the peak of each $h_{\ell m}(t)$ mode in order to be directly implemented in Eq.~\eqref{eq.:QNM mixing ansatz}. 

\section{Results} \label{Sec.: results}
\subsection{The importance of an eccentric merger-ringdown model} \label{sec.:the importance of an eccentric merger-ringdown model}
\begin{figure}[tp!]
  	\includegraphics[width=1.\linewidth]{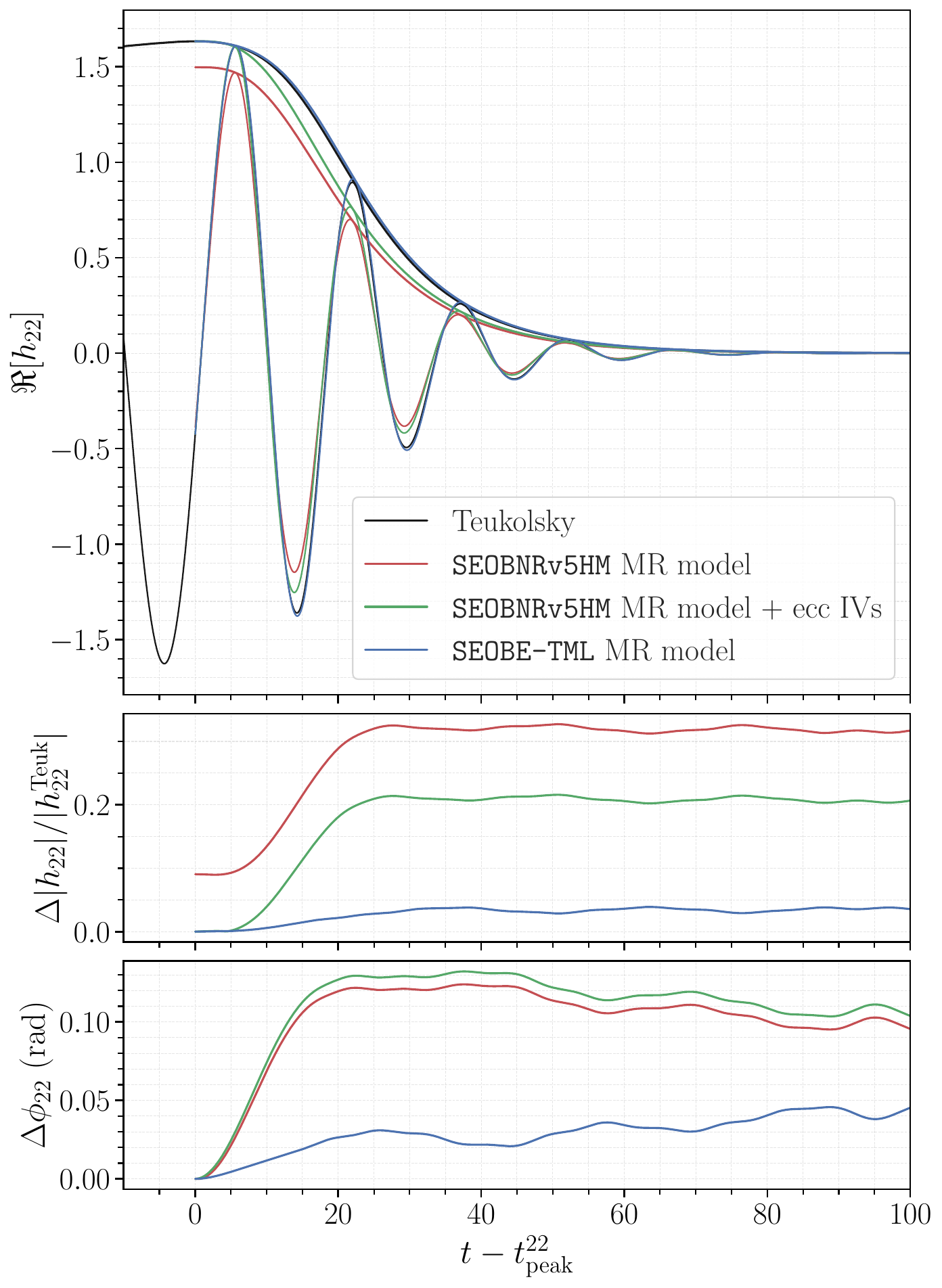}
  	\caption{Comparison of the $h_{22}$ Teukolsky waveform mode with three different EOB MR model prescriptions. The top panel shows the real part of $h_{22}$: the Teukolsky waveform is shown in black, the \texttt{SEOBNRv5HM} QC MR model attached using QC IVs at $t_{\rm peak}^{22}$ in red, the same QC MR model attached using the eccentric IVs measured from the numerical waveform in green, and the \texttt{SEOBE-TML} MR model developed in this work in blue. The system considered has $a = 0.30$ , $e_{\rm LSO} = 0.50$ and $\xi_{\rm LSO} = \pi $ and all waveforms are aligned at the peak of the mode, $t = t_{\rm peak}^{22}$. The bottom panels show the fractional amplitude difference $\Delta |h_{22}|/|h^{\rm Teuk}_{22}|$ and the phase difference $\Delta\phi_{22}$. The QC MR model exhibits clear discrepancies in the post-merger regime, which are only partially reduced when eccentric IVs are employed, while the \texttt{SEOBE-TML} MR model provides improved agreement in both amplitude and phase.}
  	\label{fig:why_we_need_to_extend_EOB_MR_model}
\end{figure}
Before going through the detailed results of the \texttt{SEOBE-TML} MR model developed in this work, we first highlight the limitations of the current MR model employed in the \texttt{SEOBNR} models, which relies exclusively on QC information even when applied to eccentric waveforms.
In Fig.~\ref{fig:why_we_need_to_extend_EOB_MR_model}, we compare the numerical dominant $(\ell,m)=(2,2)$ mode obtained by solving the Teukolsky equation in Eq.~\eqref{Teukolsky equation} with three different MR model prescriptions. Specifically, the top panel of Fig.~\ref{fig:why_we_need_to_extend_EOB_MR_model} shows the real part of the $h_{22}$ mode of the Teukolsky waveform (black curve), the \texttt{SEOBNRv5HM} MR model, fitted to QC numerical waveforms and attached using QC IVs at the attachment time $t_{\rm peak}^{22}$ (red curve), the same MR model attached using the eccentric IVs measured from the Teukolsky waveform at the attachment time (green curve), and the \texttt{SEOBE-TML} MR model introduced in this work (blue curve), obtained by fitting the free coefficients $c^{22}_{1,f}$, $c^{22}_{2,f}$, $d^{22}_{1,f}$, and $d^{22}_{2,f}$ over the $(a,b)$ parameter space, as described in Sec.~\ref{sec.:hierarchical fit}, and by matching it using the eccentric IVs measured from the Teukolsky waveform. The bottom panels display the fractional amplitude difference $\Delta |h_{22}|/|h^{\rm Teuk}_{22}|$ (first bottom panel) and the phase difference $\Delta \phi_{22}$ (second bottom panel) of the three MR models with respect to the Teukolsky waveform. In Fig.~\ref{fig:why_we_need_to_extend_EOB_MR_model}, the waveforms are aligned at the peak of the mode amplitude, i.e. at $t_{\rm peak}^{22}$, and we consider a system with $a = 0.30$ and $e_{\rm LSO} = 0.50$.

The red curve in the bottom panels shows that when the QC MR model is attached using QC IVs to model an eccentric waveform, clear discrepancies with respect to the Teukolsky waveform are already present at the attachment time, with an amplitude fractional difference of $\sim 10\%$, reaching $\sim 30\%$ for $(t-t_{\rm peak}^{22}) \ge 20$, and a phase difference growing up to $\sim 0.12$ rad after the peak of the mode. The green curve shows that, when eccentric IVs are used at the attachment time for the same QC MR model, the amplitude discrepancies are reduced to $\sim 20\%$ for $(t-t_{\rm peak}^{22}) \ge 20$; however, the phase difference still settles at $\sim 0.12$ rad after the peak of the mode. Only when the \texttt{SEOBE-TML} MR model is employed (blue curve) it maintains agreement with the Teukolsky waveform throughout the MR regime, with post-merger amplitude differences remaining below a few percent ($\le 4 \%$) and phase discrepancies $ \le 0.04 $ rad. This progressive reduction of the mismatches, from the QC MR model, to a QC MR model with eccentric IVs, and finally to the \texttt{SEOBE-TML} MR model, demonstrates that QC descriptions of the MR can be extended in order to provide a more accurate description of the MR of eccentric systems, and that an explicit eccentric calibration of the MR ansatz in Eq.~\eqref{eq.:MR ansatz} is required to accurately capture the post-merger features.

\subsection{Waveforms characterization} \label{sec.:waveforms characterization}
In order to model the MR accurately in the mass-ratio regime considered in this work, we first characterize the features of the spin-eccentric post-merger. To do so, we analyze the main aspects of the merger and post-merger waveforms by examining specific regions of the parameter space. We begin by studying the effects of varying the eccentricity $e_{\rm LSO}$ and spin $a$ while keeping the relativistic anomaly $\xi_{\rm LSO}$ fixed, and subsequently investigate the impact of changing $\xi_{\rm LSO}$ and $a$ while holding $e_{\rm LSO}$ constant.
\subsubsection{Impact of eccentricity on the merger-ringdown} \label{sec.:impact of eccentricity on the merger-ringdown}
\begin{figure*}[tp!]
\includegraphics[width=1.\linewidth]{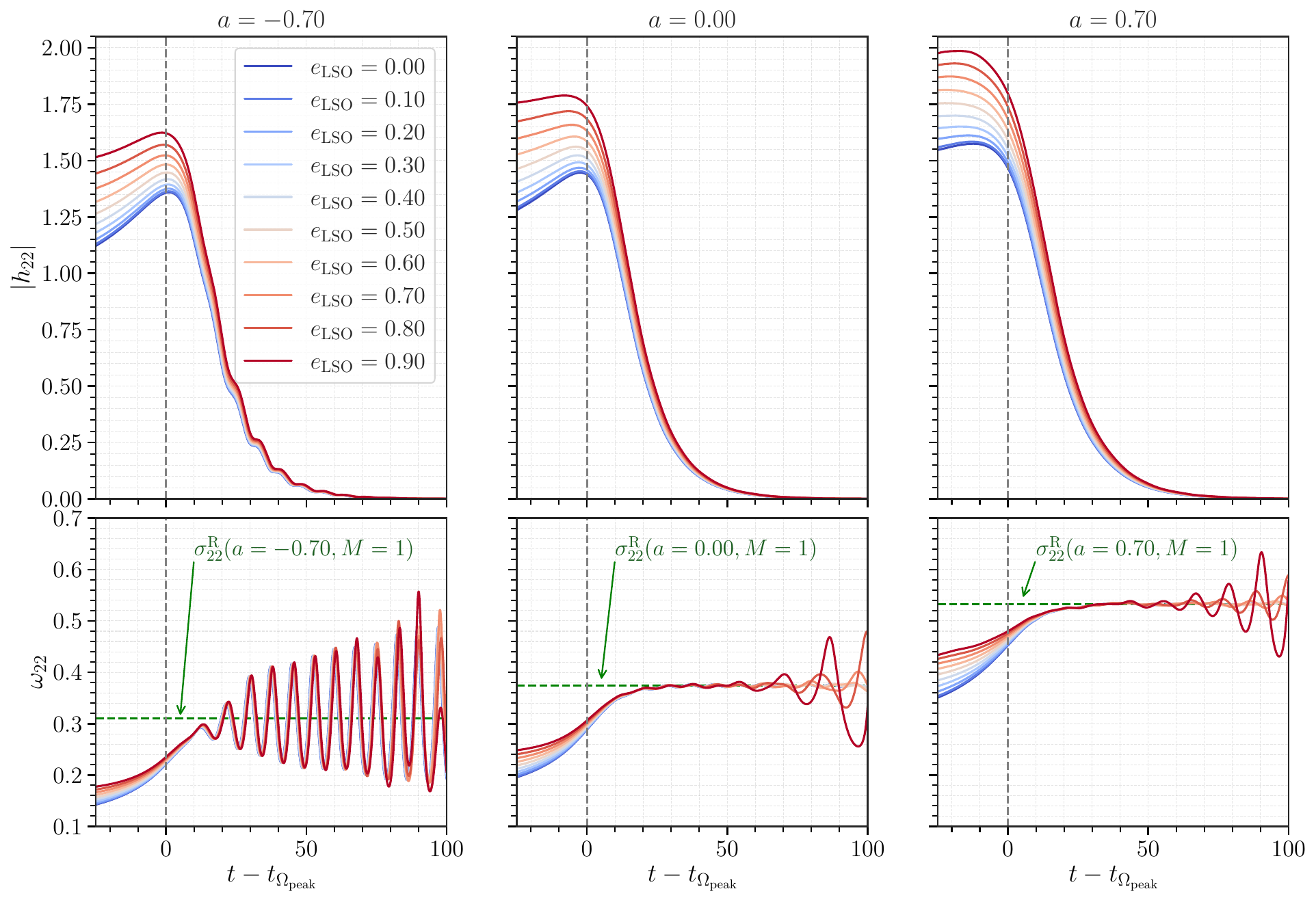}
\caption{Effect of the value of eccentricity $e_{\rm LSO}$ on the MR features of the $h_{22}$ mode. The waveform amplitude $|h_{22}|$ (top panels) and the instantaneous GW frequency $\omega_{22}$ (bottom panels) are shown as functions of time relative to the time of the peak of the orbital frequency, i.e.\ as functions of $t - t_{\Omega_{\rm peak}}$. Results are shown for three values of the spin: $a=-0.70$ (left column), $a=0.00$ (central column), and $a=0.70$ (right column). All waveforms are characterized by a relativistic anomaly $\xi_{\rm LSO}=\pi$ and span eccentricities in the range $e_{\rm LSO}\in[0.0,0.9]$. Increasing eccentricity leads to an enhancement of the value of the peak amplitude near merger and to an increasing time separation between the time of the amplitude peak $t_{\rm peak}^{22}$ and $t_{\Omega_{\rm peak}}$. In the ringdown regime, $\omega_{22}$ settles around the real part of the dominant QNM frequency $\sigma_{22}$, whose value $\sigma^{\rm R}_{22}$ is indicated in the bottom panels by dashed green horizontal lines. The oscillations observed in $\omega_{22}$ around $\sigma^{\rm R}_{22}$ have an amplitude that depends on the spin and reflect the QNM mixing introduced in Sec.~\ref{sec.:Anatomy of the ringdown}. For a fixed spin, the frequencies corresponding to different values of $e_{\rm LSO}$ largely overlap during the QNM-dominated interval, indicating that eccentricity has a weak impact on the relative excitation of the QNMs. At late times, particularly for $e_{\rm LSO}\gtrsim0.7$, enhanced oscillations in $\omega_{22}$ are observed (especially for the $a=0.00$ and $a=0.70$ cases) and are associated with earlier tail excitation and its interference with QNM contributions, as pointed out in the main text of Sec.~\ref{sec.:impact of eccentricity on the merger-ringdown}.
}
\label{fig:effects of eccentricity}
\end{figure*}
\begin{figure}[tp!]
\includegraphics[width=1.\linewidth]{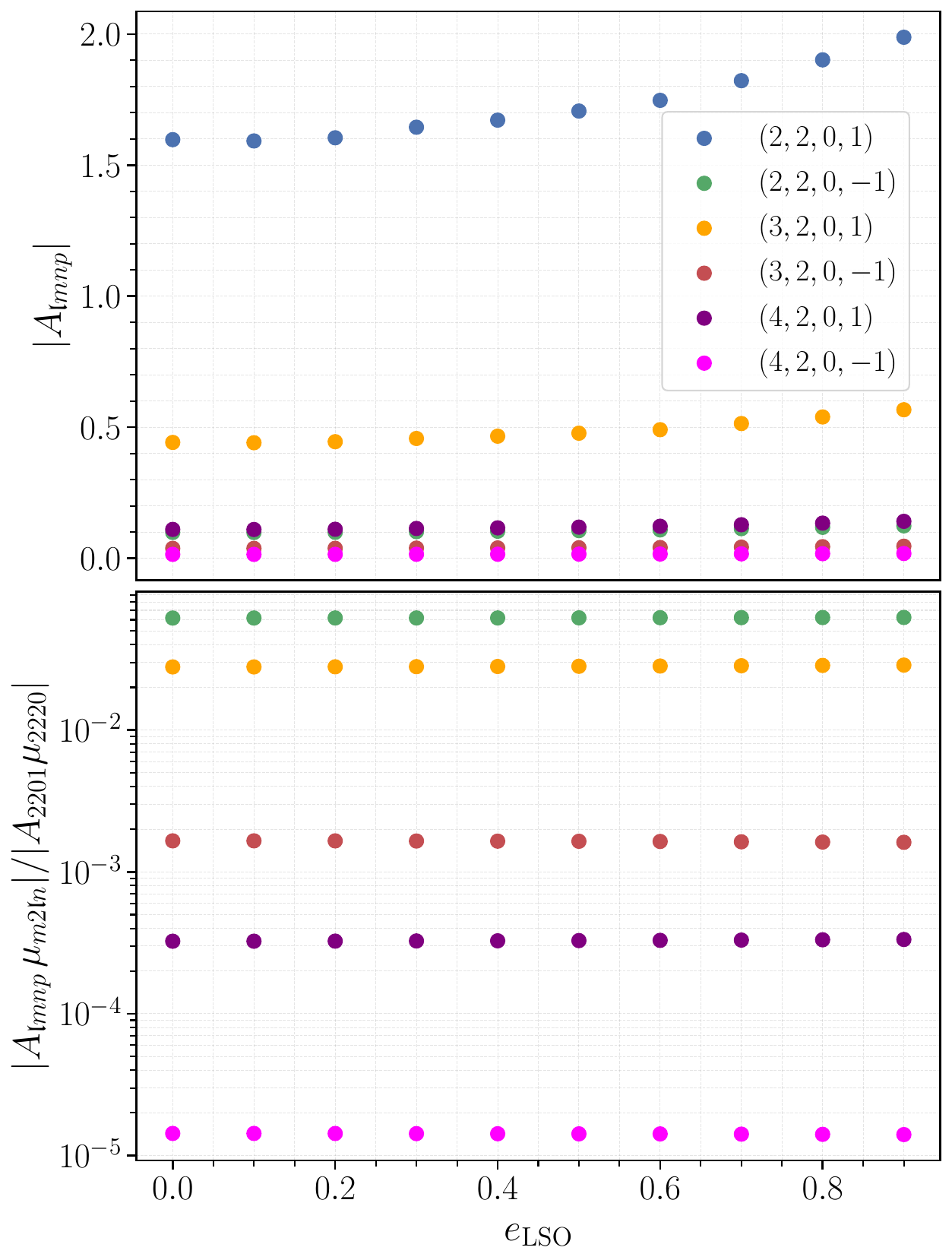}
\caption{Dependence of the QNMs excitation on $e_{\rm LSO}$, for the case $a=-0.70$ and $\xi_{\rm LSO}=\pi$.
The top panel shows the absolute values of the excitation amplitudes $|A_{\mathfrak{l} m n p}|$ introduced in Eq.~\eqref{eq: QNM spheroidal decomp} for selected fundamental prograde and retrograde QNMs with $m = 2$, plotted as functions of $e_{\rm LSO}$. In the bottom panel we plot the ratios of the coefficients $|A_{\mathfrak{l} m n p} \, \mu_{m 2 \mathfrak{l} n }(a \sigma_{\mathfrak{l} m n p})|/|A_{2201}\mu_{2 2 2 0}(a \sigma_{2 2 0 1})|$, which contain the spheroidal-spherical harmonic mixing factors $\mu_{m 2 \mathfrak{l} n }(a \sigma_{\mathfrak{l} m n p})$ for the $h_{22}$ spherical mode.
While increasing eccentricity leads to a systematic enhancement of the absolute excitation amplitudes $|A_{\mathfrak{l} m n p}|$ of all considered modes, their relative excitation remains nearly unchanged over the full range of $e_{\rm LSO}$. This indicates that eccentricity primarily affects the overall strength of QNM excitation, while leaving the hierarchy and mixing structure of the dominant ringdown modes largely unaltered. The amplitudes showed in this figure are referred at the light-ring crossing time of the TM.}
\label{fig:QNMS ampl vs eccentricity}
\end{figure}

We now examine how variations in eccentricity affect the MR waveform, focusing on their imprint on the dominant $h_{22}$ mode across different spin configurations. In Fig.~\ref{fig:effects of eccentricity}, we show how varying the eccentricity at the LSO produces a clear and systematic imprint on the MR of the $h_{22}$ mode across three spin configurations. We consider $a=-0.70$ (left column), $a=0.00$ (central column), and $a=0.70$ (right column). The top panels show the waveform amplitude $|h_{22}|$, while the bottom panels display the instantaneous GW frequency $\omega_{22}$, both plotted as functions of time relative to the peak of the orbital frequency $\Omega$ introduced in Eq.~\eqref{Ham_EOM_2}, i.e. with respect to $t_{\Omega_{\rm peak}}$. All the waveforms shown in Fig.~\ref{fig:effects of eccentricity} are characterized by a relativistic anomaly $\xi_{\rm LSO} = \pi$ at the LSO, and we span eccentricity values $e_{\rm LSO} = [0.0, 0.1, 0.2, 0.3, 0.4, 0.5, 0.6, 0.7, 0.8, 0.9]$.
The figure shows that increasing eccentricity enhances the value of the amplitude peak around merger, at $t \approx t_{\Omega_{\rm peak}}$, for all spin values, and increases the time separation between the peak of the amplitude at $t_{\rm peak}^{22}$ and the peak of the orbital frequency at $t_{\Omega_{\rm peak}}$. These features were already reported in Ref.~\cite{Albanesi:2023bgi} for the equatorial Schwarzschild eccentric case and in Ref.~\cite{Faggioli:2025hff} for the equatorial eccentric Kerr case when considering critical plunge geodesics. By inspecting the top-left panel corresponding to $a=-0.70$, we also observe the QNM mixing introduced in Sec.~\ref{sec.:Anatomy of the ringdown}. Starting from $(t - t_{\Omega_{\rm peak}}) \sim 17$, the amplitude of the $h_{22}$ mode exhibits an oscillatory behavior driven by the interference between different QNMs. 
By visual inspection, the amplitude and timescale of these oscillations remain approximately unchanged across different values of $e_{\rm LSO}$.
Although the absolute amplitudes of the individual QNMs depend on the eccentricity (see, for example, Fig.~11 of Ref.~\cite{DeAmicis:2025xuh} and Fig.~\ref{fig:QNMS ampl vs eccentricity} of this section), the relative excitation of the dominant QNMs, i.e.\ the ratios of their amplitudes $A_{\mathfrak{l}mnp}$, appear to remain approximately unchanged.
As we will show in the following, and explicitly illustrate in Fig.~\ref{fig:QNMS ampl vs eccentricity}, this expectation is indeed confirmed. 
To further assess this observation, it is instructive to examine the bottom panels of Fig.~\ref{fig:effects of eccentricity}, where we plot the instantaneous frequency $\omega_{22}$ of the mode. The frequency $\omega_{\ell m}$ of a generic $h_{\ell m}$ mode can be computed directly from the mode itself as~\cite{Taracchini:2014zpa}
\begin{equation}
\omega_{\ell m}(t) = -\Im{\Biggl[\frac{\dot{h}_{\ell m}(t) }{h_{\ell m}(t)}\Biggr]} \, ,
\end{equation}
where the dot denotes a derivative with respect to the Boyer-Lindquist time $t$, and $\Im [\cdot ]$ denotes the imaginary part.
The bottom panels show the time evolution of $\omega_{22}$ and display features consistent with those reported in previous works~\cite{Barausse:2011kb, Taracchini:2014zpa, Albanesi:2023bgi}.
By comparing the different columns, i.e.\ varying the spin, one observes that $\omega_{22}$ progressively settles around a constant value corresponding to the real part of the frequency of the most strongly excited QNM, which in this case is given by $\sigma^{\rm R}_{22}$.
We highlight the value of $\sigma^{\rm R}_{22}$ in the bottom panels of Fig.~\ref{fig:effects of eccentricity} by dashed green horizontal lines.
This behavior sets in at approximately $(t - t_{\Omega_{\rm peak}})\sim20$.
As the spin decreases, the frequency exhibits increasingly pronounced oscillations around $\sigma^{\rm R}_{22}$, reflecting the growing importance of additional QNM contributions, most notably the $(2,2,0,-1)$ retrograde mode and the prograde $(3,2,0,1)$ mode, relative to the dominant $(2,2,0,1)$ QNM.

For each considered spin configuration, we find that varying the eccentricity does not significantly affect the relative QNM content of the signal, as the frequencies corresponding to different values of $e_{\rm LSO}$ overlap closely within the QNM-dominated time interval $(t - t_{\Omega_{\rm peak}}) \in [20,80]$. 
However, for $(t - t_{\Omega_{\rm peak}}) \gtrsim 70$, waveforms characterized by larger eccentricities, i.e. $e_{\rm LSO} \ge 0.70$, exhibit enhanced oscillations of the frequency compared to those associated with smaller eccentricities. We investigated the origin of this behavior and found that it is related to the fact that, as the eccentricity increases, the tail is excited more strongly and its amplitude becomes of comparable size as the QNMs amplitude at earlier times, as reported in Refs.~\cite{Albanesi:2023bgi, Islam:2024vro, DeAmicis:2024not, Islam:2025wci}. As a consequence, the time interval over which QNM and tail contributions interfere shifts toward earlier times as $e_{\rm LSO}$ increases. The enhanced oscillations of $\omega_{22}$ observed for $e_{\rm LSO} \ge 0.70$, particularly evident for the $a = 0.00$ and $a = 0.70$ cases, are therefore associated with this interference. A dedicated discussion illustrating this behavior is provided in Appendix~\ref{Appendix: interference between QNMs and Price tails at high eccentricity}.

In order to understand how varying the eccentricity affects the relative QNM content of the signal in Fig.~\ref{fig:effects of eccentricity}, we extract and analyse the QNM amplitudes through the use of the code \texttt{qnmfinder}.
In Fig.~\ref{fig:QNMS ampl vs eccentricity} we show the results of this extraction and we plot the impact of $e_{\rm LSO}$ on the excitation of a subset of QNM contributions for the case $a=-0.70$ and $\xi_{\rm LSO}=\pi$. The upper panel displays the absolute values of the excitation amplitudes $|A_{\mathfrak{l} m n p}|$ introduced in Eq.~\eqref{eq: QNM spheroidal decomp} for a subset of prograde and retrograde fundamental QNMs with $m=2$, namely the $(2,2,0,\pm1)$, $(3,2,0,\pm1)$, and $(4,2,0,\pm1)$ modes, as functions of $e_{\rm LSO}$. The amplitudes showed in the figure are referred at the light-ring crossing time of the TM.
As the eccentricity increases, all considered amplitudes exhibit a smooth and monotonic growth, with the dominant contribution remaining the prograde $(2,2,0,1)$ mode across the entire eccentricity range. The other subdominant modes follow the same qualitative trend, indicating that eccentricity primarily rescales the overall excitation of QNM rather than altering the hierarchy among the excited modes. 
To quantify the relative contribution of the different QNMs to the spherical $h_{22}$ ringdown, in the bottom panel of Fig.~\ref{fig:QNMS ampl vs eccentricity} we plot the ratios $|A_{\mathfrak{l} m n p}\,\mu_{m 2 \mathfrak{l} n}(a \sigma_{\mathfrak{l} m n p})|/|A_{2201}\mu_{2 2 2 0}(a \sigma_{2 2 0 1})|$, which directly measure the relative weight of each spheroidal QNM contribution to the $(2,2)$ spherical harmonic mode.
Remarkably, these ratios remain nearly constant over the full range of $e_{\rm LSO}$, spanning several orders of magnitude without exhibiting any systematic eccentricity dependence. Furthermore from the bottom panel it is evident how the main mixing contributions to the ringdown of the $h_{22}$ mode come from the retrograde $(2,2,0,-1)$ and prograde $(3,2,0,1)$ modes, which corroborates our choice in how to model the QNMs mixing in Eq.~\eqref{eq.:QNM mixing ansatz}.
Moreover, this demonstrates that, although eccentricity significantly affects the absolute excitation of the QNMs, it has a negligible impact on their relative weights with respect to the dominant $(2,2,0,1)$ mode.
This behavior corroborates what we mentioned at the beginning of this section: eccentricity modulates the overall ringdown amplitude, while the structure of QNM mixing and the relative excitation pattern are largely insensitive to it. We also investigated this behaviour in the other regions of the parameter space and for the other modes considered in this work and we found the same results, that we are not explicitly showing here.

\subsubsection{Impact of relativistic anomaly on the merger-ringdown} \label{sec.:impact of relativistic anomaly on the merger-ringdown}
\begin{figure*}[tp!]
  	\includegraphics[width=1.\linewidth]{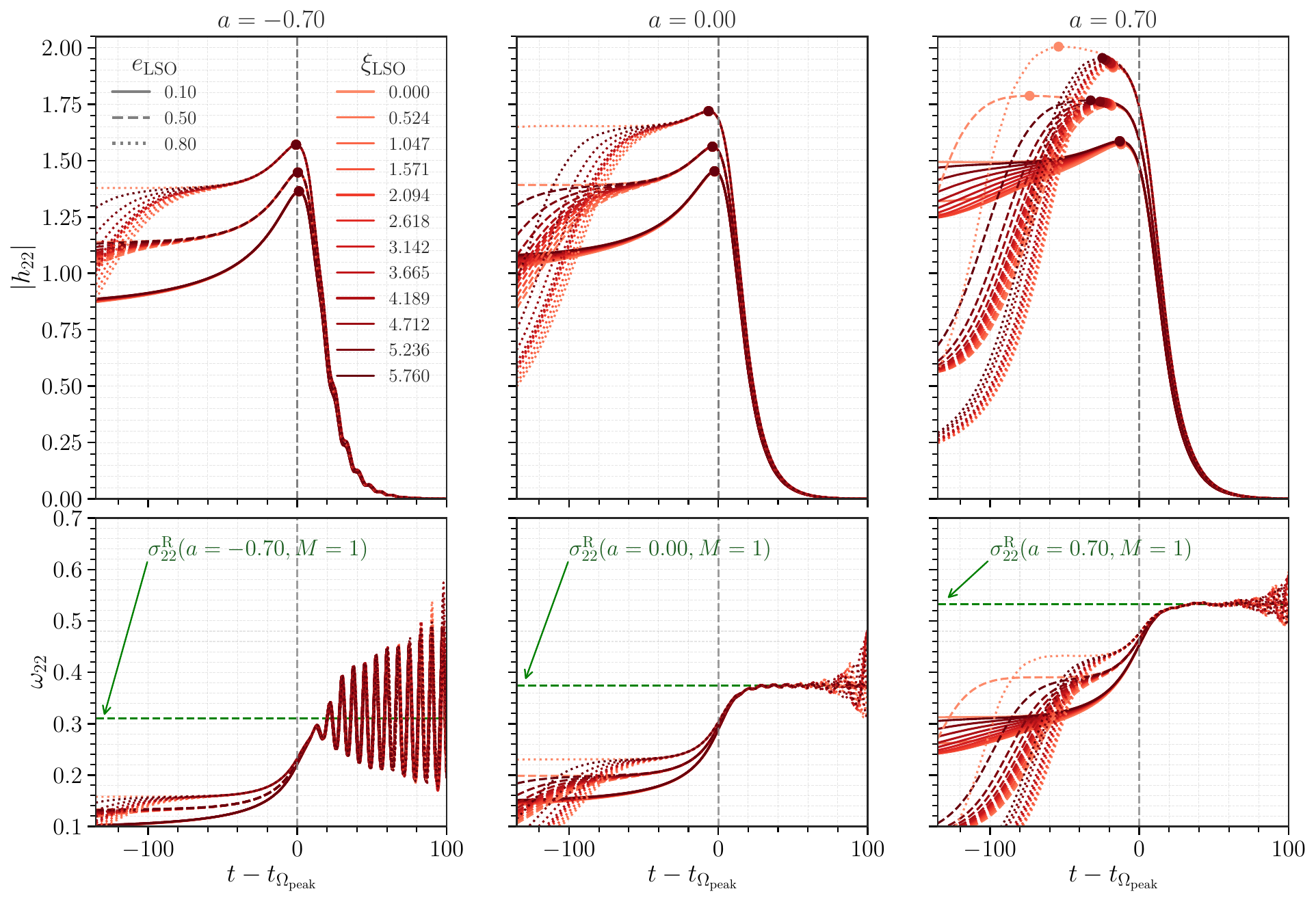}
  	\caption{Effect of the value of the relativistic anomaly $\xi_{\rm LSO}$, on the MR features of the $h_{22}$ mode. Similarly to Fig.~\ref{fig:effects of eccentricity}, we plot the waveform amplitude $|h_{22}|$ (top panels) and the instantaneous GW frequency $\omega_{22}$ (bottom panels) as functions of time relative to the peak of the orbital frequency, $t-t_{\Omega_{\rm peak}}$.
We consider three values of the spin, with $a=-0.70$ shown in the left column, $a=0.00$ in the central column, and $a=0.70$ in the right column.
For each spin configuration, we fix three different values of the eccentricity $e_{\rm LSO}$: solid lines correspond to $e_{\rm LSO}=0.10$, dashed lines to $e_{\rm LSO}=0.50$, and dotted lines to $e_{\rm LSO}=0.80$. The relativistic anomaly spans the interval $\xi_{\rm LSO}\in[0,2\pi)$ and is represented by using different shades of red, with the lightest tone corresponding to $\xi_{\rm LSO}=0.000$ and the darkest tone to $\xi_{\rm LSO}=5.760$. For the cases $a=-0.70$ and $a=0.00$, variations in $\xi_{\rm LSO}$ do not produce visible changes in either the amplitude or frequency features of the MR signal within the time interval $(t-t_{\Omega_{\rm peak}})\in[-10,100]$.
For $a=0.70$, this behavior persists at low eccentricity ($e_{\rm LSO}=0.10$), while for $e_{\rm LSO}\ge0.50$ the merger features show a dependence on $\xi_{\rm LSO}$, reflected in shifts in the timing and in differences in the height of the amplitude peak. The strongest deviations occur for values of $\xi_{\rm LSO} \simeq 0$, where an extended quasi-circularization of the trajectories at the UCO leads to a visible flattening of the waveform prior to the plunge regime. In contrast, the post-merger amplitude and frequency remain unaffected by $\xi_{\rm LSO}$. In the ringdown regime, $\omega_{22}$ settles around the real part of the dominant QNM frequency $\sigma_{22}$, whose value $\sigma^{\rm R}_{22}$ is indicated in the bottom panels by dashed green horizontal lines. The oscillations observed in $\omega_{22}$ around $\sigma^{\rm R}_{22}$ have an amplitude that depends on the spin and reflect the QNM mixing introduced in Sec.~\ref{sec.:Anatomy of the ringdown}.
}
  	\label{fig:effects of rel anomaly}
\end{figure*}
\begin{figure}[tp!]
\includegraphics[width=1.\linewidth]{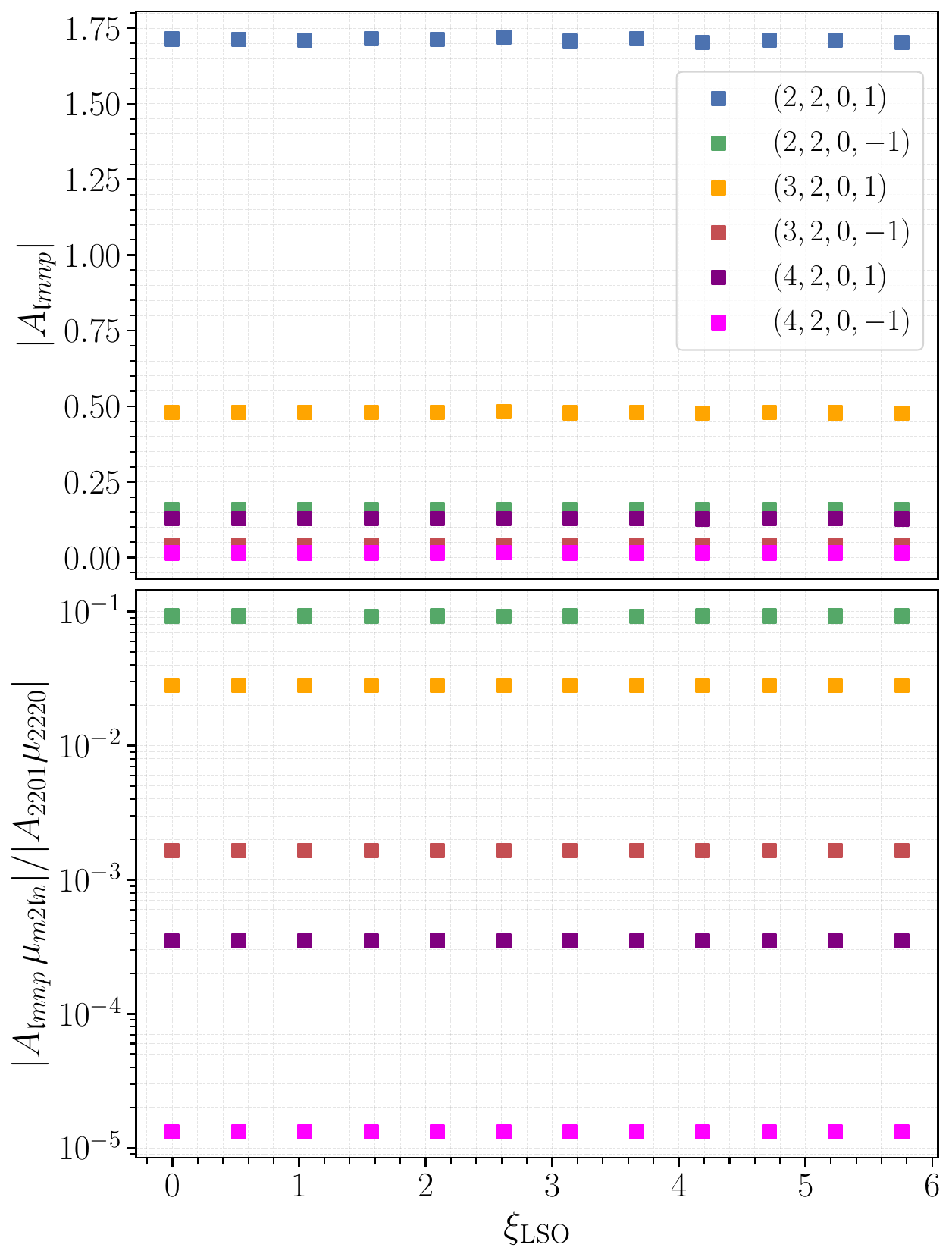}
\caption{Dependence of the QNMs excitation on $\xi_{\rm LSO}$, for the case $a=-0.70$ and $e_{\rm LSO}=0.50$.
The top panel shows the absolute values of the excitation amplitudes $|A_{\mathfrak{l} m n p}|$ introduced in Eq.~\eqref{eq: QNM spheroidal decomp} for selected fundamental prograde and retrograde QNMs with $m=2$, plotted as functions of $\xi_{\rm LSO}$.
In the bottom panel we plot the ratios $|A_{\mathfrak{l} m n p}\,\mu_{m 2 \mathfrak{l} n}(a \sigma_{\mathfrak{l} m n p})|/|A_{2201}\mu_{2 2 2 0}(a \sigma_{2 2 0 1})|$, which include the spheroidal--spherical harmonic mixing factors $\mu_{m 2 \mathfrak{l} n}(a \sigma_{\mathfrak{l} m n p})$ for the $h_{22}$ spherical mode.
Both the absolute excitation amplitudes and their relative ratios remain constant over the full interval $\xi_{\rm LSO}\in[0,2\pi)$, indicating that, for fixed spin and eccentricity, the relativistic anomaly does not play a relevant role in shaping either the overall strength or the hierarchy of QNMs excitation in the MR. The amplitudes showed in this figure are referred at the light-ring crossing time of the TM.
}
\label{fig:QNMS ampl vs rel_anomaly}
\end{figure}
After discussing the impact of eccentricity on the MR part of the waveforms, we now examine the effects of varying the relativistic anomaly on the MR features.
In this section, we analyze how variations of $\xi_{\rm LSO}$ affect the MR properties of the dominant $h_{22}$ mode, for fixed eccentricity and across different spin configurations.
Similarly to what we described in the previous section, in Fig.~\ref{fig:effects of rel anomaly} we show the dependence of the MR properties of the $h_{22}$ mode by varying the relativistic anomaly $\xi_{\rm LSO}$ for three values of the spin, $a=-0.70$ (left column), $a=0.00$ (central column), and $a=0.70$ (right column), and by fixing the eccentricity $e_{\rm LSO}$. The top panels display the waveform amplitude $|h_{22}|$, while the bottom panels show the instantaneous GW frequency $\omega_{22}$, as functions of $t-t_{\Omega_{\rm peak}}$. In the figure, we consider three fixed values of the eccentricity: solid lines correspond to $e_{\rm LSO}=0.10$, dashed lines to $e_{\rm LSO}=0.50$, and dotted lines to $e_{\rm LSO}=0.80$. For each spin value and fixed eccentricity, we span the full range $\xi_{\rm LSO}\in[0,2\pi)$. Variations of the relativistic anomaly $\xi_{\rm LSO}$ are encoded using different shades of red, with the lighter tone corresponding to $\xi_{\rm LSO}=0.0000$ and the darker tone corresponding to $\xi_{\rm LSO}=5.7600$. The vertical dashed grey line in each panel marks the reference time $t=t_{\Omega_{\rm peak}}$.

For $a=-0.70$ and $a=0.00$, the MR portion of the waveforms exhibits no visible dependence on the relativistic anomaly.
For each value of the eccentricity, the amplitude evolution around merger, within the time window $(t-t_{\Omega_{\rm peak}})\in[-10,10]$, including both the height and the timing of the amplitude peak and the subsequent ringdown decay, remains unchanged as $\xi_{\rm LSO}$ is varied.
Consistently, the corresponding instantaneous frequency $\omega_{22}$ evolution overlaps for all values of $\xi_{\rm LSO}$, with this overlap persisting into the post-merger oscillatory regime associated with QNM mixing which starts to manifest at $(t-t_{\Omega_{\rm peak}})\gtrsim17$.

For $a=0.70$, we observe that the relativistic anomaly leaves an imprint on the waveform features only in specific eccentricity regimes.
Specifically, at low eccentricity, $e_{\rm LSO}=0.10$, the waveform exhibits the same behaviour observed for the lower-spin configurations, with no visible dependence on $\xi_{\rm LSO}$ in either the amplitude or the instantaneous frequency features around merger ($(t-t_{\Omega_{\rm peak}}) \sim -10$) and in the post-merger parts.
At larger eccentricities, $e_{\rm LSO}\geq0.50$, variations of the relativistic anomaly begin to leave a visible imprint on the merger features. In particular, both the height of the amplitude peak and its time location display a clear dependence on $\xi_{\rm LSO}$, with different values of the relativistic anomaly producing distinct merger amplitudes and peak timings. For $e_{\rm LSO}=0.50$, the peak value of the $h_{22}$ amplitude varies at the level of $\sim1.5\%$, while the corresponding peak time spans the interval $(t-t_{\Omega_{\rm peak}})\in[-80,-20]$. For $e_{\rm LSO}=0.80$, the fractional variation of the amplitude peak increases to $\sim4\%$, whereas the associated peak times are distributed over the narrower interval $(t-t_{\Omega_{\rm peak}})\in[-60,-20]$.
This behaviour can be understood in light of the results of Ref.~\cite{Faggioli:2025hff}, where it was shown that, in the TP limit, for equatorial critical plunge geodesics of Kerr starting from the UCO, the waveform amplitude does not necessarily exhibit a maximum for high values of eccentricity.
While this effect is present for every spin value (see Fig.~4 of Ref.~\cite{Faggioli:2025hff}), for spin values $a\gtrsim0.30$ the associated eccentricity threshold lies within the bound regime, $e<1$.
Here the eccentricity threshold is defined as the value of $e_{\rm LSO}$ above which the waveform amplitude no longer reaches its maximum during the plunge phase, but instead peaks earlier during the transition from inspiral to plunge.
As a result, when $e_{\rm LSO}$ exceeds this threshold, the amplitude peak is reached during the inspiral--plunge transition rather than during the pure plunge.
Since the orbital dynamics of the trajectory between the LSO crossing and the onset of the plunge is affected by the relativistic anomaly, variations of $\xi_{\rm LSO}$ consequently influence both the timing and the height of the waveform amplitude peak.

We remark that this behavior becomes increasingly pronounced as $\xi_{\rm LSO}\simeq 0$. In this limit, and consistently across all the waveforms considered, we observe the effects of an extended quasi-circularization of the trajectory at the UCO. This prolonged dwelling close to the UCO manifests in the waveform as a marked flattening of the amplitude and of the frequency prior to the plunge (see the lightest shades of red in Fig.~\ref{fig:effects of rel anomaly}), delaying the growth toward the merger peak in those configurations where the amplitude maximum is reached during the pure plunge phase, namely for $a=-0.70$ and $a=0.00$ at all eccentricities, and for $a=0.70$ when $e_{\rm LSO}\le0.50$. It is important to emphasize, however, that this quasi-circularization is an exponentially fine-tuned effect~\cite{Gundlach:2012aj}. In fact, it occurs only when $\xi_{\rm LSO} \simeq 0$ and is therefore restricted to a very narrow range of relativistic anomaly values. As a result, while this behavior provides a clear illustration of how $\xi_{\rm LSO}$ can influence the late inspiral dynamics, its impact is limited to a narrow subset of trajectories.

Despite the sensitivity of the merger features to the relativistic anomaly at large spin and eccentricity, we find that the post-merger remains unaffected by the value of $\xi_{\rm LSO}$. Indeed, by inspecting the $a=0.70$ case in Fig.~\ref{fig:effects of rel anomaly}, one can observe that for $(t-t_{\Omega_{\rm peak}})\gtrsim -10$ both the waveform amplitude and the instantaneous frequency $\omega_{22}$ corresponding to different values of $\xi_{\rm LSO}$ closely overlap. This indicates that, even in the high-spin, high-eccentricity regime, the ringdown is insensitive to variations of the relativistic anomaly.

These results have direct implications for the fitting strategy adopted in this work. For $a=-0.70$ and $a=0.00$, no visible dependence on $\xi_{\rm LSO}$ is observed across the entire MR. Similarly, for $a=0.70$, the waveform morphology of the MR remains insensitive to variations of the relativistic anomaly when $e_{\rm LSO}<0.50$.
More generally, for increasing spin values the eccentricity threshold above which the merger features become sensitive to $\xi_{\rm LSO}$ is expected to decrease, as shown in Ref.~\cite{Faggioli:2025hff}, where for $a=0.90$ this threshold is found to be $e_{\rm LSO}\simeq0.10$.
However, our analysis shows that, even when $e_{\rm LSO}$ exceeds this threshold, the largest deviations of the merger features associated with variations of $\xi_{\rm LSO}$ arise only in highly fine-tuned configurations characterized by an extended quasi-circularization near the UCO. Such configurations occur within very narrow intervals of $\xi_{\rm LSO}$. When excluding these finely tuned cases, the dependence on $\xi_{\rm LSO}$ typically induces differences in the peak features at the level of $\sim2\%$.
On the basis of these observations, we do not include $\xi_{\rm LSO}$ as an independent fitting parameter in the construction of the hierarchical fit introduced in Sec.~\ref{sec.:hierarchical fit}. In doing so, we are aware of the fact that, in the region of the parameter space where $\xi_{\rm LSO}$ introduces the mild $2 \%$ shift of the peak features, the associated effects remain subdominant compared to those driven by solely varying eccentricity and spin. We thus decide to neglect these variations in our model and postpone the more challenging modeling of the effects on the merger features due to the quasi-circularization at the UCO to the future.
\begin{figure*}[tp!]
  	\includegraphics[width=1.\linewidth]{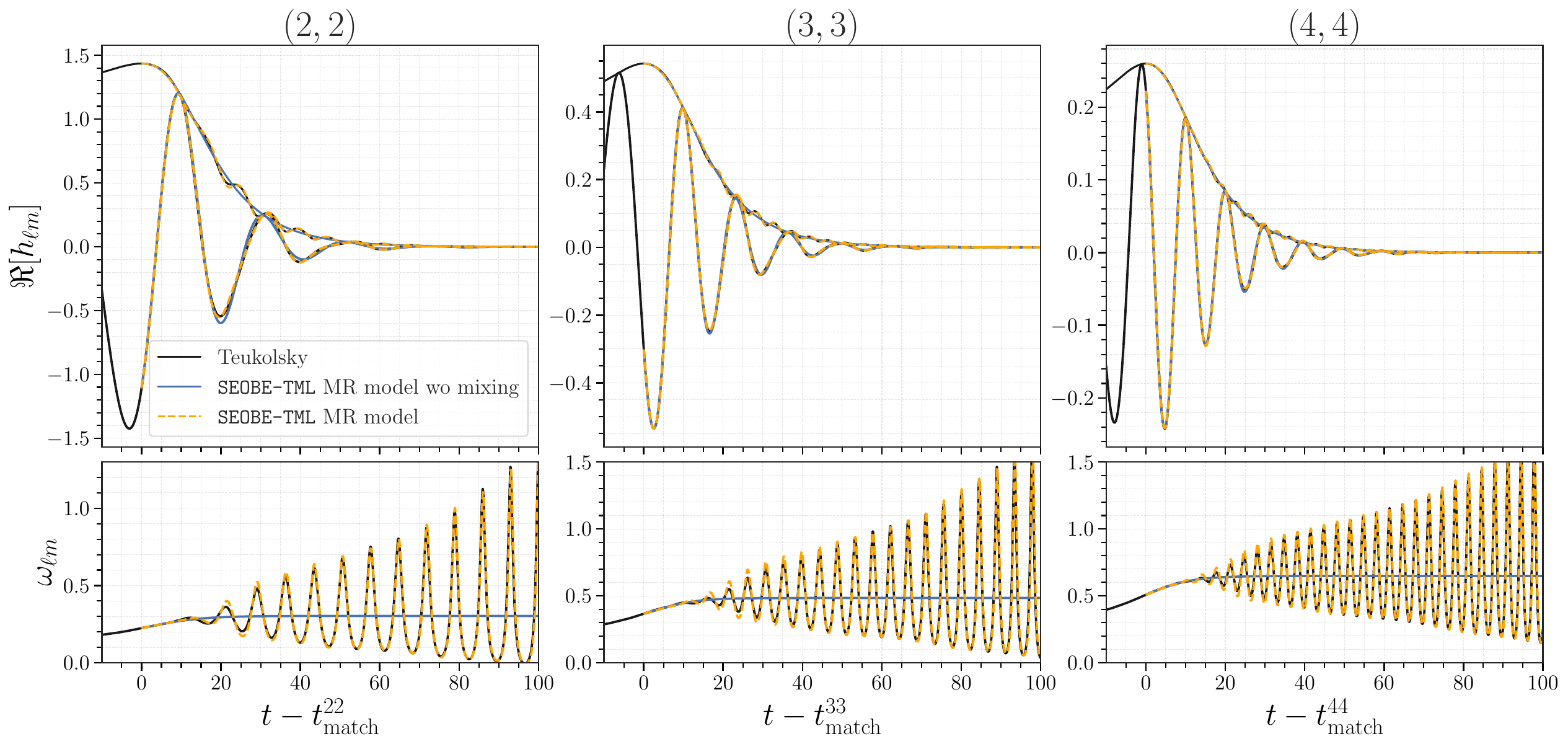}
  	\caption{
Comparison between Teukolsky waveforms and the \texttt{SEOBE-TML} MR model for a representative eccentric configuration with spin $a=-0.8$ and eccentricity $e_{\rm LSO}=0.50$. The three columns correspond to the $(2,2)$, $(3,3)$, and $(4,4)$ modes. In each column, the top panel shows the real part of the waveform, $\Re[h_{\ell m}]$, while the bottom panel displays the instantaneous frequency $\omega_{\ell m}$ as a function of $t - t^{\ell m}_{\rm match}$. The black solid curves are connected to the Teukolsky waveforms, the blue solid curves show the \texttt{SEOBE-TML} MR model without QNM mixing, and the orange dashed curves correspond to the complete model including QNM mixing. The inclusion of QNM mixing is essential to reproduce the oscillatory modulations starting from the intermediate-times of the ringdown for $(t - t^{\ell m}_{\rm match}) \ge 17$.}
  	\label{fig:ell is m waveforms with freq}
\end{figure*}

To further support this conclusion at the level of the solely ringdown physics, it is instructive to directly examine how the excitation of the individual QNMs depends on the relativistic anomaly.
In particular, we now focus on the behaviour of the QNM amplitudes themselves, in order to assess whether variations of $\xi_{\rm LSO}$ leave any imprint.
In Fig.~\ref{fig:QNMS ampl vs rel_anomaly} we explore the dependence of the QNM excitations on $\xi_{\rm LSO}$, for the fixed configuration $a=-0.70$ and $e_{\rm LSO}=0.50$.
The upper panel shows the absolute values of the excitation amplitudes $|A_{\mathfrak{l} m n p}|$ for the same representative set of QNMs considered in Fig.~\ref{fig:QNMS ampl vs eccentricity}, i.e. the fundamental prograde and retrograde $(2,2,0,\pm1)$, $(3,2,0,\pm1)$, and $(4,2,0,\pm1)$ modes, as functions of $\xi_{\rm LSO}$. The amplitudes showed in Fig.~\ref{fig:QNMS ampl vs rel_anomaly} are referred at the light-ring crossing time of the TM.
In contrast to the increasing monotonic dependence of $|A_{\mathfrak{l} m n p}|$ observed when varying the eccentricity, in this case all amplitudes remain essentially constant over the full interval $\xi_{\rm LSO}\in[0,2\pi)$. The careful reader may notice a slight dependence of $|A_{2201}|$ on $\xi_{\rm LSO}$; however, these variations are at the level of $\le 0.4\%$ and are attributable to numerical noise in both the waveforms and the QNM extraction procedure.
Although not shown explicitly here, we verified that the QNM amplitudes exhibit the same independence on $\xi_{\rm LSO}$ for other values of $a$ and $e_{\rm LSO}$, as well as for all additional modes that are extracted from the Teukolsky waveforms.
We remark that this result contrasts with the findings of Ref.~\cite{Becker:2025zzw}, where the QNM amplitudes extracted at the light-ring crossing time were found to depend on the relativistic anomaly, exhibiting a noticeable decrease near the quasi-circularization at the UCO, i.e. when $\xi_{\rm LSO} \simeq 0$. In that work, this behaviour is referred to as \textit{chifurcation}. In our analysis, we do not observe any evidence of such a dependence on $\xi_{\rm LSO}$.
We believe that this difference may primarily originate from the distinct treatment of the inspiral-to-plunge transition of the TM dynamics adopted in our work and in Ref.~\cite{Becker:2025zzw}. An additional source of discrepancy may arise from the fact that, in Ref.~\cite{Becker:2025zzw}, the eccentricity and relativistic anomaly are defined at an earlier reference point in the inspiral, rather than at the LSO as in the present work. As a consequence, trajectories characterized by different values of $\xi$ are not guaranteed to reach the LSO with the same eccentricity, which may in turn affect the inferred QNM excitation. While this effect is expected to be negligible in the small-mass-ratio regime, we cannot currently exclude its impact. Finally, the use of different QNM extraction procedures in the two works may also contribute to the observed discrepancies.
We note, however, that Ref.~\cite{Becker:2025zzw} employs the same TD Teukolsky code~\cite{Sundararajan:2007jg,Sundararajan:2008zm,Sundararajan:2010sr,Zenginoglu:2011zz,Field:2020rjr} used in our work, allowing us to exclude waveform generation as the origin of the observed differences. A more detailed investigation of this discrepancy is left to future work.

To assess whether the relativistic anomaly leaves any imprint on the relative QNM content of the spherical ringdown modes, in the bottom panel of Fig.~\ref{fig:QNMS ampl vs rel_anomaly} we plot the ratios $|A_{\mathfrak{l} m n p}\,\mu_{m 2 \mathfrak{l} n}(a \sigma_{\mathfrak{l} m n p})|/|A_{2201}\mu_{2 2 2 0}(a \sigma_{2 2 0 1})|$, which directly quantify the relative excitation of the different QNMs contributing to the $h_{22}$ mode.
These ratios are constant across the entire range of $\xi_{\rm LSO}$, spanning several orders of magnitude without exhibiting any appreciable dependence on the relativistic anomaly. This confirms that, for fixed spin and eccentricity, variations in $\xi_{\rm LSO}$ do not affect either the absolute or the relative excitation of the dominant QNMs. Together with the results obtained by varying $e_{\rm LSO}$, Fig.~\ref{fig:QNMS ampl vs rel_anomaly} provides direct evidence that the relativistic anomaly does not play a relevant role in shaping the QNMs excitation in the ringdown, reinforcing the conclusion that, in the TML, $\xi_{\rm LSO}$ can be safely neglected as an independent parameter in the modeling of equatorial spin-eccentric MR waveforms.
\subsection{The spin-eccentric merger-ringdown model} \label{sec.:the spin-eccentric merger-ringdown model}
\begin{figure*}[tp!]
  	\includegraphics[width=1.\linewidth]{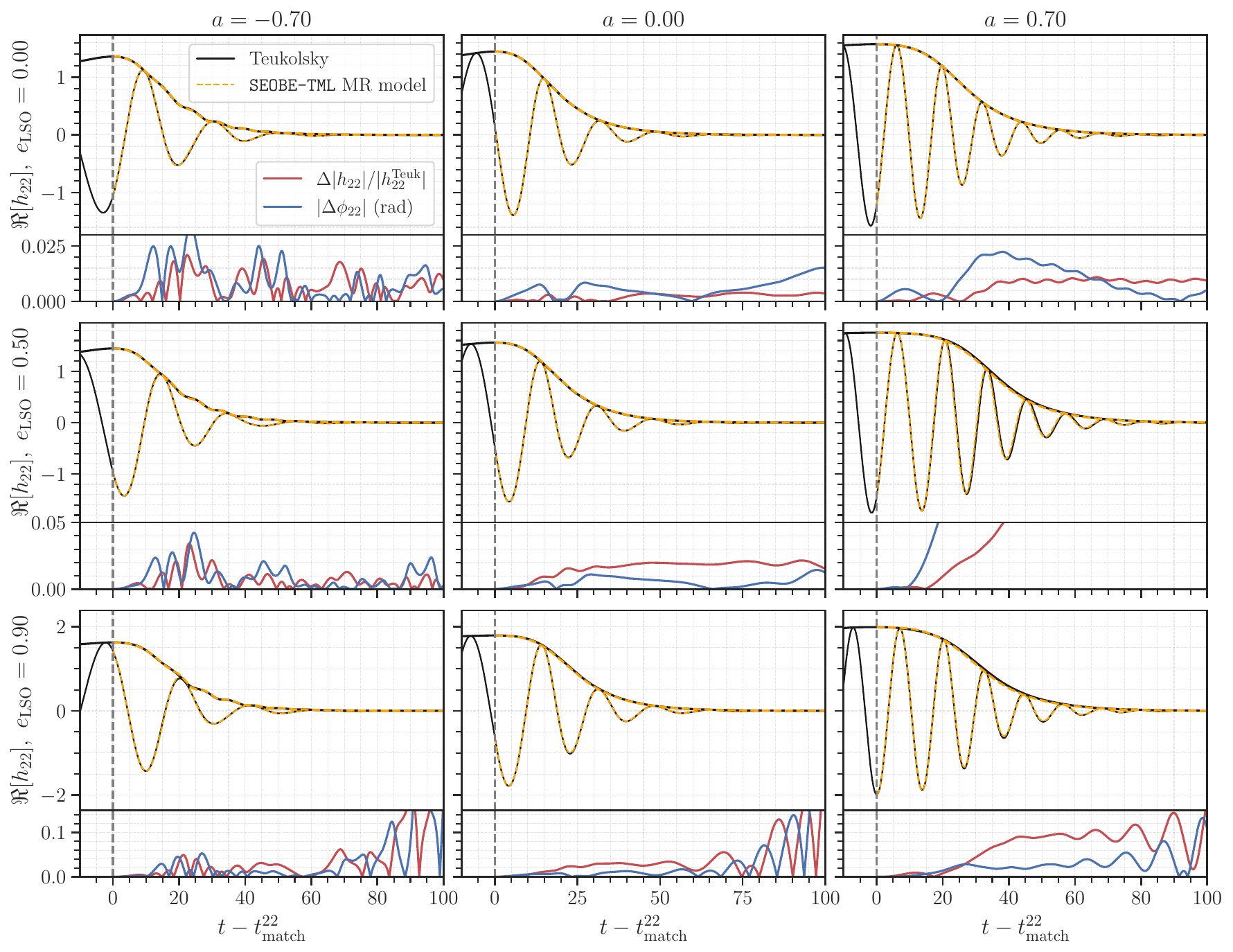}
  	\caption{Comparison between the Teukolsky $(2,2)$ waveforms and the \texttt{SEOBE-TML} MR model across different spins and eccentricities. The three columns correspond to $a=-0.70$, $a=0.00$, and $a=0.70$, while the three rows correspond to $e_{\rm LSO}=0.00$, $0.50$, and $0.90$, respectively. In each panel, the upper plot shows the real part of the waveform, $\Re[h_{22}]$, with the Teukolsky waveform (black solid line) and the \texttt{SEOBE-TML} MR model (orange dashed line). The lower plot in each panel displays the relative amplitude difference, $\Delta |h_{22}|/|h^{\rm Teuk}_{22}|$ (red curve), and the phase difference, $\Delta \phi_{22}$ (blue curve), as functions of $t - t^{22}_{\rm match}$.
}
  	\label{fig:h22_waveforms}
\end{figure*}
\begin{figure*}[tp!]
  	\includegraphics[width=1.\linewidth]{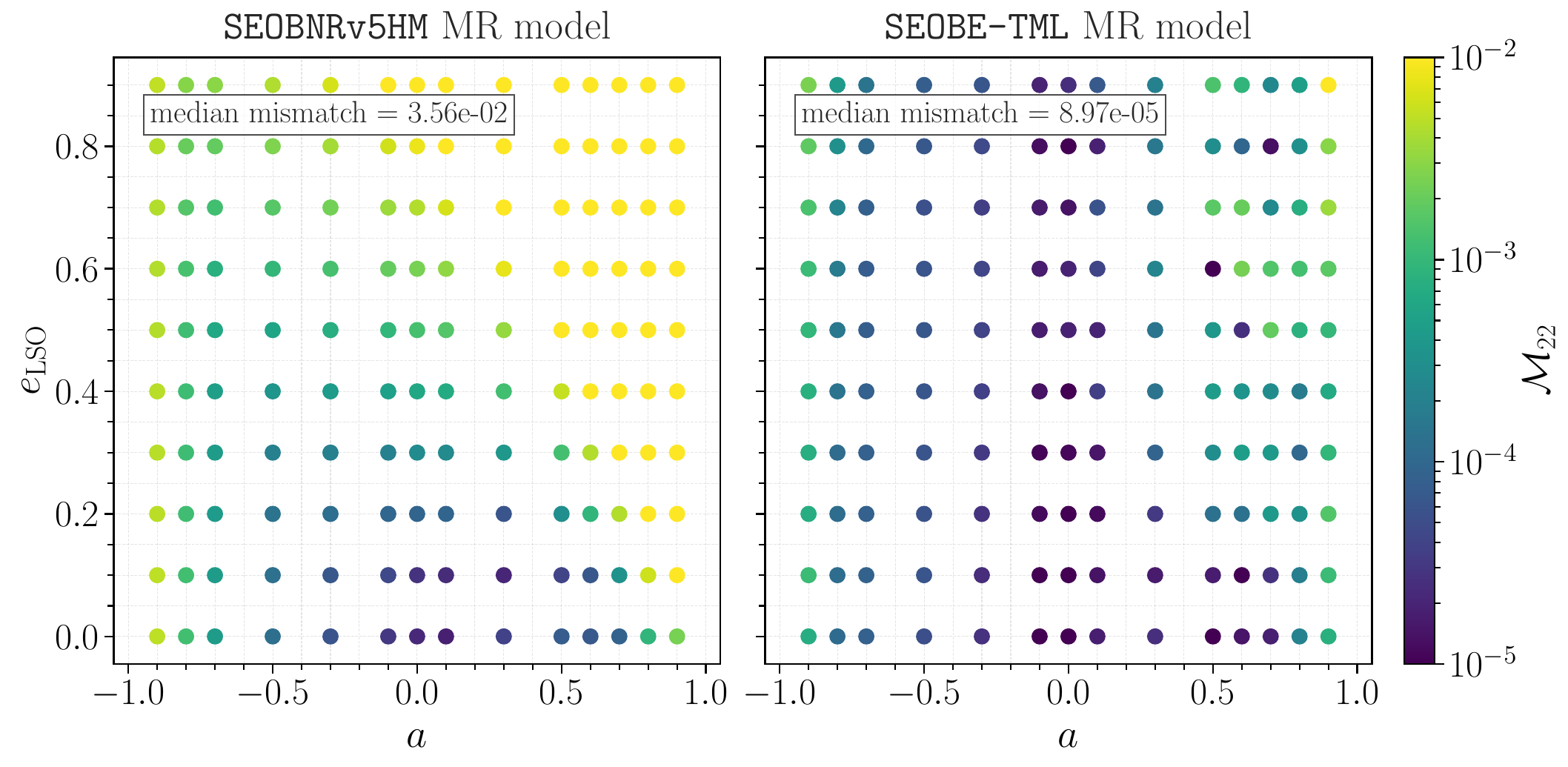}
  	\caption{Mismatch $\mathcal{M}_{22}$ between Teukolsky waveforms and MR models across the parameter space spanned by the BH spin $a$ and the eccentricity $e_{\rm LSO}$. Each dot corresponds to a waveform configuration, with the color indicating the mismatch value on a logarithmic scale. The left panel shows the results obtained with the \texttt{SEOBNRv5HM} MR model, while the right panel displays the mismatches for the \texttt{SEOBE-TML} MR model introduced in this work. The median mismatch over the explored parameter space is reported in each panel. The \texttt{SEOBE-TML} MR model achieves significantly smaller mismatches and more accuracy across spins and eccentricities compared to the \texttt{SEOBNRv5HM} MR model.}
  	\label{fig:h22_mismatches}
\end{figure*}
\begin{figure*}[tp!]
  	\includegraphics[width=1.\linewidth]{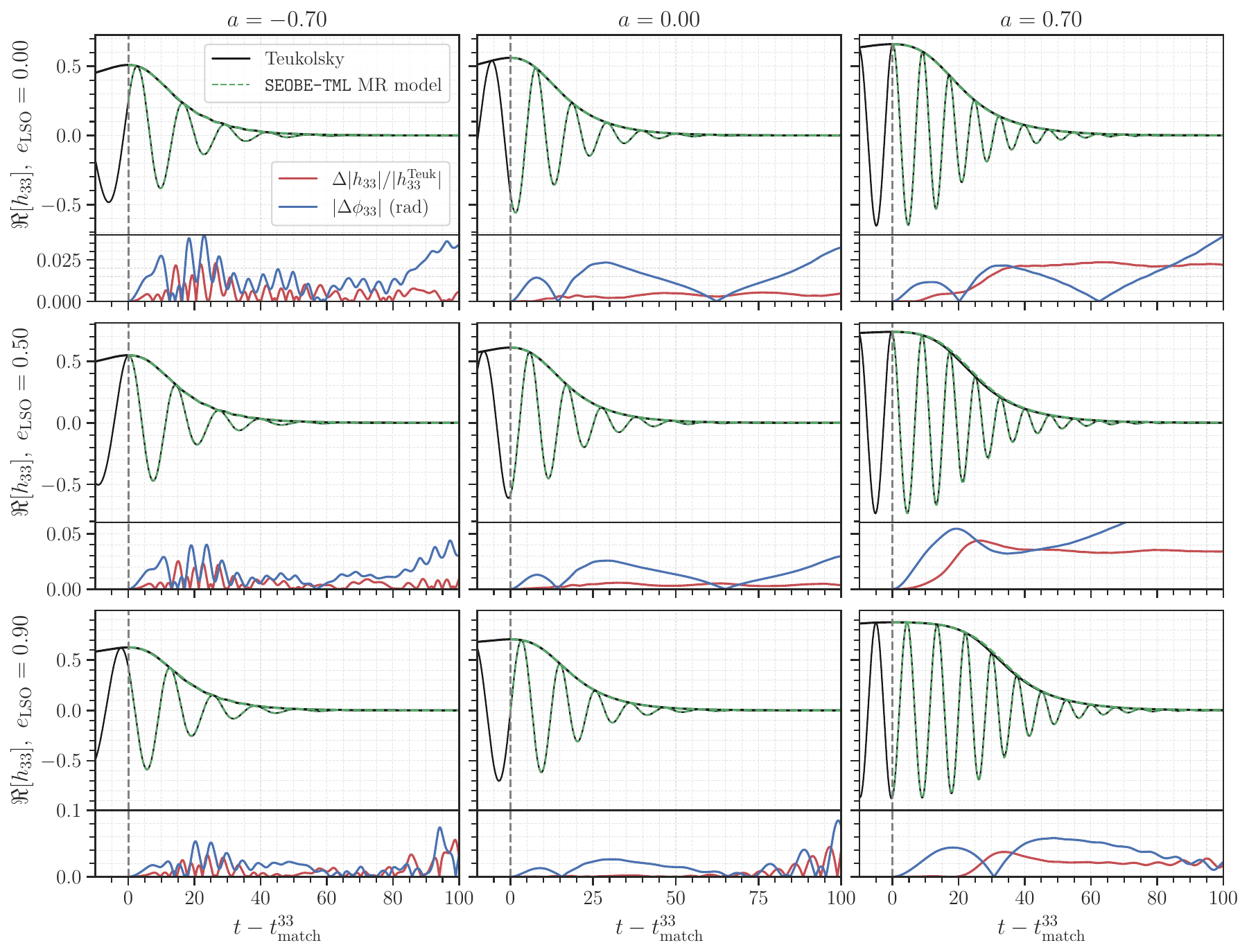}
  	\caption{Comparison between the Teukolsky $(3,3)$ waveforms and the \texttt{SEOBE-TML} MR model, considering the same configurations shown in Fig.~\ref{fig:h22_waveforms}. The real part of the Teukolsky waveform $\Re[h_{33}]$ (black) is compared with the \texttt{SEOBE-TML} MR model (green dashed). The lower panels display the relative amplitude difference $\Delta |h_{33}|/|h^{\rm Teuk}_{33}|$ (red) and the phase difference $\Delta \phi_{33}$ (blue), as functions of $t - t^{33}_{\rm match}$.}
  	\label{fig:h33_waveforms}
\end{figure*}
\begin{figure*}[tp!]
  	\includegraphics[width=1.\linewidth]{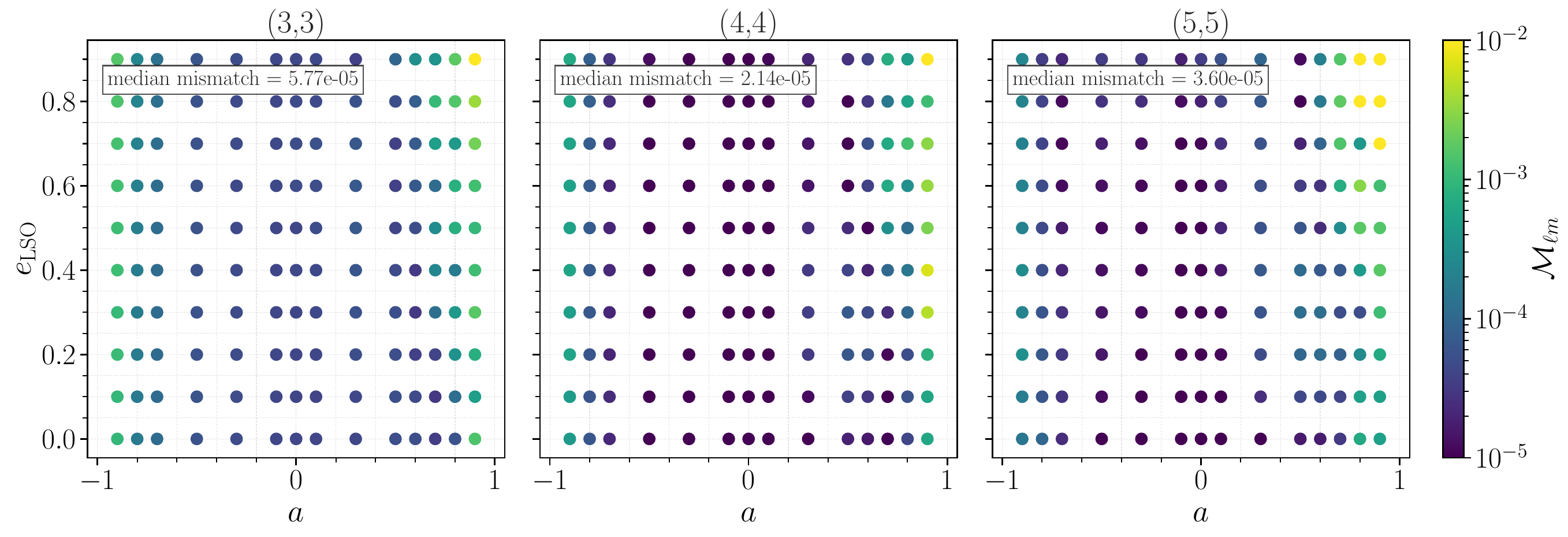}
  	\caption{
Mismatch $\mathcal{M}_{\ell m}$ between the Teukolsky waveforms and the \texttt{SEOBE-TML} MR model across the parameter space spanned by the BH spin $a$ and the eccentricity $e_{\rm LSO}$ for the higher $\ell = m$ multipoles. The three panels correspond to the $(3,3)$, $(4,4)$ and $(5,5)$ modes. Each dot represents a configuration in the $(a,e_{\rm LSO})$ parameter space, while the color encodes the value of the mismatch on a logarithmic scale. The median mismatch across the explored parameter space is reported in each panel, showing that the \texttt{SEOBE-TML} MR model maintains good accuracy also for the subdominant multipoles.}
  	\label{fig:hlm_mismatches}
\end{figure*}
In this section we provide the main results of our \texttt{SEOBE-TML} MR model. We provide an overall review of the performances of the model splitting between the $\ell = m$ and $\ell \neq m$ modes.
\subsubsection{$\ell = m$ modes} \label{Sec.: ell is m modes}
We begin by examining a representative configuration in which QNM mixing effects are more pronounced for the $\ell = m$ modes, namely a case with negative spin. This choice allows us to highlight the impact of the mode mixing modeling in a regime where it is expected to play a more significant role, before turning to a broader assessment of the model performance over the full parameter space.

In Fig.~\ref{fig:ell is m waveforms with freq} we show the comparison between the Teukolsky waveforms and the \texttt{SEOBE-TML} MR model for a representative eccentric configuration with spin $a=-0.8$ and eccentricity $e_{\rm LSO}=0.50$. The three columns correspond to the $(2,2)$, $(3,3)$, and $(4,4)$ modes, respectively. In each column, the top panel displays the real part of the waveform, $\Re[h_{\ell m}]$, while the bottom panel shows the instantaneous frequency $\omega_{\ell m}$, as functions of $t - t^{\ell m}_{\rm match}$. The black solid curves represent the Teukolsky waveforms, the blue solid curves denote the \texttt{SEOBE-TML} MR model without QNM mixing, and the orange dashed curves represent the complete \texttt{SEOBE-TML} MR model, included with QNM mixing.
For the dominant $(2,2)$ mode (left column), the MR model reproduces the overall MR behaviour of the Teukolsky waveform with good accuracy. When QNM mixing is not included, the waveform frequency and amplitude settle smoothly toward those of the dominant fundamental QNM. However, this simplified description does not capture the oscillatory modulations visible in both the amplitude and the instantaneous frequency, which arise from interference among the different QNM contributions. 
When QNM mixing is incorporated in the modeling, these oscillatory features are accurately reproduced. In particular, the frequency exhibits the characteristic modulations around its asymptotic value, and the corresponding amplitude oscillations are captured with good agreement.
The same behaviour is reflected in the subdominant $(3,3)$ (central column) and $(4,4)$ (right column) modes. In both cases, the model accurately reproduces the early part of the MR and the inclusion of QNM mixing accurately reproduces the oscillatory structure present during the late ringdown part. When QNM mixing is not included, the waveform smoothly approaches the dominant QNM behaviour, but the modulations arising from interference among multiple QNMs are not captured. 
By incorporating QNM mixing, the model successfully reproduces the subsequent intermediate-time oscillations around the asymptotic least-damped QNM frequency. This agreement is visible not only in the waveform amplitude, but also in the instantaneous frequency, where the oscillations are tracked with high accuracy. 
Although not explicitly shown in the figure, we observe the same behaviour for the $(5,5)$ mode, where the inclusion of QNM mixing similarly improves the agreement with the Teukolsky waveforms during the MR.
Overall, for the eccentric and negative-spin configuration considered here ($a=-0.8$, $e_{\rm LSO}=0.50$), the \texttt{SEOBE-TML} MR model provides an accurate description of the MR across all the considered $\ell = m$ modes. The role of QNM mixing is relevant for all the multipoles, as it is essential to reproduce the oscillatory structure in both amplitude and frequency.
We note that for the positive spins the overall agreement remains very good across all the $\ell = m$ modes. In that regime, QNM mixing effects become progressively less relevant, and the MR waveform is largely dominated by the least damped fundamental QNM at intermediate times after the attachment time. As a result, even the simplified model without explicit mixing is already able to capture the main features of the ringdown with high accuracy, and the inclusion of mixing leads to further marginal corrections.

Having illustrated the behaviour of the model for a representative configuration, we now turn to a systematic assessment of its performance over the full parameter space. We begin by focusing on the $(2,2)$ mode, which carries the largest fraction of the radiated energy.
Figure~\ref{fig:h22_waveforms} shows the comparison between the Teukolsky waveforms and the \texttt{SEOBE-TML} MR model for the dominant $(2,2)$ mode, for three representative values of the BH spin, $a=-0.70$ (left column), $a=0.00$ (central column), and $a=0.70$ (right column), and increasing eccentricity at the LSO, $e_{\rm LSO}= [0.00,\,0.50,\,0.90]$ (from top to bottom). For each configuration, we display the real part of the waveform, $\Re[h_{\ell m}]$, together with the relative amplitude difference $\Delta |h_{22}|/|h^{\rm Teuk}_{22}|$ (red) and the phase difference $\Delta \phi_{22}$ (blue), as functions of time measured from the $t^{22}_{\rm match}$. Overall, the MR model (dashed orange) shows very good agreement with the Teukolsky waveforms (black) across the majority of spins and eccentricities, accurately reproducing both the merger morphology and the subsequent ringdown part, including the oscillations due to QNMs mixing. In the QC case ($e_{\rm LSO}=0$), amplitude and phase differences remain small throughout the MR, typically at the level of $\le 2.5 \%$ for the fractional difference of the amplitude and $0.025$ rad for the phase difference. As the eccentricity increases, deviations appear after the merger, especially for $e_{\rm LSO}=0.90$, where the amplitude fractional differences can reach $10 \%$ at late times, when $(t - t_{22}^{\rm match}) \ge 80$. This disagreement is due to the interference with the late-time tails, which are contributing more at earlier times in the ringdown for high eccentricities and which are not modeled in this work. This late-time tails excitation also reflects in an increased dephasing, as shown in the bottom panels of the third row of Fig.~\ref{fig:h22_waveforms}. We also remark the fact that the case $a = 0.70$ and $e_{\rm LSO} = 0.50$ shows a particular enhanced disagreement between the MR model and the numerical waveform, due to the fact that we found particularly challenging capturing the prograde high-spin scenarios with our fits for some configurations. For this reasons in some parts of the parameter space characterized by values of the spin $a \ge 0.70$, the model presents some enhanced disagreements. However, if we do not consider the cases where the late-time excitation occurs for times $t - t^{22}_{\rm match} \le 100$, overall the model is able to capture the phenomenology of the merger and of the post-merger showing fractional differences that are on average $\le 10 \%$ and phase differences $ \le 0.1$.

To provide a metric of the performances of the \texttt{SEOBE-TML} MR model all over the parameter space we compute the mismatch, defined in Eq.~\eqref{eq.: def mismatch}, of the model with the numerical waveforms.
In Fig.~\ref{fig:h22_mismatches}, we show the $h_{22}$ mismatch, $\mathcal{M}_{22}$, across the parameter space, for two different MR models. The left panel corresponds to the mismatches of the \texttt{SEOBNRv5HM} (and \texttt{SEOBNRv5EHM}) MR model with the numerical waveforms, while the right panel shows the mismatches considering the \texttt{SEOBE-TML} MR model.
Each dot represents a configuration in the $(a, e_{\mathrm{LSO}})$ parameter space, with the color encoding the value of the mismatch $\mathcal{M}_{22}$ on a logarithmic scale. For the \texttt{SEOBNRv5HM} MR model, mismatches are systematically larger over the full parameter space, with a median value of $3.56 \times 10^{-2}$. The color distribution indicates a significant loss of accuracy, particularly at higher eccentricities and for large positive spins. Also the negative spin configurations (in particular $a \le -0.7$) are connected with higher mismatches, as the \texttt{SEOBNRv5HM} model does not contain any QNM mixing modeling.
In contrast, the \texttt{SEOBE-TML} MR model exhibits substantially smaller mismatches, with a median value of $8.97 \times 10^{-5}$. Most configurations lie in the $10^{-5}$--$10^{-4}$ range, especially for spin values $-0.8 \le a \le 0.4$, indicating an enhanced agreement across the explored spins and eccentricities. For spin values $a \ge 0.5$, we observe that the accuracy of the \texttt{SEOBE-TML} MR model begins to degrade, especially for large eccentricities. This behavior reflects the increased difficulty in modeling highly prograde spin configurations with the same level of accuracy achieved for smaller spin values discussed above. Despite this degradation, the overall mismatch level remains significantly reduced with respect to the \texttt{SEOBNRv5HM} MR model. Therefore, this figure highlights the substantial improvement achieved by the \texttt{SEOBE-TML} MR model in reproducing the $h_{22}$ waveform across the considered region of the parameter space.

Having established the overall performance of the \texttt{SEOBE-TML} MR model for the $(2,2)$ mode across the parameter space, we now turn to the higher $\ell = m$ multipoles. As a representative example, we focus in particular on the $(3,3)$ mode and inspect the behaviour of the model at the waveform level for selected configurations.
Figure~\ref{fig:h33_waveforms} shows the comparison between the Teukolsky waveforms and the \texttt{SEOBE-TML} MR model for the $(3,3)$ mode, for the same three representative values of the BH spin as in Fig.~\ref{fig:h22_waveforms}, i.e. $a=-0.70$ (left column), $a=0.00$ (central column), and $a=0.70$ (right column), and increasing eccentricity at the LSO, $e_{\rm LSO}= [0.00,\,0.50,\,0.90]$ (from top to bottom). For each configuration, we display the real part of the waveform, $\Re[h_{33}]$, together with the relative amplitude difference $\Delta |h_{33}|/|h^{\rm Teuk}_{33}|$ (red) and the phase difference $\Delta \phi_{33}$ (blue), as functions of time measured from $t^{33}_{\rm match}$. Overall, the MR model (dashed green) shows good agreement with the Teukolsky waveforms (black) across the majority of spins and eccentricities, reproducing well both the merger morphology and the subsequent ringdown behaviour, capturing the QNM mixing.
In the QC case ($e_{\rm LSO}=0$), amplitude and phase differences remain small throughout the MR, typically at the level of $\lesssim 2.5\%$ for the fractional amplitude difference and $\sim 0.025$ rad for the phase difference. However, we note that for $(t-t^{33}_{\rm match}) \gtrsim 60$ the phase difference starts to increase. This behaviour can be attributed to the fact that, although QNM mixing is captured particularly well, as shown in the second column of Fig.~\ref{fig:ell is m waveforms with freq}, a small mismatch remains between the modeled frequency oscillations and those of the numerical waveform. This discrepancy gradually accumulates in phase, leading to an increase of the dephasing, especially in the part of the ringdown that can be described as a superposition of multiple QNMs. As the eccentricity increases, deviations become more visible during the post-merger evolution. In particular, for $e_{\rm LSO}=0.90$, the phase difference tends to increase at late times, reaching values of $\sim 0.08$ rad for $(t-t^{33}_{\rm match}) \gtrsim 80$, while the amplitude fractional differences generally remain within $\sim 5\%$. 
We also remark that the configurations with positive spin, particularly $a = 0.70$, show a more pronounced disagreement between the MR model and the numerical waveform, especially for moderate and large eccentricities. This reflects the increased difficulty in accurately modeling some prograde high-spin scenarios with the current fits, as also mentioned when discussing Fig.~\ref{fig:h22_waveforms}. The model overall captures well the phenomenology of the merger and post-merger dynamics of the $(3,3)$ mode, typically yielding fractional amplitude differences $\lesssim 8\%$ and phase differences $\lesssim 0.08$ rad across the explored configurations. 

To complement the waveform comparison discussed above, we now assess the global performance of the \texttt{SEOBE-TML} MR model for the higher $\ell = m$ multipoles across the explored parameter space. In particular, we quantify the agreement with the Teukolsky waveforms through the mismatch $\mathcal{M}_{\ell m}$, allowing us to systematically evaluate the accuracy of the MR description for the $(3,3)$ mode and for the higher $(4,4)$ and $(5,5)$ multipoles.
In Fig.~\ref{fig:hlm_mismatches} we show the mismatches $\mathcal{M}_{\ell m}$ between the Teukolsky waveforms and the \texttt{SEOBE-TML} MR model across the explored parameter space for the higher $\ell = m$ multipoles, namely the $(3,3)$, $(4,4)$ and $(5,5)$ modes. Each panel displays the mismatches as a function of the BH spin $a$ and of the eccentricity $e_{\rm LSO}$. Each dot corresponds to a specific configuration in the $(a,e_{\rm LSO})$ parameter space, while the color encodes the value of the mismatch on a logarithmic scale.
Overall, the \texttt{SEOBE-TML} MR model shows very good agreement with the Teukolsky waveforms across the considered parameter space for all the examined multipoles. Most configurations exhibit mismatches in the range $\sim10^{-5}$--$10^{-4}$, indicating that the model is able to accurately reproduce the merger and the ringdown of the higher $\ell = m$ modes. The median mismatches across the parameter space are $5.77\times10^{-5}$ for the $(3,3)$ mode, $2.14\times10^{-5}$ for the $(4,4)$ mode, and $3.60\times10^{-5}$ for the $(5,5)$ mode, confirming the overall robustness of the MR description also for these subdominant modes.
As observed for the $(2,2)$ mode, the accuracy remains particularly good for spins in the range $-0.8 \le a \le 0.4$, where the majority of configurations cluster at mismatches $\lesssim10^{-4}$. For larger positive spins, especially for $a \ge 0.5$, the mismatches tend to increase, particularly at higher eccentricities. This behaviour is consistent with the trends discussed above and reflects the increased difficulty in accurately modeling highly prograde spin configurations with the same level of accuracy achieved for smaller spins. Nevertheless, even in these regions of the parameter space the mismatches remain typically below $\sim 5 \times 10^{-3}$ (excluding extreme scenarios like $a = 0.9$ with $e_{\rm LSO} \ge 0.70$) , indicating that the \texttt{SEOBE-TML} MR model is able to capture the phenomenology of the merger and post-merger dynamics of the higher $\ell = m$ modes with good accuracy across the explored configurations.
\subsubsection{$\ell \neq m$ modes}
\begin{figure*}[tp!]
  	\includegraphics[width=1.\linewidth]{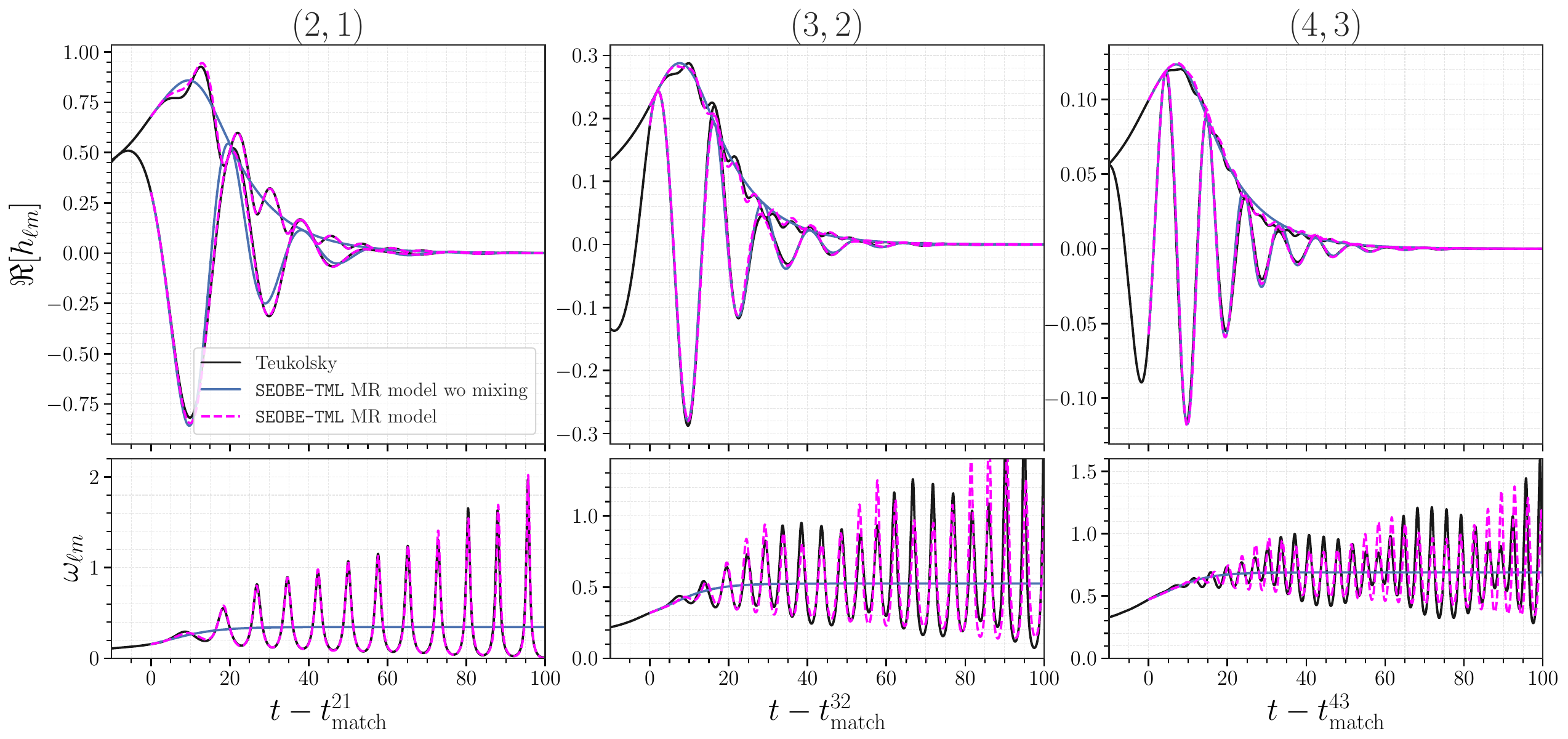}
  	\caption{Same as Fig.~\ref{fig:ell is m waveforms with freq}, but for the non-diagonal $(2,1)$, $(3,2)$, and $(4,3)$ modes (left, central, and right columns, respectively). The configuration shown is the same representative eccentric case with spin $a=-0.8$ and eccentricity $e_{\rm LSO}=0.50$. The black solid curves correspond to the Teukolsky waveforms, the blue solid curves show the \texttt{SEOBE-TML} MR model without QNM mixing, and the magenta dashed curves represent the complete model including QNM mixing. In these multipoles the oscillatory structure of the ringdown is again driven by QNM mixing, although the model reproduces the modulations with slightly reduced accuracy for the $(3,2)$ and $(4,3)$ modes compared to the $\ell=m$ modes.}
  	\label{fig:ell is not m waveforms with freq}
\end{figure*}
\begin{figure*}[tp!]
  	\includegraphics[width=1.\linewidth]{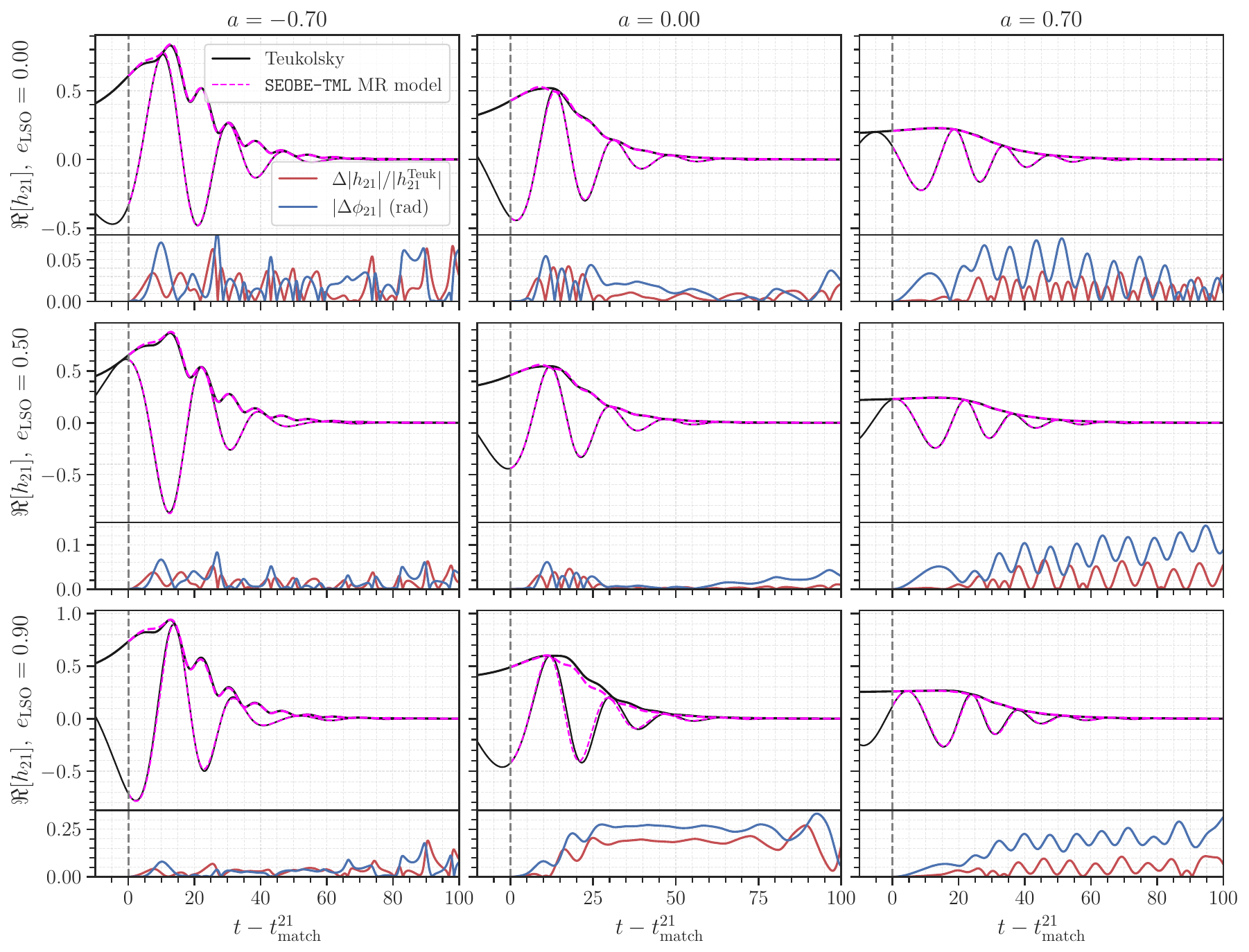}
  	\caption{Comparison between the Teukolsky $(2,1)$ waveforms and the \texttt{SEOBE-TML} MR model, considering the same configurations shown in Fig.~\ref{fig:h22_waveforms}. The real part of the Teukolsky waveform $\Re[h_{21}]$ (black) is compared with the \texttt{SEOBE-TML} MR model (magenta dashed). The lower panels display the relative amplitude difference $\Delta |h_{21}|/|h^{\rm Teuk}_{21}|$ (red) and the phase difference $\Delta \phi_{21}$ (blue), as functions of $t - t^{21}_{\rm match}$.}
  	\label{fig:h21_waveforms}
\end{figure*}
\begin{figure*}[tp!]
  	\includegraphics[width=1.\linewidth]{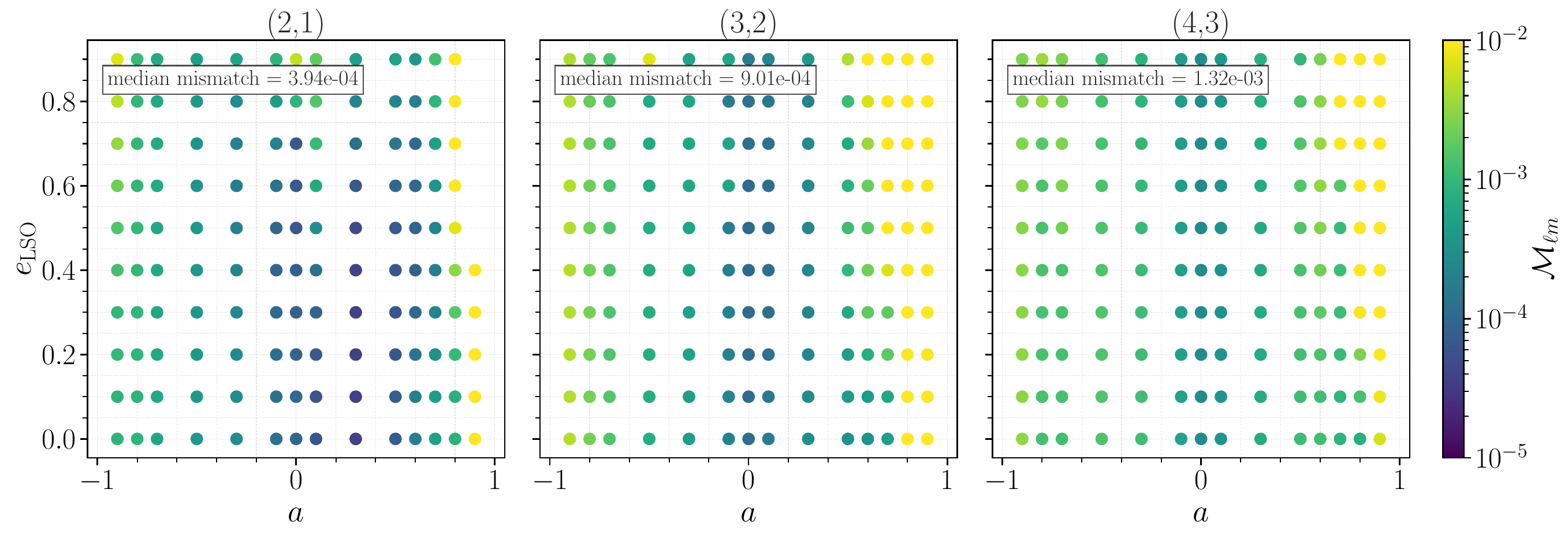
}
  	\caption{Same as Fig.~\ref{fig:hlm_mismatches}, but for the non-diagonal multipoles. The three panels correspond to the $(2,1)$, $(3,2)$, and $(4,3)$ modes. Each dot represents a configuration in the $(a,e_{\rm LSO})$ parameter space, while the color indicates the value of the mismatch $\mathcal{M}_{\ell m}$ on a logarithmic scale. The median mismatch across the explored parameter space is reported in each panel. Configurations with $a=0.9$ and $e_{\rm LSO} \ge 0.5$ for the $(2,1)$ mode are not shown, as the MR ringdown could not be fitted with the ansatz of Eq.~\eqref{eq.:MR ansatz}. These cases therefore appear as missing points in the first panel.}
  	\label{fig:hlm_mismatches_ell_is_not_m}
\end{figure*}
We now turn to the non-diagonal $\ell \neq m$ modes. For these modes the MR signal exhibits a richer structure, as their intermediate-time behaviour after the attachment time is more strongly affected by QNM mixing. As a result, the waveform amplitude and the instantaneous frequency typically display more complex oscillatory features than in the $\ell = m$ case. In the following, we first illustrate these features for a representative configuration and then assess the overall performance of the model across the parameter space.
In Fig.~\ref{fig:ell is not m waveforms with freq} we show the comparison between the Teukolsky waveforms and the \texttt{SEOBE-TML} MR model for the non-diagonal $(2,1)$, $(3,2)$, and $(4,3)$ modes, displayed in the left, central, and right columns, respectively. We consider the same representative eccentric configuration of Fig.~\ref{fig:ell is m waveforms with freq} with negative spin, $a=-0.80$ and $e_{\rm LSO}=0.50$. In each column, the top panel shows the real part of the waveform, $\Re[h_{\ell m}]$, while the bottom panel displays the instantaneous frequency $\omega_{\ell m}$, as functions of $t-t_{\rm match}^{\ell m}$. As in Fig.~\ref{fig:ell is m waveforms with freq}, the black solid curves denote the numerical waveforms, the blue solid curves correspond to the \texttt{SEOBE-TML} MR model without QNM mixing, and the magenta dashed curves represent the complete \texttt{SEOBE-TML} MR model including QNM mixing.

For the $(2,1)$ mode (left column), the MR waveform exhibits pronounced oscillations in the waveform amplitude and a modulated instantaneous frequency. The model without QNM mixing captures the overall decay toward the late-time ringdown, but it misses the sequence of oscillatory features that characterize the signal after the attachment time. In particular, its frequency approaches the asymptotic value too smoothly and therefore does not reproduce the sharp modulations visible in the Teukolsky waveform. Once QNM mixing is included, the agreement improves substantially: the model reproduces the oscillatory pattern in the waveform with good accuracy and, at the same time, tracks the sequence of peaks in the instantaneous frequency associated with the strong interference among the contributing $(2,1,0,-1)$ and $(2,2,0,1)$ QNMs.

A similar behaviour is observed for the $(3,2)$ and $(4,3)$ modes, shown in the central and right columns. In both cases, the waveform displays an oscillatory ringdown structure that is not described by a single least-damped QNM. Indeed, when QNM mixing is not included, the \texttt{SEOBE-TML} MR model (blue curve) reproduces the broad behaviour of the waveform, but the modulations due to the interference with other QNMs are significantly underestimated, for both the amplitude and the instantaneous frequency. By contrast, when QNM mixing is incorporated (purple dashed), the model partly reproduces the modulations in both the waveform amplitude and the instantaneous frequency. However, the agreement is not as accurate as for the other modes discussed above. This behaviour was also observed in the QC case in Ref.~\cite{Nishimura:2026nse}. We attribute this reduced accuracy to the simplified QNM mixing prescription adopted in our model. In particular, the model currently includes the mirror QNM $(\ell,m,0,-1)$ and the $(\ell-1,m,0,1)$ contribution. It is likely that, especially at later times, the $(\ell-1,m,0,-1)$ QNM also plays a relevant role in the interference pattern. Since this additional contribution is not included in the current implementation, the resulting beating structure in the waveform and in the instantaneous frequency cannot be reproduced with the same level of precision as in the other modes. As we will show in Fig.~\ref{fig:hlm_mismatches_ell_is_not_m}, this reduced accuracy also translates into larger mismatches for these modes.

Overall, for this eccentric and negative-spin configuration, the \texttt{SEOBE-TML} MR model is able to capture the main features of the MR also for the non-diagonal $(2,1)$, $(3,2)$, and $(4,3)$ modes, provided that QNM mixing is included. For these multipoles, the role of mixing is particularly important, as it generates the complex oscillatory structure observed in both the waveform amplitude and the instantaneous frequency during the intermediate and late stages of the ringdown. While the model reproduces these features with good accuracy for the $(2,1)$ mode, the agreement becomes somewhat less precise for the $(3,2)$ and $(4,3)$ modes, reflecting the limitations of the simplified QNM mixing prescription currently adopted. Nevertheless, the model is still able to reproduce the overall phenomenology of the MR signal, capturing the main beating patterns produced by the interference among the dominant QNM contributions.

So far we illustrated the behaviour of the model for a representative configuration. We now turn to a systematic assessment of its performance across the parameter space for the non-diagonal modes. We begin by focusing on the $(2,1)$ mode.
Figure~\ref{fig:h21_waveforms} shows the comparison between the Teukolsky waveforms and the \texttt{SEOBE-TML} MR model for the $(2,1)$ mode, for the same three representative values of the BH spin as in Fig.~\ref{fig:h22_waveforms}, i.e. for $a=-0.70$ (left column), $a=0.00$ (central column), and $a=0.70$ (right column), and increasing eccentricity at the LSO, $e_{\rm LSO}=[0.00,\,0.50,\,0.90]$ (from top to bottom). For each configuration, we display the real part of the waveform, $\Re[h_{21}]$, together with the relative amplitude difference $\Delta |h_{21}|/|h^{\rm Teuk}_{21}|$ (red) and the phase difference $\Delta \phi_{21}$ (blue), as functions of time measured from $t^{21}_{\rm match}$. 

Overall, the MR model (dashed magenta) shows good agreement with the numerical waveforms (black) across a large portion of the parameter space, capturing the main features of the ringdown. In particular, the model is able to reproduce the oscillatory structure of the waveform amplitude and of the instantaneous frequency associated with the interference among the QNM contributions discussed above, which for the $(2,1)$ mode correspond to the $(2,1,0,-1)$ and $(2,2,0,1)$ QNMs.

In the QC case ($e_{\rm LSO}=0$), the agreement between the model and the numerical waveforms is generally good, with amplitude fractional differences $\le 5\%$ and phase differences $\le 0.06$ rad throughout the MR. As the eccentricity increases, deviations become more visible after the merger, especially for $e_{\rm LSO}=0.90$, where, for the $a = 0.00$ case, the fractional amplitude differences can grow up to $\sim 25 \%$ at intermediate and late times of the ringdown. This behaviour is associated with the increased difficulty of accurately capturing all the waveform features across the full parameter space within the hierarchical fitting procedure adopted in this work.
We also observe that configurations with large positive spin can display enhanced discrepancies with respect to the Teukolsky waveforms. This trend indicates that large prograde spin configurations remain more challenging to model within the current fitting strategy, a behaviour already observed for the $(2,2)$ mode. In these cases, the model still captures the overall morphology of the signal, although larger amplitude and phase differences can appear at intermediate and late times. These results should therefore be regarded as a first step toward a more accurate modeling of the MR signal in the presence of both eccentricity and spin effects.

To further quantify the accuracy of the \texttt{SEOBE-TML} MR model for the non-diagonal multipoles, we evaluate its performance across the explored parameter space by computing the mismatches $\mathcal{M}_{\ell m}$ with respect to the numerical waveforms. As done for the $\ell = m$ modes in Sec.~\ref{Sec.: ell is m modes}, this provides a global measure of the agreement between the two descriptions and allows us to assess how well the model reproduces the merger and ringdown signals of the $(2,1)$, $(3,2)$, and $(4,3)$ modes over the full set of configurations.
In Fig.~\ref{fig:hlm_mismatches_ell_is_not_m} we summarize the results of this comparison. Each panel shows the mismatches for one of the non-diagonal mode as a function of the BH spin $a$ and of the eccentricity at the LSO, $e_{\rm LSO}$. Each point corresponds to a specific configuration in the $(a,e_{\rm LSO})$ parameter space, while the color scale indicates the value of the mismatch on a logarithmic scale.
Across the explored parameter space the model generally reproduces the numerical waveforms with a large fraction of configurations which exhibit mismatches in the range $\sim10^{-4}$--$10^{-3}$, indicating that the MR description captures the main features of the signal also for these non-diagonal modes. The median mismatches over the parameter space are $3.94 \times10^{-4}$ for the $(2,1)$ mode, $9.01 \times10^{-4}$ for the $(3,2)$ mode, and $1.32 \times10^{-3}$ for the $(4,3)$ mode.

Consistent with the waveform comparisons discussed above in Fig.~\ref{fig:ell is not m waveforms with freq}, the $(2,1)$ mode typically shows the smallest mismatches, whereas the $(3,2)$ and $(4,3)$ multipoles tend to display somewhat larger values. This trend reflects the more intricate oscillatory structure of these modes, which originates from the interference among multiple QNM contributions and is only partially captured by the mixing prescription currently implemented in the model.
We note that for the $(2,1)$ mode we were not able to successfully fit the MR ringdown with the ansatz in Eq.~\eqref{eq.:MR ansatz} for configurations with large positive spin, $a = 0.9$, and eccentricities $e_{\rm LSO} \ge 0.5$. As a consequence, these configurations are not included in the mismatch computation and therefore appear as missing points in the first panel of Fig.~\ref{fig:hlm_mismatches_ell_is_not_m}. These systems are characterized by pericenter passages of the small mass that occur very close to the event horizon, probing extremely strong-field regions of the spacetime. In this regime, the waveform develops complex features around the merger that we were not able to capture with the current modeling ansatz. In Appendix~\ref{Appendix: (2,1) mode high spin}, we discuss these configurations in more detail and provide a possible physical interpretation of the behaviour observed in the numerical waveforms.
More pronounced discrepancies appear in the prograde high-spin region of the parameter space. In particular, for the $(3,2)$ mode the mismatches exceed $9\times10^{-3}$ whenever the spin satisfies $a \ge 0.6$ and the eccentricity $e_{\rm LSO} \ge 0.3$. An even worse trend is observed for the $(4,3)$ mode, where mismatches above this level appear already for prograde spins $a \ge 0.5$ with $e_{\rm LSO} \ge 0$. This behaviour reflects the increased difficulty of accurately modeling configurations that simultaneously involve large prograde spins and significant eccentricity with non-trivial QNM mixing features within the hierarchical fitting strategy adopted in this work. While this represents a limitation of the current model that will require further refinement, it should be regarded as a first step toward consistently incorporating the combined effects of spin and eccentricity in the modeling of the MR signal, paving the way for future extensions of this approach to the comparable-mass regime.

\section{Summary of the results and sources of error in the model} \label{Sec: summary of the results and sources of error}
\begin{table*}
    \setlength{\tabcolsep}{18pt}
    \centering
    \begin{ruledtabular}
    \begin{tabular}{ccccc}
        \multirow{2}{*}{$(\ell,m)$}
        & \multirow{2}{*}{median $\mathcal{M}_{\ell m}$}
        & \multirow{2}{*}{mean $\mathcal{M}_{\ell m}$}
        & \multicolumn{2}{c}{maximum mismatch} \\
        \cline{4-5}
        & & & $\max \mathcal{M}_{\ell m}$ & $(a,e_{\rm LSO})_{\rm max}$ \\
        \midrule
        (2,2) & $8.97 \times 10^{-5}$ & $6.98 \times 10^{-4}$ & $4.54 \times 10^{-2}$ & $(0.9,0.9)$ \\
        (2,1) & $3.94 \times 10^{-4}$ & $6.27 \times 10^{-3}$ & $2.15 \times 10^{-1}$ & $(0.9,0.4)$ \\
        (3,3) & $5.77 \times 10^{-5}$ & $1.74 \times 10^{-3}$ & $2.02 \times 10^{-1}$ & $(0.9,0.9)$ \\
        (3,2) & $9.01 \times 10^{-4}$ & $3.36 \times 10^{-2}$ & $8.09 \times 10^{-1}$ & $(0.8,0.9)$ \\
        (4,4) & $2.14 \times 10^{-5}$ & $1.61 \times 10^{-3}$ & $1.88 \times 10^{-1}$ & $(0.9,0.9)$ \\
        (4,3) & $1.32 \times 10^{-3}$ & $2.02 \times 10^{-2}$ & $5.28 \times 10^{-1}$ & $(0.8,0.8)$ \\
        (5,5) & $3.60 \times 10^{-5}$ & $2.44 \times 10^{-3}$ & $1.82 \times 10^{-1}$ & $(0.9,0.9)$ \\
    \end{tabular}
    \end{ruledtabular}
    \caption{Median, mean, and maximum mismatches for the different waveform modes over the parameter space considered. The last column reports the parameter-space point $(a,e_{\rm LSO})_{\rm max}$ at which the maximum mismatch for each mode occurs.}
    \label{tab:mode_mismatches}
\end{table*}

Here we provide a summary of the accuracy of the MR model developed in this article. In Table~\ref{tab:mode_mismatches}, we report, for each modeled spherical-harmonic $(\ell, m)$ mode, the median and mean values of the mismatch $\mathcal{M}_{\ell m}$ over the parameter space, together with the maximum mismatch and the corresponding values of $(a,e_{\rm LSO})$ at which it occurs. The median mismatches remain below $1.35\times10^{-3}$ for all the modes considered. In particular, the $(2,2)$, $(3,3)$, $(4,4)$, and $(5,5)$ modes exhibit median mismatches below $10^{-4}$, whereas somewhat larger values are obtained for the $(2,1)$, $(3,2)$, and $(4,3)$ modes. A similar hierarchy is observed for the mean mismatches, which remain below $10^{-2}$ for all modes except the $(3,2)$ and $(4,3)$ modes. The largest mismatches are predominantly found in the region of the parameter space characterized by large positive spins $(a \ge 0.8)$ and high eccentricities $(e_{\rm LSO} \ge 0.4)$, consistently with the trends observed in the mismatch distributions presented in Fig.~\ref{fig:h22_mismatches}, Fig.~\ref{fig:hlm_mismatches} and Fig.~\ref{fig:hlm_mismatches_ell_is_not_m}. This region therefore represents the most challenging regime for the present model.

Several sources may contribute to the mismatch between the modeled and numerical waveforms. A first source is the model-form error associated with the functional ansatz adopted to describe the dependence of the MR parameters on the orbital parameters. Although the chosen fitting functions accurately reproduce the numerical data over most of the parameter space, their finite flexibility may limit their ability to capture the more complex behavior observed in the high-spin, high-eccentricity regime.

A further source of error arises from neglecting the relativistic anomaly $\xi_{\rm LSO}$ as an additional parameter in the fits. As discussed in Sec.~\ref{sec.:impact of relativistic anomaly on the merger-ringdown} and depicted in Fig.~\ref{fig:effects of rel anomaly}, the last-peak features exhibit a negligible dependence on $\xi_{\rm LSO}$ throughout most of the parameter space. However, a non-negligible dependence emerges for sufficiently large positive spins and high eccentricities. Since the present model parametrizes the MR quantities only in terms of $a$ and $e_{\rm LSO}$, this residual dependence cannot be captured and may contribute to the reduced accuracy observed in this region of the parameter space.

Additional uncertainties are associated with the extraction of the QNM excitation amplitudes used to construct the ringdown model. As described in Sec.~\ref{sec.:QNMs mixing modeling}, these amplitudes are obtained by fitting the numerical waveforms with the \texttt{qnmfinder} code. The extraction procedure is itself affected by uncertainties related, for instance, to the choice of the fitting interval and to the separation of multiple, simultaneously excited QNM contributions. Such uncertainties can propagate to the fitted MR parameters and, consequently, to the final modeled waveform.

Finally, an additional source of modeling error is related to the finite set of QNMs included to describe the mode mixing. In the present model, we retain only the fundamental QNMs and restrict the mixing contributions to the dominant modes identified in Sec.~\ref{sec.:QNMs mixing modeling}. While this choice provides an accurate and economical description of the ringdown over most of the parameter space, subdominant QNMs and higher overtones are neglected. Their omission may contribute to the residual discrepancies, particularly in the regions where the relative excitation of additional QNMs becomes more significant.

\section{Conclusions} \label{Sec: conclusions}

In this work, we characterized and modeled the MR of GWs modes emitted by a small mass plunging and merging into a Kerr BH on eccentric equatorial orbits. The waveforms were generated using a TD Teukolsky code and parametrized in terms of the spin $a$ of the central Kerr BH, the eccentricity at the last stable orbit $e_{\rm LSO}$, and the relativistic anomaly at the LSO $\xi_{\rm LSO}$. 
We considered a small mass $\mu = 10^{-3}$ orbiting a Kerr BH with spin values $-0.9 \le a \le 0.9$, and characterized by $0.0 \le e_{\rm LSO} \le 0.90$ and $0 \le \xi_{\rm LSO} < 2 \pi$. We focused on the characterization and modeling of the $(2,2)$, $(3,3)$, $(4,4)$, $(5,5)$, $(2,1)$, $(3,2)$, and $(4,3)$ $-2$ spin-weighted spherical harmonic modes.

By inspecting the waveform morphology across the explored parameter space, we confirmed that eccentricity plays a significant role in shaping the merger part of the signal. Its impact on the ringdown is less pronounced, but still present, influencing features such as the relative contribution of the different QNMs with respect to the least-damped mode. On the other hand, the relativistic anomaly at the LSO affects the merger morphology only within a restricted portion of the parameter space in the small-mass regime. In particular, for a large region of $(a,e_{\rm LSO})$ values, the waveform properties at merger appear largely insensitive to the value of the relativistic anomaly measured at the LSO, while noticeable differences arise only for specific configurations. We did not find any dependence of the relativistic anomaly measured at the LSO on the ringdown features, like the QNMs amplitudes. These results clarify the relative role of the different orbital parameters in determining the structure of the MR signal in the TML and provide guidance for modeling them within the EOB waveform framework.

Based on this characterization, we developed a phenomenological description of the MR waveform that incorporates the main physical features observed in the numerical signals. In particular, we constructed the \texttt{SEOBE-TML} MR model, which extends current QC MR prescriptions by explicitly including the effects of eccentricity and by modeling the behaviour associated with QNMs mixing. 
We validated this MR model by comparing it with the Teukolsky waveforms across the considered region of the parameter space. We found that the model reproduces well both the early and intermediate times of the ringdown for the dominant $(2,2)$ mode, as well as for higher $\ell = m$ modes such as $(3,3)$, $(4,4)$ and $(5,5)$. In particular, the model is also able to capture the modulations in amplitude and frequency induced by QNMs mixing and thus provides an accurate description of both the amplitude and frequency evolution during the post-merger phase. 
To quantify the global accuracy of the model, we computed mismatches between the \texttt{SEOBE-TML} MR waveforms and the numerical signals across the explored parameter space. For the $(2,2)$ mode, the mismatches are significantly reduced compared to current QC MR prescriptions, reaching typical values in the range $\sim 10^{-5}$--$10^{-4}$ over most of the parameter space. Similar levels of accuracy are obtained for the higher $\ell = m$ modes. We also observed that the model performs particularly well for spins in the range $-0.8 \lesssim a \lesssim 0.4$, while the accuracy mildly degrades for highly prograde configurations ($a \gtrsim 0.5$), especially at large eccentricities.

For the non-diagonal modes $(2,1)$, $(3,2)$ and $(4,3)$ the model is also able to reproduce the main features of the MR signal. In particular, the agreement with the numerical waveforms remains good for the $(2,1)$ mode across most of the explored parameter space, while the $(3,2)$ and $(4,3)$ modes show somewhat reduced accuracy, reflecting the increased complexity of the mixing pattern and the limitations of the simplified prescription adopted in the present model. 
Quantitatively, most configurations exhibit mismatches in the range $\sim10^{-4}$--$10^{-3}$.  These results indicate that the \texttt{SEOBE-TML} MR model is able to capture the overall phenomenology of the ringdown also for the non-diagonal modes, although the accuracy degrades in regions of the parameter space characterized by large prograde spins and significant eccentricity, where the mismatches can reach values of order $5 \times 10^{-2}$, especially for the $(3,2)$ and $(4,3)$ modes.

The results presented in this article provide a first step toward the construction of eccentric MR models in the TML. Several directions can be pursued to further improve and extend this work. 
A natural next step is to refine the fitting procedure introduced in this work in order to further optimize the accuracy of the model across the explored parameter space. In particular, improving the modeling in specific regions of the parameter space characterized by large prograde spins and significant eccentricity will be increasingly important. Also, an improvement concerning the non-diagonal modes will be important, as these multipoles exhibit larger mismatches in the proposed model.
Another important development will be the extension of the present modeling strategy to the comparable-mass regime within the \texttt{SEOBNR} family of waveform models. Once a sufficient number of NR simulations of spin-aligned eccentric binaries become available to adequately cover the parameter space, the approach developed here could be generalized by incorporating the mass ratio as an additional dimension in the fitting procedure. This would enable the extension of current QC \texttt{SEOBNR} MR prescriptions to consistently include the effects of eccentricity.
Finally, it will be important to investigate the ringdown of waveforms generated by generic orbits of Kerr, thus including inclined configurations. Extending the present analysis to inclined eccentric trajectories would enable the construction of a fully generic MR model of a TM orbiting in the Kerr spacetime.

\textbf{Addendum} - During the final stages of this work we
became aware of a study of eccentric coalescences in the
TML being carried out by Albanesi et al.~\cite{Albanesi:2026},
which conducts similar studies.

\section*{Acknowledgments}
The authors are grateful to Adrian Abac, Simone Albanesi, Devin Becker, Francisco Blanco, Thibault Damour, Scott Hughes, Guillaume Lhost, Aldo Gamboa, Elisa Maggio, Alessandro Nagar, Peter James Nee, Nami Nishimura, Maria de Lluc Planas and Lorenzo Pompili for the insightful and constructive discussions. G.F. is particularly grateful to Nami Nishimura for the fruitful interactions during the development of this work.
MvdM acknowledges financial support by 
the VILLUM Foundation (grant no. VIL37766),
the DNRF Chair program (grant no. DNRF162) by the Danish National Research Foundation and the MPI for Gravitational Physics, and the European Union’s Horizon ERC Synergy Grant “Making Sense of the Unexpected in the Gravitational-Wave Sky” grant agreement no.\ GWSky–101167314. 
G.K. acknowledges support from US National Science Foundation grants No. DMS-2309609 and PHY-2307236. Simulations were performed on the UMass-URI UNITY HPC/AI  supercomputer supported by the Massachusetts Green High Performance Computing Center (MGHPCC).

This work makes use of the Black Hole Perturbation Toolkit~\cite{BHPToolkit}.

\appendix
\section{Additional details on the hierarchical fitting procedure}
\label{sec: hierarchical fit results}
\begin{figure}[tp!]
\includegraphics[width=1.\linewidth]{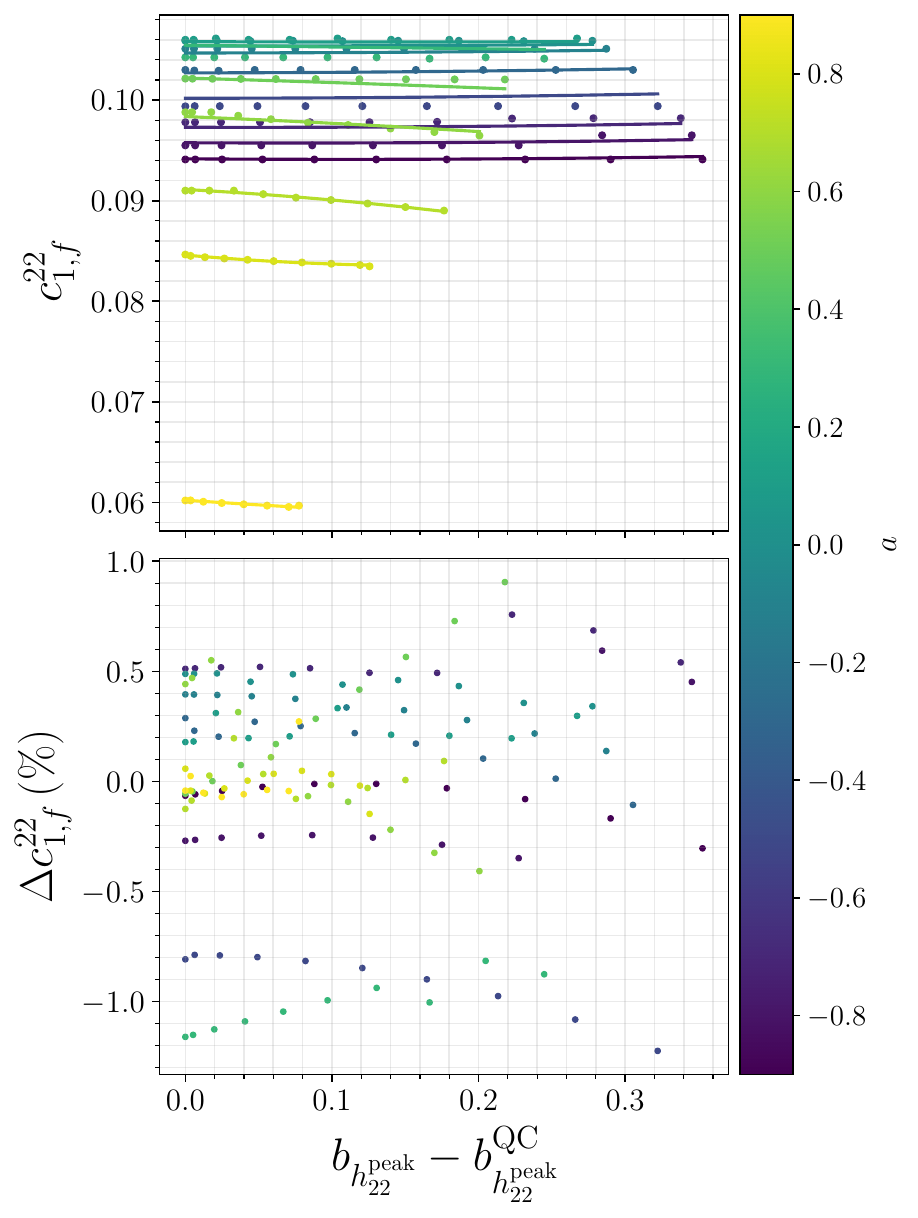}
\caption{Example of the hierarchical fit for one of the parameters entering the MR ansatz, namely $c^{22}_{1,f}$. 
Top panel: values of the coefficient obtained from the individual waveform fits (colored dots) as a function of the parameter $b$, together with the corresponding hierarchical fit given by the rational ansatz of Eq.~\eqref{eq.:hierarchical fit ansatz} (colored solid lines). The dependence on the spin $a$ is indicated by the different colors. 
Bottom panel: fractional residuals (in percentage) between the hierarchical fit and the coefficients obtained from the best fits done on the numerical waveforms.}
\label{fig: c1f example residuals}
\end{figure}
\begin{figure*}[tp!]
\includegraphics[width=1.\linewidth]{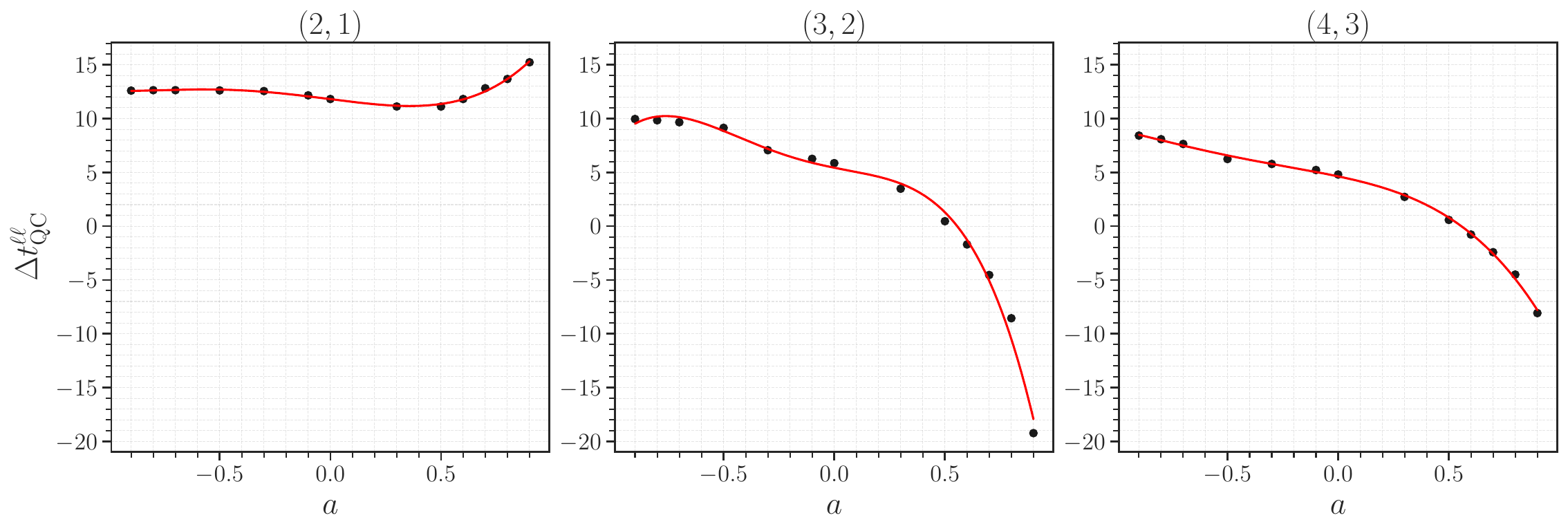}
\caption{Spin dependence of the offset $\Delta t^{\ell \ell}_{\rm QC}$ for the non-diagonal modes $(2,1)$, $(3,2)$, and $(4,3)$. The black dots are obtained from the numerical waveforms, while the red solid lines represent 4-th order polynomial fits used to model the spin dependence of the offset.}
\label{fig: deltat_ll_QC}
\end{figure*}
\begin{figure*}[tp!]
\includegraphics[width=1.\linewidth]{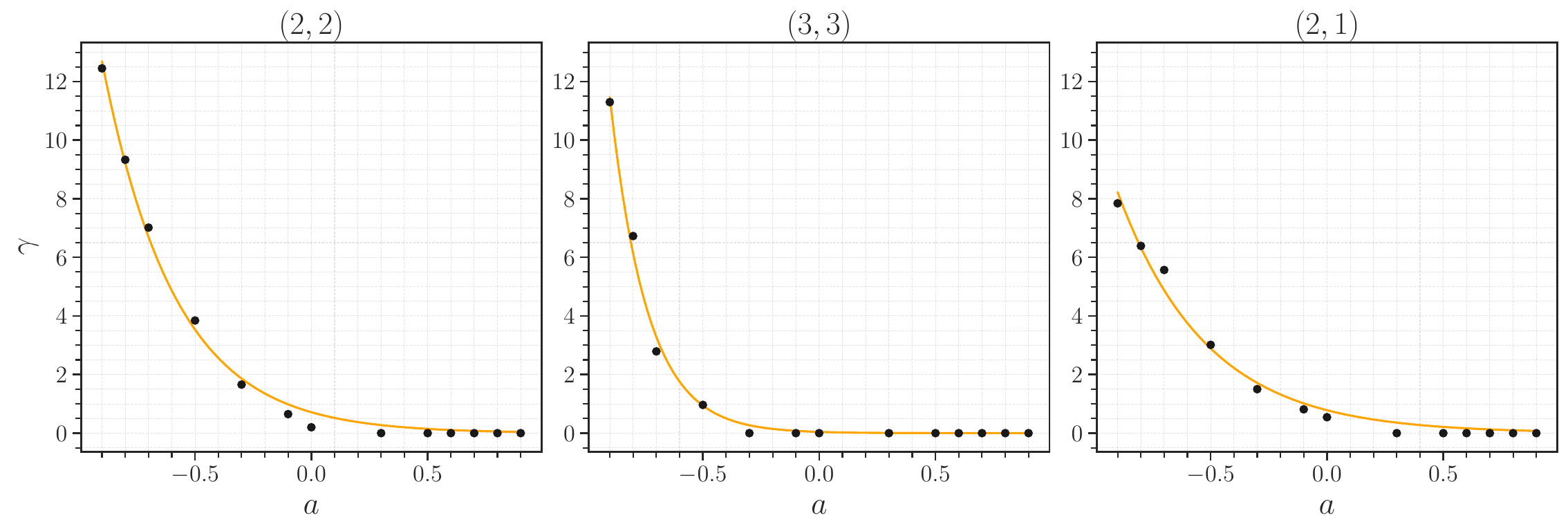}
\caption{
Spin dependence of the parameter $\gamma$ for the modes $(2,2)$, $(3,3)$, and $(2,1)$. The black dots represent the values of $\gamma$ that optimize the mismatch defined in Eq.~\eqref{eq.: def mismatch}, while the orange solid lines represent exponential fits.
}
\label{fig: gamma_exp_fit}
\end{figure*}
\begin{figure*}[tp!]
\includegraphics[width=1.\linewidth]{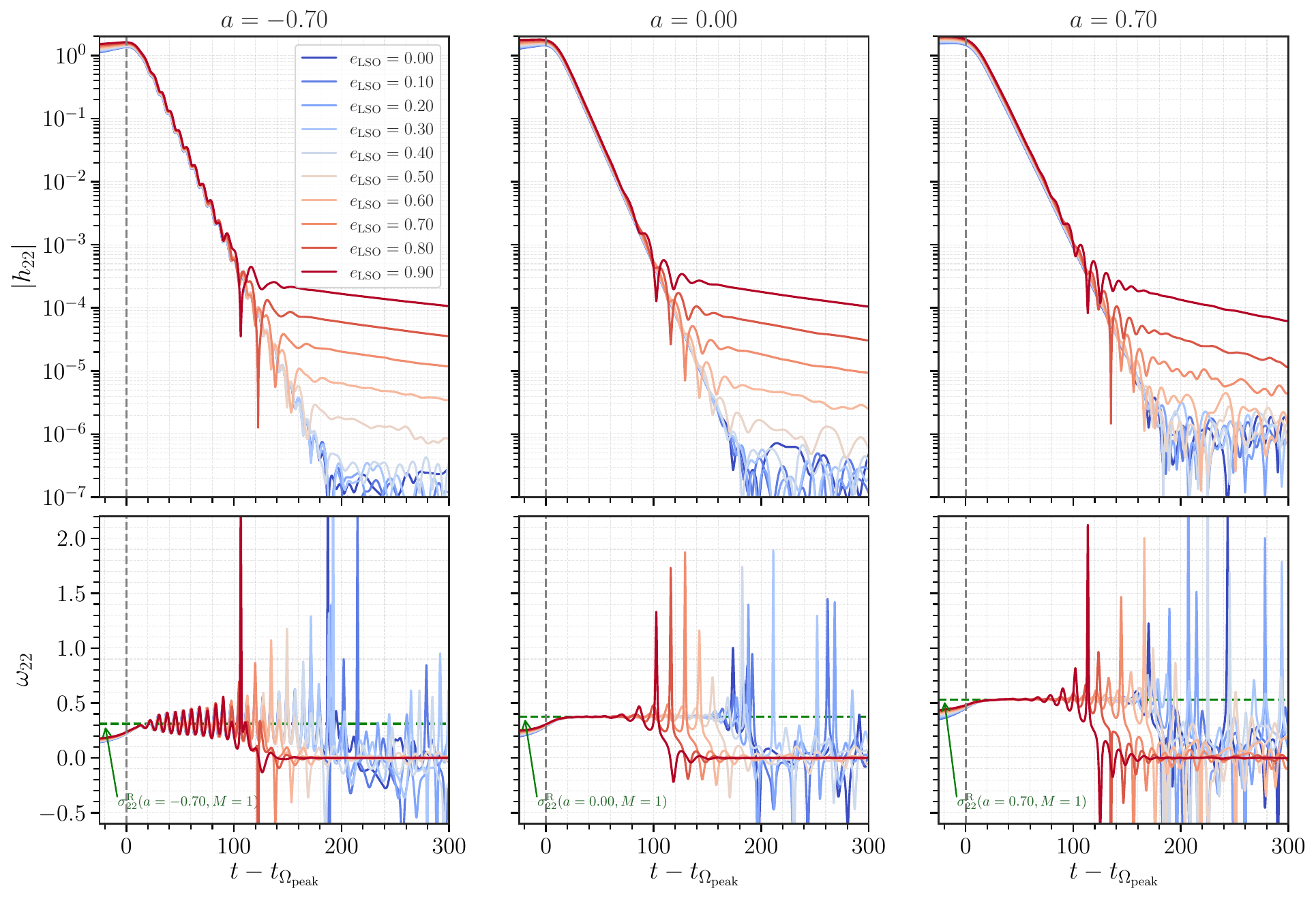}
\caption{Same as Fig.~\ref{fig:effects of eccentricity}. The waveform amplitudes are plotted on a logarithmic scale in order to highlight the regimes where the Price-tail contribution becomes dominant. From this plot it is evident that, as the eccentricity increases, the time at which the QNM contribution and the Price-tail contribution start to interfere occurs progressively earlier, as discussed in Sec.~\ref{sec.:impact of eccentricity on the merger-ringdown}.}.
\label{fig: Price tails effects of eccentricity}
\end{figure*}
\begin{figure*}[tp!]
\includegraphics[width=1.\linewidth]{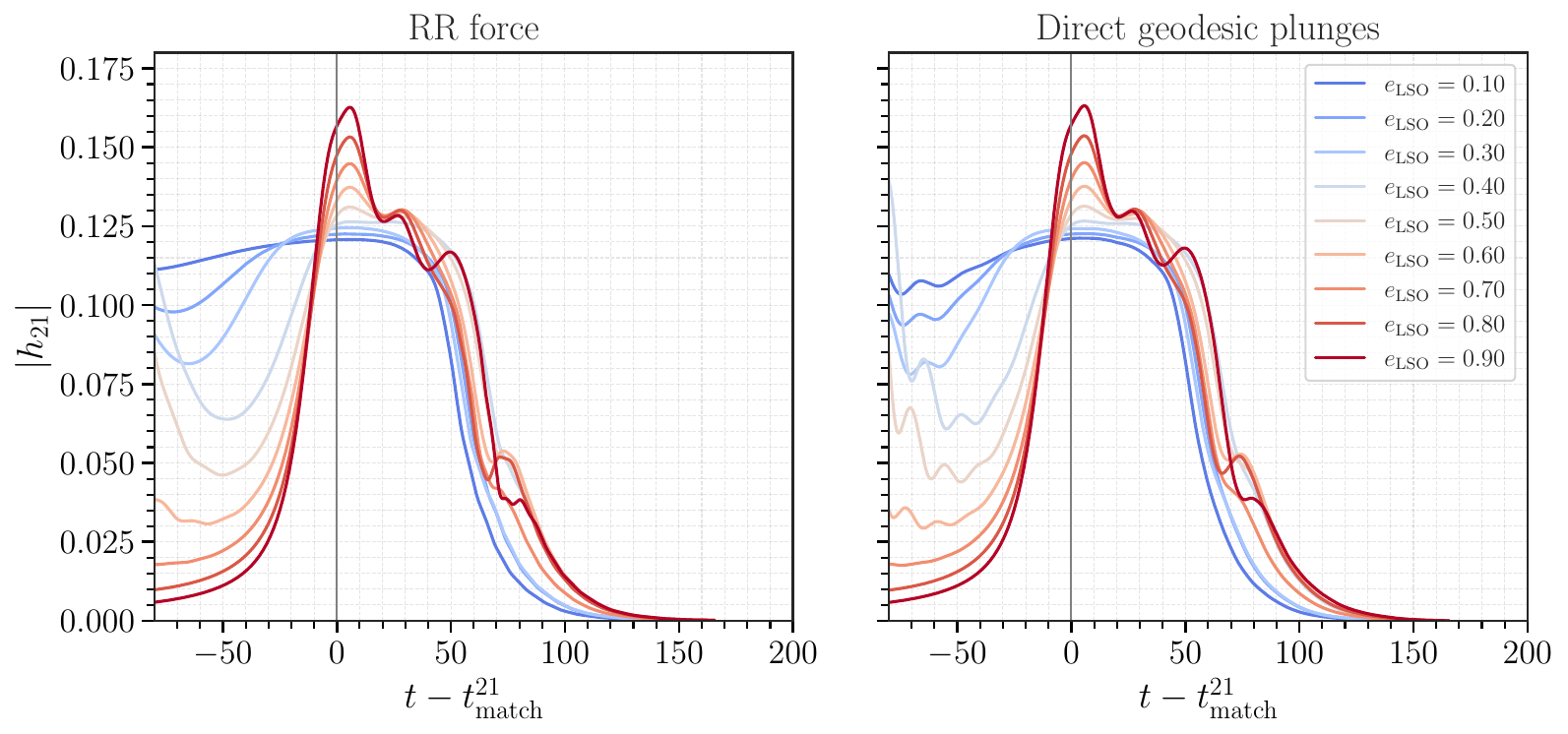}
\caption{Comparison of the $(2,1)$ waveform for highly prograde configurations with $a=0.9$ and increasing eccentricity. The left panel shows the waveforms obtained using the trajectories employed in this work, where the TM evolves under the RR force. Starting from $e_{\rm LSO}=0.5$, the waveform features close to the peak develop a phenomenology that differs from the simpler behaviour typically observed for positive spins. In order to test whether these features are caused by inaccuracies in the RR force in the strong-field regime, we generate waveforms using bound direct geodesic plunges whose energy corresponds to that of the bound geodesic configuration associated with the same $e_{\rm LSO}$ considered for the RR-force-driven trajectories, 
with an additional offset $\Delta \mathcal{E}_{\rm UCO}$ above the UCO chosen such that, for each configuration, it matches the value of $\Delta \mathcal{E}_{\rm UCO}$ at the onset of the plunge of the corresponding RR-force-driven inspiral shown in the left panel.
The right panel shows the corresponding waveforms. The qualitative structure near the last peak of the $(2,1)$ mode remains essentially unchanged, indicating that the observed phenomenology is not driven by the RR force prescription, although it affects the quantitative details of the waveform.
}.
\label{fig: h21 high prograde spins}
\end{figure*}
In this section we provide additional details on the hierarchical fitting procedure adopted for the construction of the \texttt{SEOBE-TML} model.
As described in Sec.~\ref{sec.:hierarchical fit}, the fitting strategy is performed in two steps. First, for each numerical waveform corresponding to a point in the parameter space $(b,a)$, we determine the optimal values of the coefficients $c^{\ell m}_{1,f}$, $c^{\ell m}_{2,f}$, $d^{\ell m}_{1,f}$, and $d^{\ell m}_{2,f}$ by fitting the MR ansatz in Eq.~\eqref{eq.:MR ansatz} directly to the Teukolsky waveform. This procedure provides a set of best-fit coefficients associated with each individual waveform in the dataset. 
In the second step, these coefficients are fitted as functions of the parameters $(b,a)$ using the rational-function ansatz introduced in Eq.~\eqref{eq.:hierarchical fit ansatz}. We recall that the parameter $b$ used in the hierarchical fit is defined as the offset of the impact parameter $\tilde b=\mathcal{L}/\mathcal{E}$ with respect to its QC value $b = \tilde{b} - \tilde{b}_{\rm QC}$.
In practice, the quantity $b$ is evaluated at the time corresponding to the peak of the $(2,2)$ mode, $t=t^{22}_{\rm peak}$. This choice provides a physically meaningful reference time that can be directly identified from the waveform and therefore facilitates a consistent extension of the present fitting strategy to waveform datasets that do not rely on trajectory information.
In order to ensure that the rational ansatz remains well defined over the explored parameter space, i.e. it does not exhibit spurious poles, we impose that the denominator
\begin{equation}
D^{\ell m}(b,a)=C^{\ell m}_{3}(a)+C^{\ell m}_{4}(a)b
\end{equation}
never vanishes in the fitting domain. In practice, during the fitting procedure we require
\begin{equation}
C^{\ell m}_{3}(a)+C^{\ell m}_{4}(a)b > 0
\qquad
\forall (b,a) \in \mathcal{D},
\end{equation}
where $\mathcal{D}$ denotes the region of the $(b,a)$ parameter space covered by the dataset. The hierarchical fit of the parameters is performed through a least-square method.

In Fig.~\ref{fig: c1f example residuals} we illustrate an example of the hierarchical fit for one of the parameters entering the MR ansatz, i.e. $c^{22}_{1,f}$ . In the top panel we show the values of the coefficient obtained from the individual waveform fits (coloured dots) together with the corresponding hierarchical fit (coloured solid lines) as a function of the parameters $(b,a)$. The dependence on the parameter $b$ is in the x-axis, while the dependence on the spin $a$ is made explicit through different colors. The bottom panel displays the fractional residuals (in percentage) between the fitted model and the coefficients obtained from the waveform best fits. 
Another parameter which enters our model and that requires a fit is the quantity $\Delta t^{\ell \ell}_{\rm QC}$ defined in Eq.~\eqref{eq:t_match}. Figure~\ref{fig: deltat_ll_QC} shows the behaviour of this time offset as a function of the spin $a$ for all the non-diagonal modes $(2,1)$, $(3,2)$, and $(4,3)$. This quantity represents the time shift, measured in the QC case, between the peak of the non-diagonal $(\ell,m)$ mode and the peak of the corresponding $(\ell, \ell)$ mode. This offset is used to define the attachment of the MR model for the non-diagonal modes.

The smooth behaviour of $\Delta t^{\ell \ell}_{\rm QC}$ as a function of $a$ supports the use of a simple phenomenological fit to model this quantity. We employ a 4th order polynomial.
Finally, we also provide some examples of the fits of the quantity $\gamma$ entering the activation function defined in Eq.~\eqref{eq.: def activation function}
Figure~\ref{fig: gamma_exp_fit} shows the behaviour of $\gamma$ as a function of $a$ for the modes $(2,2)$, $(3,3)$, and $(2,1)$.
Across the explored parameter space, $\gamma$ exhibits a smooth dependence on the spin, with a clear trend that varies among the different modes. In particular, the magnitude of $\gamma$ tends to increase as the system moves toward highly retrograde spin configurations, reflecting the growing importance of tuning this parameter in these regimes. 

The regular behaviour of $\gamma(a)$ suggests that it can be accurately modelled using a simple phenomenological prescription. In this work, we adopt an exponential ansatz of the form $\gamma(a) = A e^{B a}$, which provides a good representation of the numerical data across the full spin range and for all the considered modes (we are not showing the other modes' result explicitly). This choice allows us to capture the main trends of the parameter while keeping the model as simple as possible.
The explicit expressions of all the fitted coefficients entering the MR model are provided in the Supplemental Material.

\section{Interference between QNMs and Price tails at high eccentricity} \label{Appendix: interference between QNMs and Price tails at high eccentricity}

In this Appendix we illustrate how the contribution of the late-time Price tails affects the ringdown signal as the orbital eccentricity increases. As mentioned in Sec.~\ref{sec.:Anatomy of the ringdown}, at sufficiently late times after the peak of the modes, the ringdown is no longer dominated by the QNMs, but instead transitions to a power-law decay produced by backscattering off the curved spacetime, the so-called Price tails~\cite{Price:1972pw}. 
Figure~\ref{fig: Price tails effects of eccentricity} shows the waveform amplitudes for different values of the eccentricity at the last stable orbit as in Fig.~\ref{fig:effects of eccentricity}. The amplitudes are plotted on a logarithmic scale in order to clearly highlight the regime in which the power-law tail becomes comparable to the QNM contribution. As the eccentricity increases, the time at which the QNM-dominated signal and the Price-tail contribution start to interfere occurs progressively earlier in time in the ringdown.
This behaviour explains the features discussed in Sec.~\ref{sec.:impact of eccentricity on the merger-ringdown}, where we observed the appearance of irregular oscillations in the instantaneous frequency for highly eccentric configurations ($e_{\rm LSO} \ge 0.7$). In particular, for $e_{\rm LSO}=0.90$ these oscillations start to appear already around $t-t_{\rm peak}\sim 80$. As shown in Fig.~\ref{fig: Price tails effects of eccentricity}, this corresponds precisely to the time when the exponentially damped QNM contribution becomes comparable to the emerging power-law tail. The resulting interference between these two components produces the modulations observed in the instantaneous frequency.
These results corroborate the interpretation discussed in Sec.~\ref{sec.:impact of eccentricity on the merger-ringdown}: the oscillatory behaviour observed at late times for highly eccentric configurations is not associated with additional QNM contributions, but rather with the onset of the Price-tail regime, which, at fixes time after the peak of the mode, becomes increasingly significant as the eccentricity grows.

\section{$(2,1)$ mode: the prograde high spin and high eccentricity case} \label{Appendix: (2,1) mode high spin}
In this Appendix we discuss the peculiar behaviour observed in the $(2,1)$ mode for highly prograde configurations with large eccentricity. In particular, for $a=0.9$ and $e_{\rm LSO}\gtrsim0.5$ we find that, close to the peak, the $(2,1)$ mode exhibits a phenomenology that differs from the standard behaviour typically observed for positive-spin configurations. As shown in Fig.~\ref{fig: h21 high prograde spins}, when using the trajectories generated employing the RR force considered in this work (Eq.~\eqref{eq:full RR force}), the waveform develops additional structures near the last peak of the $(2,1)$ mode as the eccentricity increases.
Since these configurations correspond to orbital motion of the TM that probes strong-field regions of the spacetime, one possible concern is that the observed behaviour could be influenced by inaccuracies in the RR force used to generate the trajectories in these regimes. In the present model the RR force is constructed from resummed PN fluxes, whose accuracy is known to degrade when the TM approaches the event horizon. This motivates the question of whether the features observed near the peak of the $(2,1)$ mode could be a consequence of inaccuracies in the RR prescription.

To determine whether the observed behaviour originates from the RR force, we perform an experiment in which the effect of the RR force is removed. Specifically, we generate a set of bound direct geodesic plunges using the Black Hole Perturbation Toolkit~\cite{BHPToolkit}. 
Each configuration is chosen such that the energy of the direct plunge corresponds to that of the bound geodesic configuration associated with the same $e_{\rm LSO}$ considered for the RR-force-driven trajectories, with an additional offset $\Delta \mathcal{E}_{\rm UCO}$ chosen to match the maximum value of $\Delta \mathcal{E}_{\rm UCO}$ reached at the onset of the plunge by the corresponding RR-force-driven inspiral shown in the left panel of Fig.~\ref{fig: h21 high prograde spins}. This choice reproduces the energy excess above the UCO after the LSO crossing for the inspirals considered in this work, providing a physically motivated proxy for the final stage of the EOB evolution before plunge.

The waveforms produced through the bound direct plunges are shown in the second panel of Fig.~\ref{fig: h21 high prograde spins}. We find that the qualitative features observed near the last peak of the $(2,1)$ mode remain essentially unchanged. In particular, the waveform still displays the same non-trivial structure close to the peak, despite the absence of the RR force in the construction of the trajectory. This indicates that the phenomenon is not caused by the specific RR prescription adopted in our model, although the RR dynamics can affect the precise quantitative details of the waveform.
These results suggest that the behaviour of the $(2,1)$ mode for highly prograde and eccentric configurations is instead associated with the strong-field dynamics of the TM before plunging and with the possible excitation of different QNM contributions in this regime that affect the features of the last peak of the mode. We also note that the waveforms we show in Fig.~\ref{fig: h21 high prograde spins} exhibit oscillatory features ("wiggles") for times $-70 \le (t - t^{21}_{\rm match} ) \le -40$, which arise from the excitation of QNMs during the last pericenter passage of the TM. These excitations are physical, as discussed in Refs.~\cite{Thornburg:2019ukt,Rifat:2019fkt}.

\bibliography{references.bib}

\end{document}